\documentclass[twocolumn,prd,aps,superscriptaddress,preprintnumbers,showpacs,nofootinbib,eqsecnum,amsfonts,amsmath]{revtex4-1}
\usepackage{placeins}
\usepackage{color}
\usepackage{calc}
\usepackage{amsmath,amssymb,graphicx,mathrsfs}
\usepackage{tensor}
\usepackage{bm}
\usepackage[varg]{txfonts}
\usepackage[colorlinks, pdfborder={0 0 0}]{hyperref}
\usepackage{float}
\usepackage{dcolumn}
\usepackage{physics}
\usepackage[nolist,nohyperlinks]{acronym}
\usepackage{xspace}
\usepackage[abs]{overpic}
\usepackage{pict2e}
\usepackage{enumitem}
\usepackage[usenames,dvipsnames]{xcolor}
\usepackage[utf8]{inputenc}
\usepackage{acronym}
\usepackage{gensymb}
\usepackage{romannum}
\usepackage{subfigure}
\usepackage{cleveref} 
\usepackage[normalem]{ulem}
\usepackage{longtable}
\usepackage{mathtools}

\newcommand{\AEI}{\affiliation{Max Planck Institute for Gravitational Physics (Albert Einstein Institute), Am M\"uhlenberg 1, Potsdam 14476, Germany}}
\newcommand{\Maryland}{\affiliation{Department of Physics, University of Maryland, College Park, MD 20742, USA}}

\def\be{\begin{equation}}
\def\ee{\end{equation}}
\def\bea{\begin{eqnarray}}
\def\eea{\end{eqnarray}}
\newcommand{\bes}{\begin{subequations}}
\newcommand{\ees}{\end{subequations}}

\newcommand{\doubleline}{\hline \hline}

\allowdisplaybreaks

\begin{document}

\pagenumbering{arabic} 

\title{Enriching the Symphony of Gravitational Waves from Binary Black Holes \newline by Tuning Higher Harmonics}

%\title{Inspiral-merger-ringdown waveform model of spinning, nonprecessing binary black holes \ab{with higher harmonics} 
%in the effective-one-body formalism}

\author{Roberto Cotesta}
\email{roberto.cotesta@aei.mpg.de}
\AEI

\author{Alessandra Buonanno}
\AEI \Maryland

\author{Alejandro Boh\'e}
\AEI

\author{Andrea Taracchini}
\AEI

\author{Ian Hinder}
\AEI

\author{Serguei Ossokine}
\AEI

%\date{\today}

\begin{abstract}
  For the first time, we construct an inspiral-merger-ringdown waveform model within the
  effective-one-body formalism for spinning, nonprecessing binary
  black holes that includes gravitational modes beyond the dominant 
$(\ell,|m|) = (2,2)$ mode, specifically $(\ell,|m|)=(2,1),(3,3),(4,4),(5,5)$. 
Our multipolar waveform model incorporates 
  recent (resummed) post-Newtonian results for the inspiral and
  information from 157 numerical-relativity simulations, and 13
  waveforms from black-hole perturbation theory for the
  (plunge-)merger and ringdown. We quantify the improvement in accuracy when including 
higher-order modes by computing the faithfulness of the waveform model against the numerical-relativity 
waveforms used to construct the model. We define the faithfulness as the match maximized over 
time, phase of arrival, gravitational-wave polarization and sky position of the waveform model,
and averaged over binary orientation, gravitational-wave
  polarization and sky position of the numerical-relativity waveform. When the waveform
  model contains only the $(2,2)$ mode, we find that the averaged
  faithfulness to numerical-relativity waveforms containing all modes
  with $\ell \leq$ 5 ranges from 90\% to 99.9\% for binaries with
  total mass $20-200 M_\odot$ (using the Advanced LIGO's design noise
  curve). By contrast, when the $(2,1),(3,3),(4,4),(5,5)$ modes are also included in
  the model, the faithfulness improves to 99\% for all but four
 configurations in the numerical-relativity catalog, for which the
  faithfulness is greater than 98.5\%. Starting from the complete inspiral-merger-ringdown model, we develop also a (stand-alone) 
waveform model for the merger-ringdown signal, calibrated to numerical-relativity waveforms, which 
can be used to measure multiple quasi-normal modes.
The multipolar waveform model can be extended to include spin-precessional effects, and will be employed 
in upcoming observing runs of Advanced LIGO and Virgo.
\end{abstract}

\maketitle

\section{Introduction}
\label{sec:Intro}

The Advanced LIGO detectors \cite{TheLIGOScientific:2014jea} have reported, so far, 
the observation of five gravitational-wave (GW) signals from coalescing binary
black holes (BBHs): GW150914 \cite{Abbott:2016blz}, GW151226
\cite{Abbott:2016nmj}, GW170104 \cite{Abbott:2017vtc}, GW170608 \cite{Abbott:2017gyy}, GW170814
\cite{Abbott:2017oio} (observed also by the Virgo detector
\cite{TheVirgo:2014hva}), and one GW signal from a coalescing binary
neutron star (BNS) \cite{TheLIGOScientific:2017qsa}. 
The modeled search for GWs from binary systems and the extraction of binary parameters, 
such as the masses and spins, are based on the matched-filtering technique~\cite{Nitz:2017svb,Veitch:2014wba,Usman:2015kfa,Canton:2014ena,Cannon:2011vi,Cannon:2012zt}, 
which requires accurate knowledge of the waveform of the incoming signal. 
During the first two observing runs (O1 and O2), the Advanced LIGO and Virgo 
modeled-search pipelines employed, for binary total masses below $4 M_\odot$, templates~\cite{Sathyaprakash:1991mt} 
built within the post-Newtonian (PN) approach~\cite{Arun:2008kb,Buonanno:2009zt,Mishra:2016whh,Blanchet:2013haa}, 
and, for binary total masses in the range $4\mbox{--} 200 M_\odot$, templates developed using the effective-one-body (EOB) formalism calibrated 
to numerical-relativity (NR) simulations~\cite{Buonanno:1998gg,Buonanno:2000ef,Damour:2008qf,Damour:2009kr,Barausse:2009xi,Taracchini:2013rva,Purrer:2015tud,Bohe:2016gbl}. 
For parameter-estimation analyses~\cite{TheLIGOScientific:2016pea,TheLIGOScientific:2016wfe,TheLIGOScientific:2017qsa,Veitch:2014wba} and tests of General Relativity (GR) ~\cite{TheLIGOScientific:2016src}, PN~\cite{Arun:2008kb,Buonanno:2009zt,Mishra:2016whh}, 
EOBNR~\cite{Pan:2013rra,Taracchini:2013rva,Bohe:2016gbl,Babak:2016tgq} and 
also inspiral-merger-ringdown phenomenological (IMRPhenom) waveform models~\cite{Husa:2015iqa,Khan:2015jqa,Hannam:2013oca} were used.

The -2 spin-weighted spherical harmonics comprise a convenient basis into which one can decompose the two polarizations of GWs.
The spinning, nonprecessing EOBNR waveform model~\cite{Bohe:2016gbl}  employed in searches and parameter-estimation studies during the O2 run 
(henceforth, \texttt{SEOBNRv4} model),  
only used the dominant $(\ell,|m|) = (2,2)$ mode to build the gravitational polarizations.
This approximation was accurate enough for detecting and inferring astrophysical information of the sources observed during O2 (and also O1), as discussed in Refs.~\cite{Littenberg:2012uj,Brown:2012nn,Capano:2013raa,Harry:2017weg,Varma:2014jxa,Graff:2015bba,Varma:2016dnf,Bustillo:2016gid,Abbott:2016wiq}.

Because of the expected increase in sensitivity during the third observing run (O3), which is planned to start in the Fall of 2018, 
some GW signals are expected to have 
much larger signal-to-noise ratio (SNR) with respect to the past, and may lie in regions of parameter space so far unexplored (e.g., 
more massive and/or higher mass-ratio systems than observed in O1 and O2). This poses an excellent 
opportunity to improve our knowledge of astrophysical and gravitational properties of the sources, but it also requires more accurate 
waveform models to be able to take full advantage of the discovery and inference potential.
More accurate waveform models would be
useful, as well, from the detection point of view to further increase the effective volume reached by the search, in
particular for regions of the parameter space where the approximation of restricting to the 
(2,2) mode starts to degrade~\cite{Brown:2012nn,Capano:2013raa,Harry:2017weg}. Following these motivations, we build here an improved version of the 
\texttt{SEOBNRv4} waveform model that includes the modes $(\ell,|m|) = (2,1),(3,3),(4,4),(5,5)$ beyond the dominant $(2,2)$ mode (henceforth, 
\texttt{SEOBNRv4HM} model). 
Similar work was done for the nonspinning case for the EOBNR waveform model of Ref.~\cite{Pan:2011gk} (henceforth, \texttt{EOBNRv2HM} model), and for the nonspinning and spinning, nonprecessing \texttt{IMRPhenom} models in Refs.~\cite{Mehta:2017jpq, London:2017bcn}. 

In building the \texttt{SEOBNRv4HM} model we incorporate new informations from PN calculations~\cite{Marsatetal2017,Fujita:2012cm}, 
from NR simulations (produced with the (pseudo) Spectral Einstein code (\texttt{SpEC}) \cite{Chu:2015kft} of the Simulating
eXtreme Spacetimes (\texttt{SXS}) project and the \texttt{Einstein Toolkit} code \cite{Zilhao:2013hia,Loffler:2011ay}), and also from merger-ringdown 
waveforms computed in BH perturbation theory solving the Teukolsky equation~\cite{Barausse:2011kb,Taracchini:2014zpa}.  
The NR waveforms are described in Refs.~\cite{Mroue2013,Kumar:2016dhh,Chu:2015kft,Kumar:2015tha,Lovelace:2010ne,Scheel:2014ina,Mroue:2013xna,Bohe:2016gbl}, 
and summarized in Appendix~\ref{sec:NRcatalog}. They were also employed to build the \texttt{SEOBNRv4} waveform model in 
Ref.~\cite{Bohe:2016gbl} 
(see Sec.~\Romannum{3} therein). However, here, we do not use the BAM simulation 
\texttt{BAMq8s85s85} \cite{Bruegmann:2006at,Husa:2007hp}, because the higher-order modes are not available to us. Thus, for the same binary configuration, 
we produce a new NR simulation using the \texttt{Einstein Toolkit} code and extract higher-order modes (henceforth, \texttt{ET:AEI:0004}).

As by product of the \texttt{SEOBNRv4HM} model, we obtain a (stand-alone) merger-ringdown model~\cite{Baker:2008mj,Damour:2014yha,London:2014cma,Nagar:2016iwa,Bohe:2016gbl,London:2018gaq}, tuned to the NR and Teukolsky-equation waveforms, which can be employed to extract multiple quasi-normal modes from GW signals, and test General Relativity~\cite{Dreyer:2003bv,Berti:2005ys,Meidam:2014jpa,Yang:2017zxs}.

The paper is organized as follows. In
  Sec.~\ref{sec:motivations} we use the NR waveforms at our disposal
  to quantify the importance of higher harmonics in presence of
  spins. In Sec.~\ref{sec:faithfulness} we determine, taking also into account 
  the error in NR waveforms, which gravitational modes are
  crucial to achieve at least $\sim 99\%$ accuracy. In Sec.~\ref{sec:eob_formalism} we
  develop the multipolar EOB waveform model, and describe how to
  enhance its performance by including information from NR simulations
  and BH perturbation theory. We also highlight the construction and use 
of the multipolar (stand-alone) merger-ringdown model. 
In Sec.~\ref{sec:comparison} we compare
  the newly developed \texttt{SEOBNRv4HM} model to 157 NR waveforms. In
  Sec.~\ref{sec:concl} we summarize our main conclusions, and outline
  possible future work. Finally, in Appendices \ref{app:modes},
  \ref{app:NQCfits} and \ref{app:ringdownfits} we provide interested
  readers with explicit expressions of all quantities entering the
  higher-order modes of the \texttt{SEOBNRv4HM} model, and point out the 
presence of numerical artifacts in the (4,4) and (5,5) modes of some 
NR simulations. For convenience, we summarize in Appendix~\ref{sec:NRcatalog} the NR
  waveforms used in this paper. In Appendix~\ref{sec:EOBNRv2HM} we
  also compare the model \texttt{SEOBNRv4HM} with the nonspinning
  \texttt{EOBNRv2HM} waveform model, developed in
  2011~\cite{Pan:2011gk}. Finally in Appendix~\ref{sec:time_domain} we
  compare the \texttt{SEOBNRv4HM} model with an NR waveform in time
  domain.

In this paper we adopt the geometric units $G = c = 1$.

\section{Motivations to model higher-order modes for binary black holes}
\label{sec:motivations}

\begin{figure}[h]
  \centering
  \includegraphics[width=0.5\textwidth]{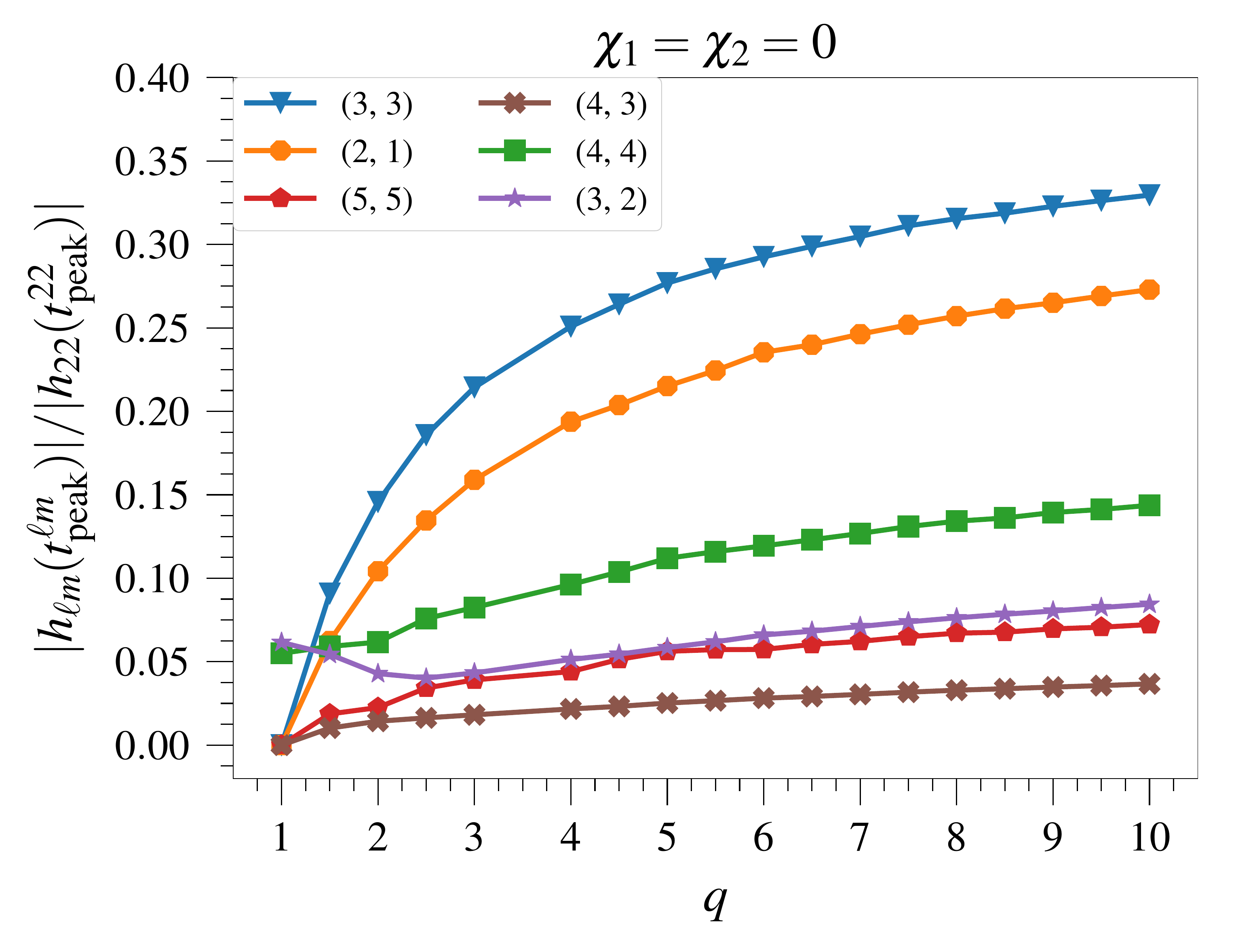}
\caption{Amplitude ratio between the $(\ell,m)$ mode and the dominant $(2,2)$ mode, both 
evaluated at their peak, as function of the mass ratio. 
We use only nonspinning NR waveforms. 
(Note that the markers represent the NR data, and we connect them by a line). We note that the importance of a given higher-order mode with respect to the dominant one is not controlled only by the amplitude ratio between the two, but also by the -2 spin-weighted spherical harmonic associated to the mode (see Eq.~(\ref{eq:spHarmDec})).}
\label{fig:rel_amp_hm_0spin}
\end{figure}

In this section we describe the spherical-mode decomposition of the gravitational polarizations and 
discuss the motivations for building an inspiral-merger-ringdown waveform model (\texttt{SEOBNRv4HM}) 
with higher harmonics for spinning BHs. 

Henceforth, we denote the binary's total mass with $M = m_1 + m_2$, and choose the body's masses $m_1$ and $m_2$ such that the mass ratio $q = m_1/m_2 \geq 1$. Since we consider only spinning, nonprecessing BHs (i.e., 
spins aligned or antialigned with the direction perpendicular to the orbital plane $\mathbf{\hat{L}})$, we only 
have one (dimensionless) spin parameter for each BH, $\chi_{1,2}$, defined as $\mathbf{S_{\mathrm{1,2}}}= 
\chi_{1,2} m_{1,2}^2 \mathbf{\hat{L}}$, where $\mathbf{S_{\mathrm{1,2}}}$ are the BH's spins ($-1 \leq \chi_{1,2} \leq 1$).

The observer-frame's gravitational polarizations read 
\begin{equation}
\label{eq:spHarmDec}
h_+(\iota,\varphi_0;t) - i \ h_x(\iota,\varphi_0;t) = \sum_{\ell=2}^\infty \sum_{m=-\ell}^\ell \tensor[_{-2}]{Y}{_{\ell m}}(\iota,\varphi_0) \ h_{\ell m}(t),
\end{equation}
where we denote with $\iota$ the inclination angle (computed with respect to the direction perpendicular to the 
orbital plane), $\varphi_0$ the azimuthal direction to the observer, and 
$\tensor[_{-2}]{Y}{_{\ell m}}(\iota,\varphi_0)$'s the -2 spin-weighted spherical harmonics. For spinning, nonprecessing BHs' 
we have $h_{\ell m} = (-1)^\ell h^*_{\ell -m}$. Thus, without loss of generality, we 
restrict the discussion to $(\ell,m)$ modes with $m>0$. 

\begin{figure*}
    \centering
    \begin{minipage}{0.5\textwidth}
        \centering
        \includegraphics[width=1.\textwidth]{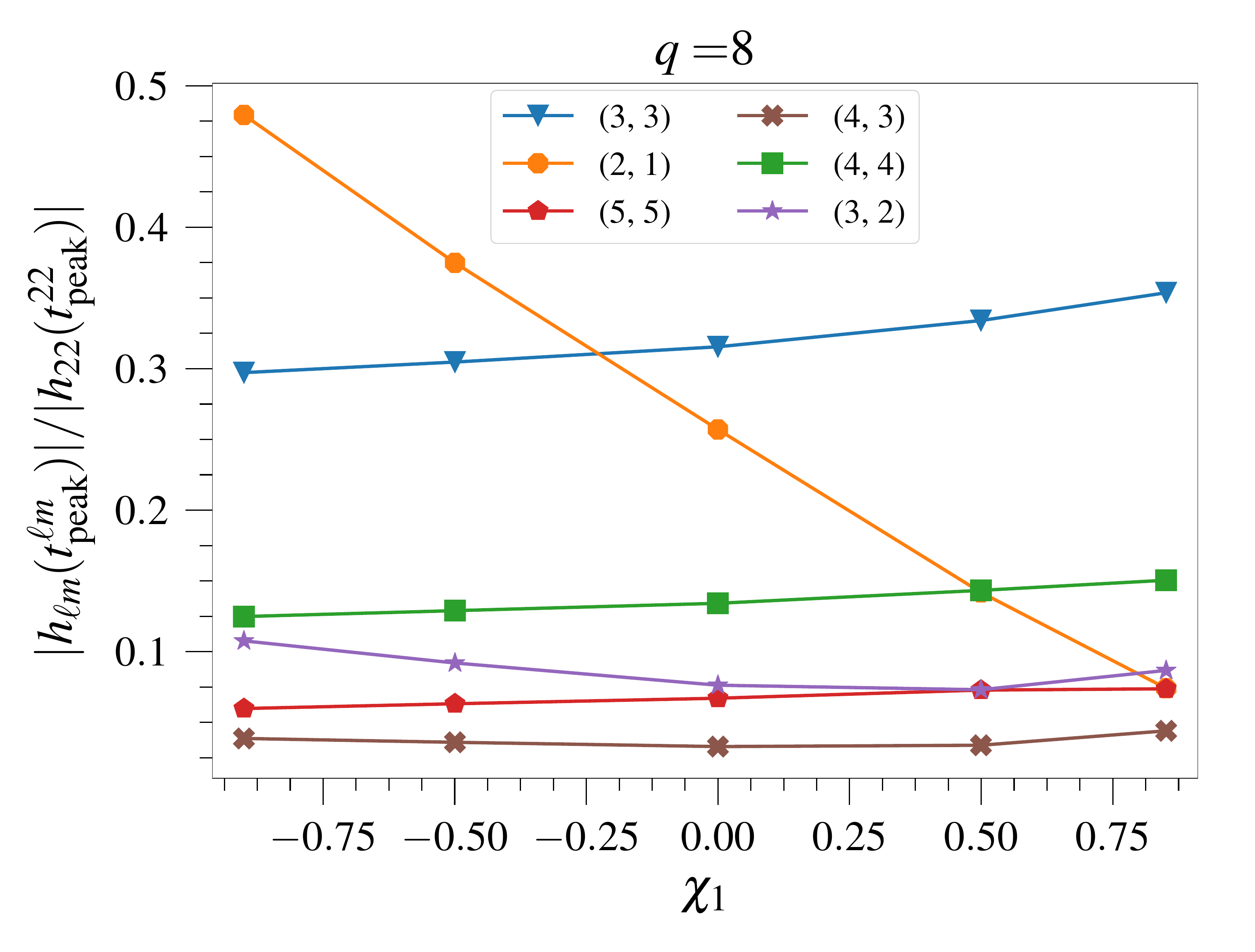} % first figure itself
    \end{minipage}\hfill
    \begin{minipage}{0.5\textwidth}
        \centering
        \includegraphics[width=1.\textwidth]{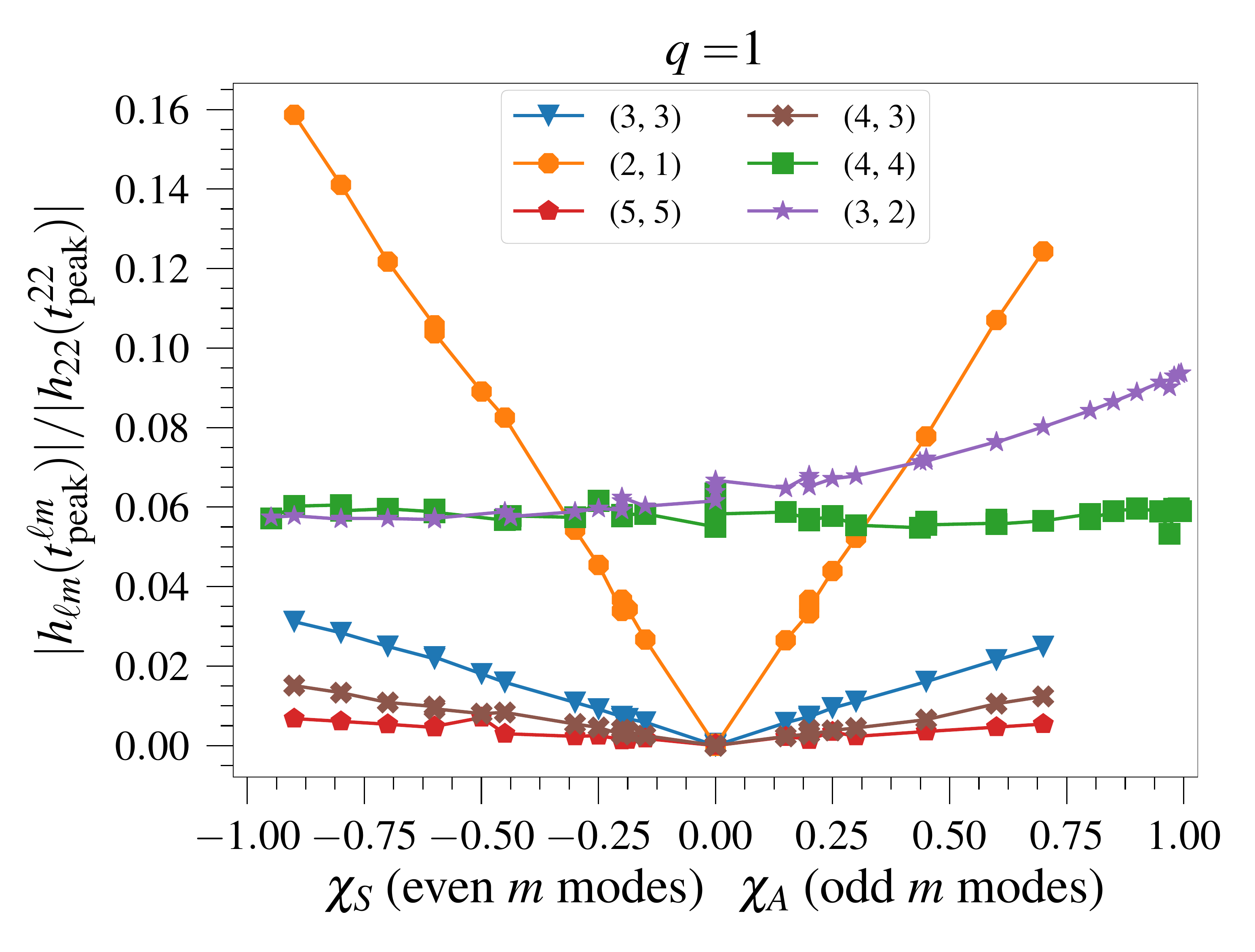} % second figure itself
    \end{minipage}
\caption{Amplitude ratio between the $(\ell,m)$ mode and the dominant $(2,2)$ mode, both 
evaluated at their peak. In the left (right) panel we plot these quantities for mass ratio $q = 8$  versus 
the spin of the heavier BH ($q=1$ versus $\chi_A = (\chi_1-\chi_2)/2$ for modes with odd $m$, and 
$\chi_S = (\chi_1 + \chi_2)/2$ for modes with even $m$). The markers represent the NR data, and we connect them by a line.}
\label{fig:rel_amp_hm_spin}
\end{figure*}

As we shall discuss below, for face-on/face-off binary configurations, 
the dominant mode is the $(\ell, m) = (2,2)$ mode. For generic binary orientations the modes $(\ell, m) 
\neq (2,2)$ could be comparable to the $(2,2)$ mode. Nevertheless, we will loosely refer to $(\ell, m) \neq (2,2)$ as subdominat modes; 
sometime we also refer to them as higher-order modes or higher harmonics, even if they include the $(2,1)$ mode.

Several authors in the literature have investigated the impact of
  neglecting higher-order modes for detection and parameter
  estimation. From the detection perspective, Refs.~\cite{Brown:2012nn,
    Pekowsky:2012sr,Varma:2014jxa,Capano:2013raa} showed that neglecting 
  higher-order modes in nonspinning gravitational waveforms can cause a loss in detection volume bigger
  than $10 \%$ when the mass ratio $q \geq 4$ and total mass $M \geq
  100 M_\odot$. To overcome this issue, Ref.~\cite{Harry:2017weg} suggested 
  a new method to search for GW signals with templates that include higher modes,  
  increasing the search sensitivity up to a factor of 2 in volume 
for high mass-ratio, and high total-mass binaries. While those works consider only nonspinning systems, 
the authors of Ref.~\cite{Bustillo:2015qty} show that for spinning
  systems, the loss in detection volume due to neglecting 
  higher-order modes is smaller with respect to the nonspinning
  case. This happens because the spin parameters provide an additional degree
  of freedom that templates with only the dominant $(2,2)$ mode can employ
  to better match signals containing higher-order modes.

From the parameter-estimation perspective, as discussed in
Ref.~\cite{Varma:2014jxa}, for nonspinning systems with mass ratio $q
\geq 4$ and total masses $M \geq 150 M_\odot$ the systematic error due
to neglecting higher-order modes is larger than the $1\sigma$
statistical error for signals with signal-to-noise ratio (SNR) of
8. Signals with a larger SNR yield smaller statistical errors and, the
constraints discussed before become more stringent~\cite{Littenberg:2012uj}. Indeed even for equal-mass systems,
where the higher-order modes are expected to be negligible, if the
signal has an SNR of 48, the systematic error from neglecting 
higher-order modes can be bigger than the statistical error~\cite{Littenberg:2012uj}. 
(The SNRs above refer to Advanced LIGO’s “zero-detuned
high-power” design sensitivity curve~\cite{Shoemaker:2010}).

Here we briefly review known results, and highlight some features that will be exploited below when  
building the \texttt{SEOBNRv4HM} waveform model.

In Fig.~\ref{fig:rel_amp_hm_0spin} we show the ratio between the largest subdominant $(\ell,m)$ modes and the 
$(2,2)$ mode amplitudes, evaluated at their peak, $t_{\mathrm{peak}}^{\ell m}$ and $t_{\mathrm{peak}}^{22}$, respectively,  
as function of mass ratio for all the nonspinning waveforms in our NR catalog. We note that the well-known 
mode hierarchy $(\ell, m) = (2,2), (3,3),(2,1),(4,4),(3,2),(5,5),(4,3)$ changes when approaching the
equal-mass (equal-spin) limit (see, e.g., Ref.~\cite{Healy:2013jza}). 
Indeed, in this limit all modes with odd $m$ have to vanish  in order to enforce 
the binary's symmetry under rotation $\varphi_0 \rightarrow \varphi_0 + \pi$. Thus, when 
$\nu \rightarrow 1/4$ ($\chi_{1}=\chi_2$), the $(3,2)$ and $(4,4)$ modes become the 
most important subdominant modes.
In Fig.~\ref{fig:rel_amp_hm_spin} we show how the modes' hierarchy in 
the nonspinning case (see Fig.~\ref{fig:rel_amp_hm_0spin}) changes when BH's spins are included. 
In particular, in the left panel of Fig.~\ref{fig:rel_amp_hm_spin} we fix the mass ratio to $q = 8$ and
plot the relative amplitude of the modes as function of the spin of the more massive BH. Note that for $q = 8$ 
all NR waveforms in our catalog (with the exception of \texttt{ET:AEI:0004}, $q = 8, \, \chi_1 = \chi_2 = 0.85$)  
have the spin only on the more massive BH.  We see that the relative amplitude of 
the modes $(3,3),(4,4),(3,2),(5,5),(4,3)$ depends weakly on the spins, except for the $(2,1)$ mode. Indeed, 
for $\chi_1 \gtrsim 0.5$, the $(2,1)$ mode becomes smaller than the $(4,4)$ mode and
for $\chi_1 \gtrsim 0.75$ is as small as the modes $(3,2),(5,5)$. On the other side, 
for $\chi_1 \lesssim -0.25$ the mode $(2,1)$ is larger than the $(3,3)$
mode.  We find that for smaller mass ratios the effect of $\chi_2$ (i.e., the spin of the
lighter BH), becomes more important. In particular, for a fixed value
of $\chi_1$ the amplitude ratio $|h_{\ell m}(t_{\mathrm{peak}}^{\ell
  m})|/|h_{2 2}(t_{\mathrm{peak}}^{2 2})|$ for the modes $(3,3),(4,4),(5,5)$ 
decreases with increasing $\chi_2$, while the ratio increases for the modes $(2,1),(3,2),(4,3)$.

The special case of equal-mass systems, $q = 1$, is discussed in the right panel of 
Fig.~\ref{fig:rel_amp_hm_spin}. Here we show the
amplitude ratio between the $(\ell,m)$ mode and the dominant $(2,2)$ mode, 
both evaluated at their peak, as function of $\chi_A
= (\chi_1-\chi_2)/2$ for modes with odd $m$ and as function of
$\chi_S = (\chi_1 + \chi_2)/2$ for modes with even $m$. As discussed
before, the modes with odd $m$ vanish for equal-mass, equal-spins
configurations ($\chi_A = 0$) from symmetry arguments and, the
amplitude ratio grows proportionally to $|\chi_A|$ for these modes. In
particular, we note that in this case the $(2,1)$ mode
behaves differently from the other modes, undergoing a much more significant
growth in the amplitude ratio. Regarding the modes with even $m$, we notice 
that whereas the $(4,4)$ mode is nearly
constant as function of $\chi_S$ in the spin range considered, the 
$(3,2)$ mode increases as a function of $\chi_S$ in the same range.
The amplitude of the $(2,1)$ mode has a stronger dependence 
on the spins with respect to the other modes because in its PN expansion the spin term enters at a lower
relative order (see Eqs. (38a)--(38i) in Ref.~\cite{Pan:2010hz}). A similar spin-dependence 
was found in Ref.~\cite{Kamaretsos:2012bs} for the amplitudes ratio $(A_{\ell m}/A_{2 2})$ of the quasi-normal mode oscillations.

  Finally, it is worth emphasizing that in understanding the 
  relevance of subdominant modes for the observer, it is important
  to take into account the -2 spin-weighted spherical-harmonic factor
  $\tensor[_{-2}]{Y}{_{\ell m}}(\iota,\varphi_0)$ that enters
  Eq.~\eqref{eq:spHarmDec}, notably its dependence on the angles
  $(\iota,\varphi_0)$. Indeed, the -2 spin-weighted spherical harmonic
  associated to the dominant mode starts from a maximum in the face-on
  orientation ($\iota = 0$) and decreases to a minimum at edge-on
  ($\iota = \pi/2$). On the other hand, the spherical harmonics favour the
  higher-order modes with respect to the dominant one in orientations
  close to edge-on where ${ |\tensor[_{-2}]{Y}{_{\ell m}}(\iota
    \rightarrow \pi/2)|/|\tensor[_{-2}]{Y}{_{2 2}}(\iota \rightarrow
    \pi/2)| > 1}$.  Furthermore, a direct inspection of the harmonic factor shows that 
  the modes $(3,2),(4,3)$ are suppressed (i.e., ${ |\tensor[_{-2}]{Y}{_{\ell
        m}}(\iota)|/|\tensor[_{-2}]{Y}{_{2 2}}(\iota)| < 1}$) for a larger region
  in $\iota$ than for the modes $(3,3),(2,1),(4,4),(5,5)$. For this reason the contribution of the
  former to the gravitational polarizations is limited to a smaller number of orientations with respect
  to the latter.

\section{Selecting the most-important higher-order modes for modeling}
\label{sec:faithfulness}

\begin{figure*}
    \centering
    \begin{minipage}{0.5\textwidth}
        \centering
        \includegraphics[width=1.0\textwidth]{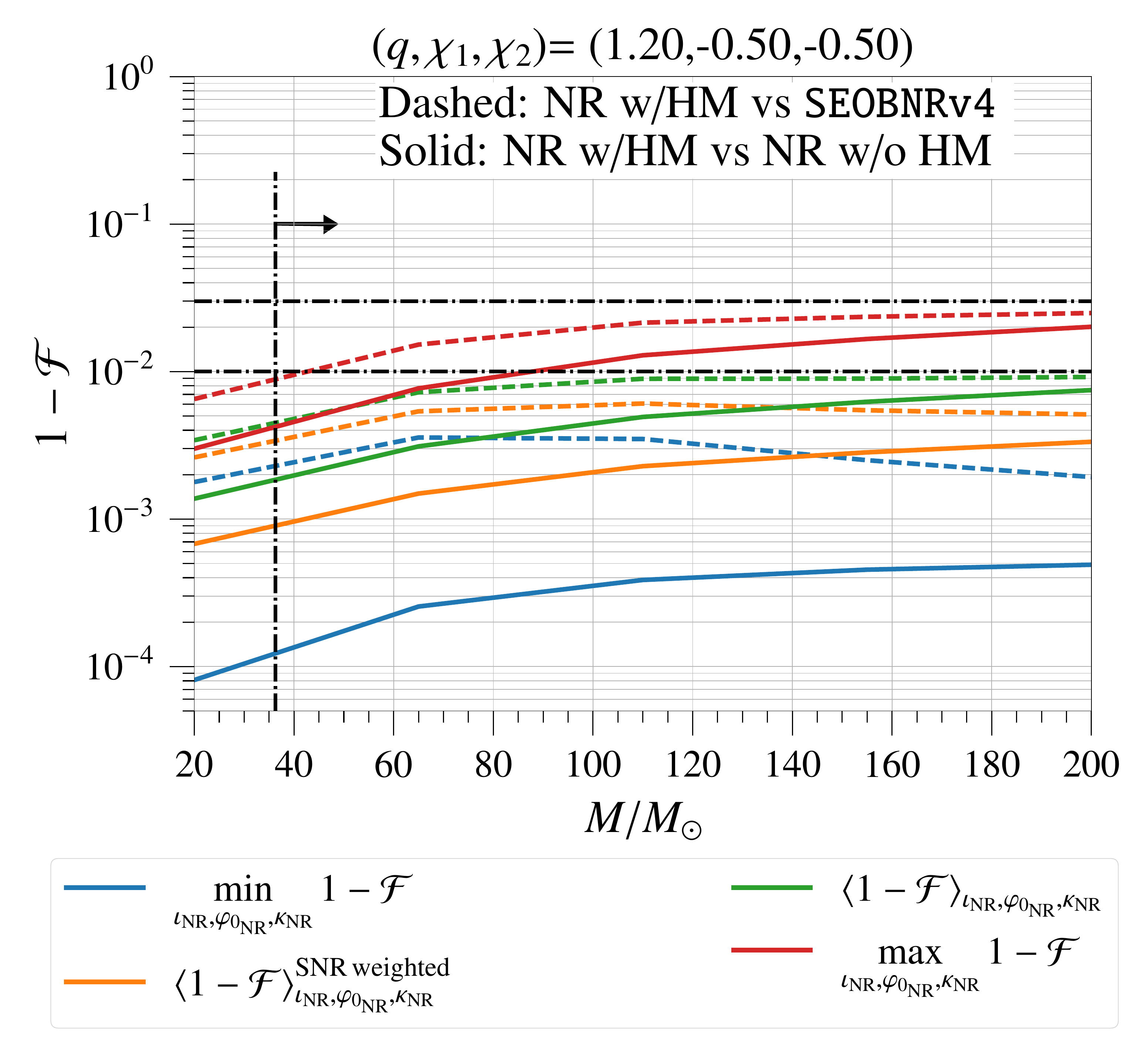} % first figure itself
    \end{minipage}\hfill
    \begin{minipage}{0.5\textwidth}
        \centering
        \includegraphics[width=1.0\textwidth]{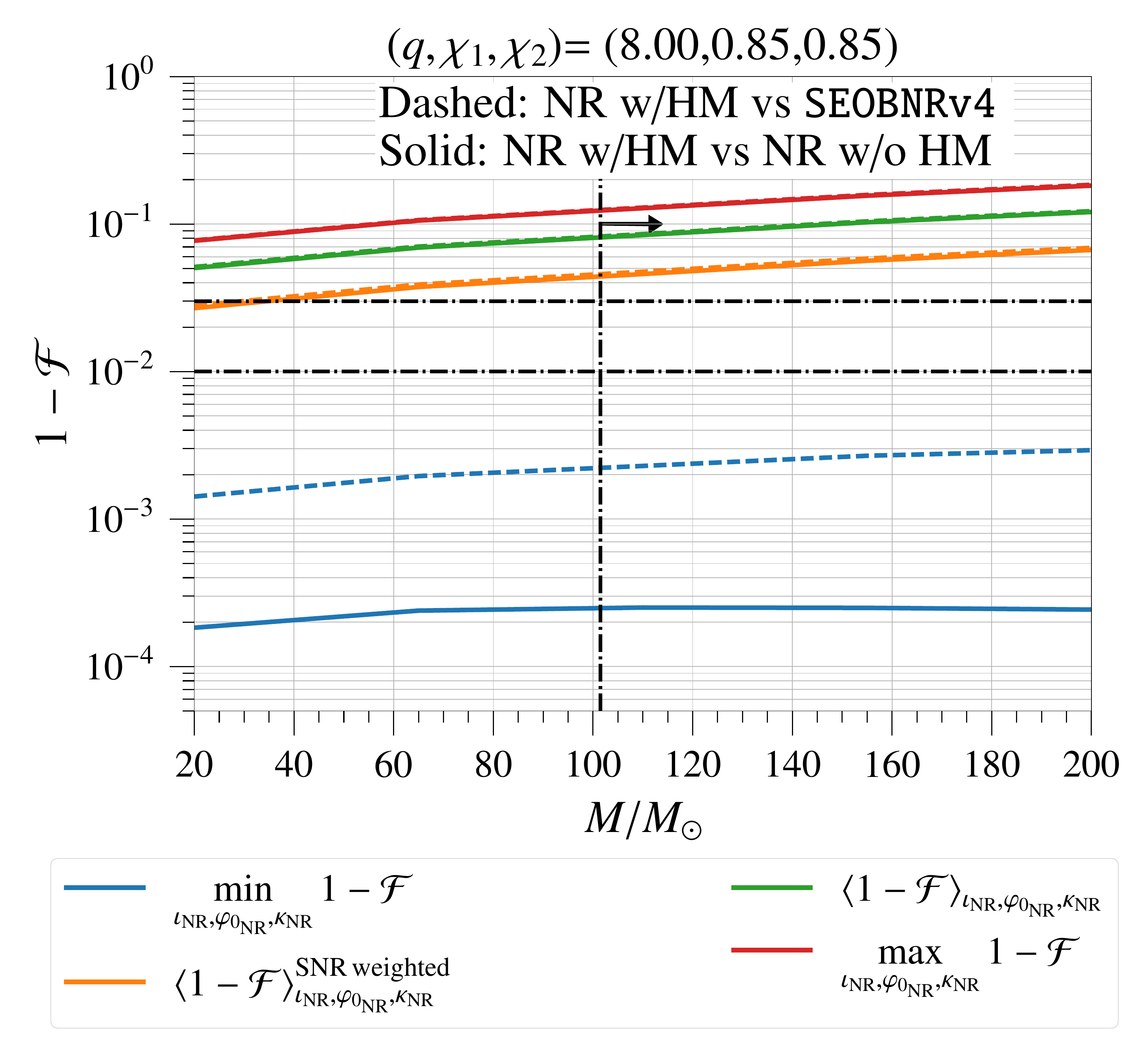} % second figure itself
    \end{minipage}
\caption{Unfaithfulness $(1-\mathcal{F})$ for the configurations $(q = 1.2, \,\chi_1 = -0.5,\,\chi_2=-0.5)$ (left panel) and $(q = 8, \,\chi_1 = 0.85,\,\chi_2= 0.85)$ (right panel)  in the mass range $20 M_\odot \leq M \leq 200 M_\odot$. In dashed the results for the \texttt{SEOBNRv4} model and in solid the results for the NR waveform containing only the dominant mode, both against the NR waveform with the modes $(\ell \leq 5, \, m \neq 0)$. The minimum of the unfaithfulness (blue curves) correspond to a face-on orientation.      We also show the unfaithfulness averaged over the three angles $\iota_{\textrm{NR}},{\phi_0}_{\textrm{NR}},\kappa_{\textrm{NR}}$ (green curves) and weighted by the cube of the SNR (orange curves). Finally the minimum of the unfaithfulness (red curves) which in practice correspond to edge-on and minimized over the other two angles. The vertical dotted-dashed black line is the smallest mass for which the $(\ell, m) = (2,1)$ mode is entirerly in the Advanced LIGO band. The $(\ell, |m'|)$ mode is entirerly in the Advanced LIGO band starting from a mass $m'$ times the mass associated with the $(\ell, m) = (2,1)$ mode.  The horizontal dotted-dashed black lines represent the values of $1\%$ and $3\%$ unfaithfulness.}
\label{fig:unfaith_mass}
\end{figure*}  

In this section we first introduce the faithfulness function as a tool to assess the closeness of two waveforms when higher-order modes 
are included. Then, we use it to estimate how many gravitational modes we need to model in order not to loose 
more than $10\%$ in event rates when rectricting to the binary's configurations in the NR catalog 
at our disposal. We also determine the loss in faithfulness of the NR waveforms due to numerical error.  

The GW signal measured from a spinning, nonprecessing and noneccentric 
BBH is characterized by 11 parameters, namely the masses of the two bodies
$m_{1}$ and $m_{2}$, the (constant) projection of the spins in the
direction perpendicular to the orbital plane, $\chi_{1}$ and $\chi_{2}$,
the angular position of the line of sight measured in the source's
frame $(\iota, \varphi_0)$ (see Eq. \eqref{eq:spHarmDec}), the sky
location of the source in the detector frame $(\theta, \phi)$, the
polarization angle $\psi$, the luminosity distance of the source
$D_{\mathrm{L}}$ and the time of arrival $t_{\mathrm{c}}$. 
The signal measured by the detector takes the form:
\begin{align}
\label{eq:det_strain}
h \equiv & F_+(\theta,\phi,\psi) \ h_+(\iota,\varphi_0, D_{\mathrm{L}}, \boldsymbol{\xi},t_{\mathrm{c}};t) \nonumber \\
&+ F_\times(\theta,\phi,\psi)\ h_\times(\iota,\varphi_0, D_{\mathrm{L}}, \boldsymbol{\xi},t_\mathrm{c};t)\,,
\end{align}
where for convenience we introduce $\boldsymbol{\xi} \equiv (m_{1}, m_{2}, \chi_{1}, \chi_{2})$.
The functions $F_+(\theta,\phi,\psi)$ and $F_\times(\theta,\phi,\psi)$ are the antenna patterns~\cite{Sathyaprakash:1991mt,Finn:1992xs}:
\begin{align}
F_+(\theta,\phi,\psi) &= \frac{1+ \cos^2(\theta)}{2} \ \cos(2\phi) \ \cos(2\psi)+\\ \nonumber
& -\cos(\theta) \ \sin(2\phi)\ \sin(2\psi),\\ 
F_\times(\theta,\phi,\psi) &= \frac{1+ \cos^2(\theta)}{2} \ \cos(2\phi) \ \sin(2\psi)+\\  \nonumber 
&+ \cos(\theta) \ \sin(2\phi)\ \cos(2\psi).
\end{align}
Equation \eqref{eq:det_strain} can be rewritten as:
\begin{align}
h \equiv & \mathcal{A}(\theta,\phi)\big[\cos\kappa(\theta,\phi,\psi) \ h_+(\iota, \varphi_0, D_{\mathrm{L}}, \boldsymbol{\xi}, t_{\mathrm{c}};t) \nonumber \\
&+ \sin\kappa(\theta,\phi,\psi) \ h_\times (\iota, \phi, D_{\mathrm{L}}, \boldsymbol{\xi},t_{\mathrm{c}};t) \big],
\end{align}
where $\kappa(\theta,\phi,\psi)$ is the \textit{effective polarization}~\cite{Capano:2013raa} defined in the region $[0, 2\pi)$ as:
\begin{equation}
e^{i \kappa(\theta,\phi,\psi)} = \frac{F_+(\theta,\phi,\psi) + i F_\times(\theta,\phi,\psi)}{\sqrt{F_+^2(\theta,\phi,\psi) + F_\times^2(\theta,\phi,\psi)}},
\end{equation}
while $\mathcal{A}(\theta,\phi)$ reads:
\begin{equation}
\mathcal{A}(\theta,\phi) = \sqrt{F_+^2(\theta,\phi,\psi) + F_\times^2(\theta,\phi,\psi)}\,.
\end{equation}
We stress that $\mathcal{A}(\theta,\phi)$ does not depend on $\psi$ despite the fact $F_+$ and $F_\times$ depend on it.
Henceforth, to simplify the notation we suppress the dependence of $\kappa$ on $(\theta,\phi,\psi)$.  
Given a GW signal $h_{\mathrm{s}}$ and a template waveform $h_{\mathrm{t}}$, we define the faithfulness as~\cite{Capano:2013raa,Harry:2016ijz}
\begin{equation}
\label{eq:faith}
\mathcal{F}(\iota_{\textrm{s}},{\varphi_0}_{\textrm{s}},\kappa_{\textrm{s}}) \equiv  \max_{t_c, {\varphi_0}_{\mathrm{t}}, \kappa_{\textrm{t}}} \left[\left . \frac{ \left( h_{\mathrm{s}},\,h_{\mathrm{t}} \right)}{\sqrt{ \left( h_{\mathrm{s}},\,h_{\mathrm{s}} \right) \left( h_{\mathrm{t}},\,h_{\mathrm{t}} \right)}}\right \vert_{\substack{\iota_{\mathrm{s}} = \iota_{\mathrm{t}} \\\boldsymbol{\xi}_{\mathrm{s}} = \boldsymbol{\xi}_{\mathrm{t}}}} \right ],
\end{equation}
where parameters with the subscript ``s'' (``t'') refer to the signal (template) waveform. 
The inner product is defined as~\cite{Sathyaprakash:1991mt,Finn:1992xs}:
\begin{equation}
\left( a, b\right) \equiv 4\ \textrm{Re}\int_{f_\textrm{l}}^{f_\textrm{h}} df\,\frac{\tilde{a}(f) \ \tilde{b}^*(f)}{S_n(f)},
\end{equation}
where a tilde indicates the Fourier transform, a star the complex
conjugate and $S_n(f)$ is the one-sided power spectral density (PSD)
of the detector noise, and we employ the Advanced LIGO’s “zero-detuned
high-power” design sensitivity curve~\cite{Shoemaker:2010}.  The integral is evaluated between the
frequencies $f_{\textrm{l}} = 20 \mathrm{Hz}$ and $f_{\textrm{h}} = 3
\mathrm{kHz}$. When the signal is an NR waveform that starts
  (ends) at a higher (lower) frequency than $f_{\textrm{l}}$
  ($f_{\textrm{h}}$), we choose the starting (ending) frequency of the NR waveform.
Note that the dependence on the luminosity distance $D_{\mathrm{L}}$ disappears in
  Eq.~\eqref{eq:faith} because template and signal are normalized in that expression. In principle, 
we could define the faithfulness in Eq.~\eqref{eq:faith} maximizing also over the inclination angle 
$\iota_\textrm{t}$. This would certainly increase the faithfulness. However, as we have discussed in the previous section,
the inclination angle $\iota$ affects considerably how higher-order modes impact the signal, 
thus we find more appropriate to investigate the waveform model in the worst situation in which we 
do not allow any bias in the measurement of the inclination angle.

The maximizations over $t_c$ and ${\varphi_0}_{\mathrm{t}}$ in Eq.~\eqref{eq:faith} are computed numerically,
  while the maximization over $\kappa_{\mathrm{t}}$ is done analytically
  following the procedure described in Ref. ~\cite{Capano:2013raa}
  (see Appendix A).  When $h_{\mathrm{t}}$ does not include
  higher-order modes, the maximization over the effective polarization
  $\kappa_{\mathrm{t}}$ in Eq. \eqref{eq:faith} becomes degenerate
  with the maximization over ${\varphi_0}_{\mathrm{t}}$ and we recover the usual 
definition of faithfulness.

The faithfulness given in Eq.~\eqref{eq:faith} depends on the signal parameters
$(\iota_{\textrm{s}},{\varphi_0}_{\textrm{s}},\kappa_{\textrm{s}})$. To understand 
how the faithfulness varies as function of those parameters, 
we introduce the minimum, maximum, average and average
weighted with the SNR unfaithfulness
$[1-\mathcal{F}(\iota_{\textrm{s}},{\varphi_0}_{\textrm{s}},\kappa_{\textrm{s}})]$
over these parameters, namely~\cite{Buonanno:2002fy,Capano:2013raa,Harry:2016ijz}:
\begin{align}
\min_{\iota_{\mathrm{s}},{\varphi_0}_{\mathrm{s}},\kappa_{\mathrm{s}}}(1 -\mathcal{F}) \equiv &  1 - \max_{\iota_{\mathrm{s}},{\varphi_0}_{\mathrm{s}},\kappa_{\mathrm{s}}}\mathcal{F}(\iota_{\textrm{s}},{\varphi_0}_{\textrm{s}},\kappa_{\textrm{s}}) \label{eq:min_unfaith}\,, \\
\max_{\iota_{\mathrm{s}},{\varphi_0}_{\mathrm{s}},\kappa_{\mathrm{s}}}(1 -\mathcal{F}) \equiv &  1 - \min_{\iota_{\mathrm{s}},{\varphi_0}_{\mathrm{s}},\kappa_{\mathrm{s}}}\mathcal{F}(\iota_{\textrm{s}},{\varphi_0}_{\textrm{s}},\kappa_{\textrm{s}}) \label{eq:max_unfaith}\,, 
\end{align}
\begin{widetext}
\begin{align}
\langle
1-\mathcal{F}\rangle_{\iota_{\mathrm{s}},{\varphi_0}_{\mathrm{s}},\kappa_{\mathrm{s}}} \equiv & 1 - \frac{1}{8\pi^2}\int_{0}^{2\pi} d\kappa_{\mathrm{s}} \int_{-1}^{1} d(\cos\iota_s) \int_{0}^{2\pi} d{\varphi_0}_{\mathrm{s}} \ \mathcal{F}(\iota_{\textrm{s}},{\varphi_0}_{\textrm{s}},\kappa_{\textrm{s}})\,, \label{eq:avg_unfaith}\\
\langle
1-\mathcal{F}\rangle_{\iota_{\mathrm{s}},{\varphi_0}_{\mathrm{s}},\kappa_{\mathrm{s}}}^{\mathrm{SNRweighted}} \equiv & 1 - \sqrt[3]{\frac{\int_{0}^{2\pi} d\kappa_ {\mathrm{s}} \int_{-1}^{1} d(\cos\iota_s) \int_{0}^{2\pi} d{\varphi_0}_{\mathrm{s}} \ \mathcal{F}^3(\iota_{\textrm{s}},{\varphi_0}_{\textrm{s}},\kappa_{\textrm{s}}) \ \mathrm{SNR}^3(\iota_{\textrm{s}},{\varphi_0}_{\textrm{s}},\kappa_{\textrm{s}})}{\int_{0}^{2\pi} d\kappa_{\mathrm{s}} \int_{-1}^{1} d(\cos\iota_s) \int_{0}^{2\pi} d{\varphi_0}_{\mathrm{s}} \ \mathrm{SNR}^3(\iota_{\textrm{s}},{\varphi_0}_{\textrm{s}},\kappa_{\textrm{s}})}} \label{eq:wavg_unfaith}\,, \\ \nonumber
\end{align}
\end{widetext}
where the $\mathrm{SNR}(\iota_{\textrm{s}},{\varphi_0}_{\textrm{s}},\theta_\textrm{s}, \phi_\textrm{s},\kappa_{\textrm{s}},{D_{\mathrm{L}}}_{\mathrm{s}},
\boldsymbol{\xi}_\mathrm{s},{t_c}_\mathrm{s})$ is defined as:
\begin{equation}
\mathrm{SNR}(\iota_{\textrm{s}},{\varphi_0}_{\textrm{s}},\theta_\textrm{s}, \phi_\textrm{s}, \kappa_{\textrm{s}},{D_{\mathrm{L}}}_{\mathrm{s}},\boldsymbol{\xi}_\mathrm{s},{t_c}_\mathrm{s}) \equiv \sqrt{\left(h_{\mathrm{s}},h_{\mathrm{s}}\right)}.
\end{equation}
We note that for the average unfaithfulness weighted with the SNR in Eq.~\eqref{eq:wavg_unfaith}, we drop in the SNR the 
explicit dependence 
on ${\cal A}(\theta,\phi)$ and $D_{\mathrm{L}}$, because they cancel out. It is important to highlight that the unfaithfulness weighted with the cube 
of the SNR is a conservative upper limit of the fraction of detection volume lost. Indeed, weighting the unfaithfulness with the SNR 
takes into account that, at a fixed distance, configurations closer to an edge-on orientation have a smaller SNR with respect to configurations closer to a face-on orientation, therefore they are less likely to be observed. The definitions of minimum, maximum and averaged unfaithfulness in Eqs.~\eqref{eq:min_unfaith}-\eqref{eq:avg_unfaith} are similar to those in Ref.~\cite{Babak:2016tgq}, with the difference that in the latter they  minimize, maximize and average also over the source orientation $\iota_\mathrm{s}$. The average weighted with the SNR in Eq.~\eqref{eq:wavg_unfaith} was introduced in Ref.~\cite{Buonanno:2002fy} and used for a similar purpose also in Ref.~\cite{Harry:2016ijz}.

In the following we shall show results  
where all the averages are computed assuming an isotropic distribution for the source orientation and sky position.

\begin{figure}[h]
\centering
\includegraphics[width=0.5\textwidth]{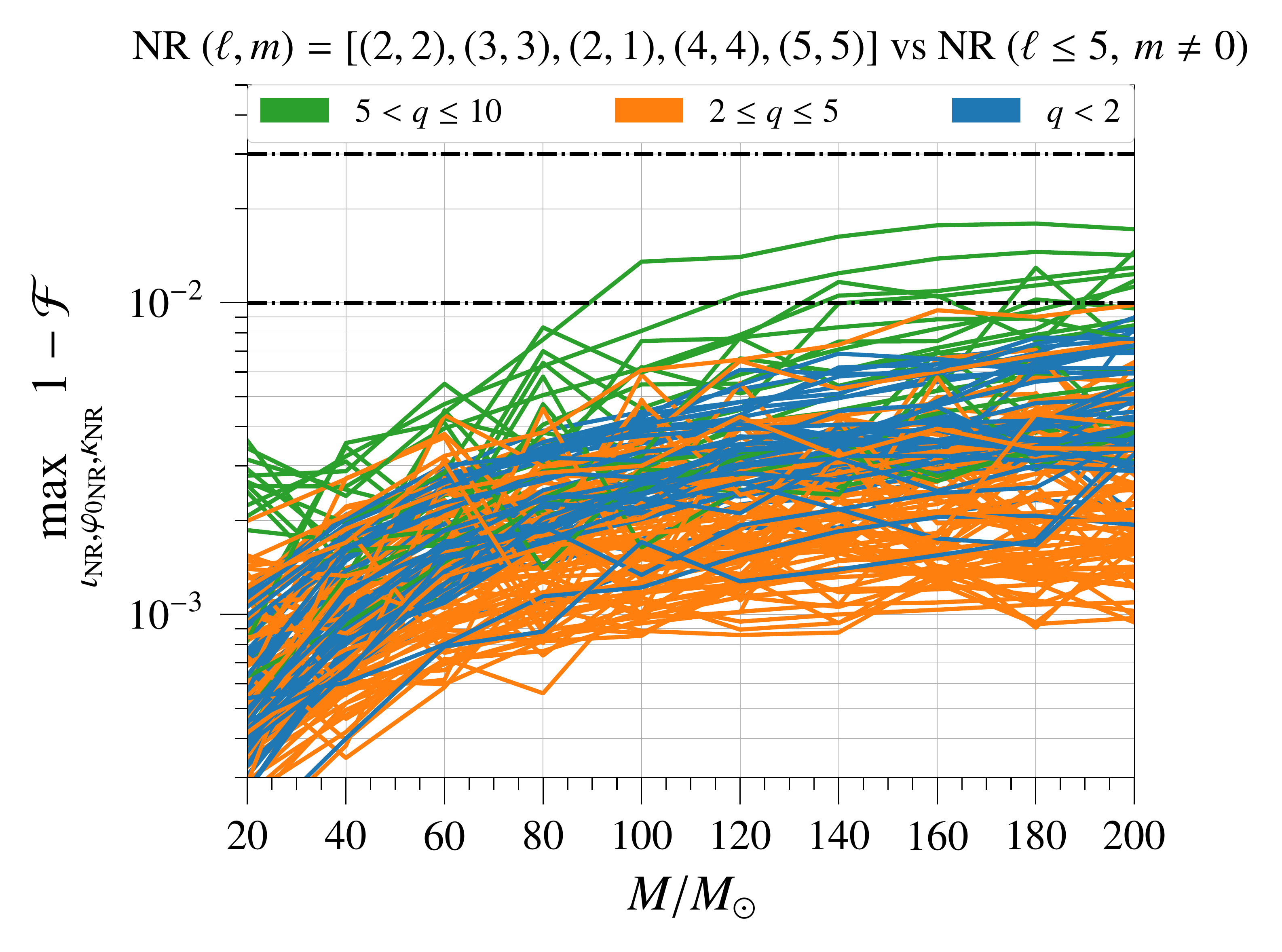}
\caption{Maximum of unfaithfulness $(1-\mathcal{F})$ over the three
  angles
  $(\iota_{\textrm{NR}},{\varphi_0}_{\textrm{NR}},\kappa_{\textrm{NR}})$
  as a function of the total mass, in the range $20 M_\odot \leq M
  \leq 200 M_\odot$ of the NR waveform with
  $(2,2),(2,1),(3,3),(4,4),(5,5)$ modes against NR waveform with
  $(\ell \leq 5, m \neq 0)$ modes. The maximum 
unfaithfulness is typically reached for edge-on orientations. The
  jaggedness of the curves is caused by the numerical noise present in
  higher-order modes that are less resolved in the NR simulations. We
  find that this feature is not present when these noisy modes are
  removed from the calculation of the faithfulness.}
\label{fig:unfaith_NR55vsNRmax}
\end{figure}

Using the aforementioned definitions (\ref{eq:min_unfaith})--(\ref{eq:wavg_unfaith}), we compute the unfaithfulness 
assuming that the signal is an NR waveform with modes $(\ell \leq 5, m \neq 0)$~\footnote{Since the
  nonoscillating $m=0$ modes are not well reproduced by NR
  simulations and their contribution is small, we do not include them
  in these calculations. We find that the contribution of the modes with $\ell \geq
  6$ is neglibigle.}, and the template is either an NR waveform 
or a \texttt{SEOBNRv4} waveform with only the $(2,2)$ mode. 

In the left panel of Fig.~\ref{fig:unfaith_mass}
we show results for the simulation \texttt{SXS:BBH:0610} having $q = 1.2,
\,\chi_1 = -0.5,\,\chi_2=-0.5$. Given the small mass ratio, 
we do not expect the higher-modes to play an important role. Indeed 
both the NR with only the dominant mode and the \texttt{SEOBNRv4} model have averaged
unfaithfulness $<1\%$ in the mass range $20 M_\odot \leq M \leq 200
M_\odot$. In both cases the unfaithfulness is maximum for an
edge-on orientation and is $<3\%$. Conversely the minima of the unfaithfulness occur for a
face-on configuration and they are always much smaller than $1\%$.
The situation is very different in the right panel of Fig.~\ref{fig:unfaith_mass} 
where we consider the simulation \texttt{ET:AEI:0004} that has 
larger mass ratio and spins: $q = 8, \,\chi_1 = \chi_2=0.85$. 
In this case the minima of the unfaithfulness correspond to a face-on orientation where the
higher-order modes are negligible and for this reason both NR with
only the dominant mode and the \texttt{SEOBNRv4} model have unfaithfulness 
smaller than $1\%$. By contrast, the results for the maximum of the
unfaithfulness correspond to an edge-on orientation and they are
equally large for the NR with only the dominant mode and for the 
\texttt{SEOBNRv4} model. They have unfaithfulness in the range $[10\%,20\%]$
for masses $20 M_\odot \leq M \leq 200 M_\odot$. In this case also the
averaged unfaithfulness are large, in the range $[5\%,15\%]$ and $[3\%,8\%]$ 
for the weighted averages.  

Thus, for this high mass-ratio configuration the
error from neglecting higher-order modes supersedes the modeling error
of the dominant mode when the orientation is far from face-on/face-off. This is not surprising because the \texttt{SEOBNRv4} waveform model was constructed requiring $1\%$ of maximum unfaithfulness against the NR waveforms when only the $(2,2)$ mode 
was included~\cite{Bohe:2016gbl}.

Only by properly including the largest subdominant modes can one hope to achieve an
unfaithfulness of the waveform model below $1\%$~\footnote{We notice that using a waveform model with 
unfaithfulness smaller than $3\%$ (or $1\%$ depending on the
features of the template bank) is a sufficient condition for a
template bank to have a loss in event rates due to modeling
error and discreteness of the template bank smaller than $10\%$
(e.g., see Ref.~\cite{Buonanno:2009zt})}. Which subdominnat modes should we include
to achieve such an accuracy? To address this question, we compute the faithfulness between
NR waveforms including the modes $(2,2),(2,1),(3,3),(4,4),(5,5)$ and 
NR waveforms including only the $(\ell \leq 5, m \neq 0)$ modes. We find that the unfaithfulness
averaged over the three angles $(\iota_{\textrm{NR}},{\varphi_0}_{\textrm{NR}},\kappa_{\textrm{NR}})$
ranges between $0.01\% \lessapprox (1-\mathcal{F}) \lessapprox 0.5\%$
for the total mass interval $20 M_\odot \leq M \leq 200
M_\odot$. Thus, we conclude that the modes $(2,2),(2,1),(3,3),(4,4),(5,5)$ 
are sufficient to model the full GW signal if we want to achieve 
an average unfaithfulness smaller than $1\%$. Furthermore, we note that these modes 
are not enough to ensure that the maximum of the unfaithfulness is smaller than
$1\%$. In fact, for some of the configurations with higher mass ratio,
the unfaithfulness is slightly larger than $1\%$ in the mass range $20
M_\odot \leq M \leq 200 M_\odot$, as it is clear from the plot in
Fig.~\ref{fig:unfaith_NR55vsNRmax}. The maximum unfaithfulness
decreases, almost reaching the requirement of being below $1\%$ for
all the waveforms in the catalog, if we add also the more subdominant
modes $(3,2),(4,3)$.  However, given that the overall improvement
in the maximum of unfaithfulness when including also the modes
$(3,2),(4,3)$ is small (of the order of a few $0.1 \%$)
with respect to the results obtained using only the $(2,2),(2,1),(3,3),(4,4),(5,5)$ 
modes, it is worth comparing this improvement with the estimation of the maximum of the unfaithfulness
due to the numerical error of the NR waveforms. The numerical errors
we consider are numerical truncation error
\cite{Hinder:2013oqa,Kumar:2015tha} and waveform extrapolation error
\cite{Hinder:2013oqa,Kumar:2015tha,Taylor:2013zia}. For our NR catalog, we estimate the
numerical truncation error computing the maximum of the unfaithfulness
between the same NR waveforms with the same modes (i.e., 
$(2,2),(2,1),(3,3),(4,4),(5,5)$), but with different resolutions, notably 
the highest (maximum) resolution and the second highest. The
waveform extrapolation error is estimated in the same way, but
employing different extrapolation orders (i.e., $N = 2$ and $N = 3$). 
We find that the contribution of each of these errors to the maximum of the unfaithfulness is in the
range $[0.1\%, 1\%]$ for the total mass interval $20 M_\odot \leq M
\leq 200 M_\odot$~\footnote{The unfaithfulness averaged over the three
  angles
  $(\iota_{\textrm{NR}},{\varphi_0}_{\textrm{NR}},\kappa_{\textrm{NR}})$ due
  to numerical errors is much smaller than $1\%$. The
  reason is that the main contribution to this average unfaithfulness
  is the numerical error of the dominant mode. The latter is 
  much smaller than $1\%$, as well. This conclusion is in agreement
  with Ref.~\cite{Kumar:2015tha} where the authors studied the
  numerical errors of the dominant mode for a subset of the waveforms in
  our NR catalog.}.
 
Since adding the modes $(3,2),(4,3)$ is a non trivial task because of the mode mixing between spherical and spheroidal harmonics~\cite{Buonanno:2006ui,Kelly:2012nd, Berti:2014fga,London:2014cma}, and considering that their contribution is at the same level of the numerical error of the NR waveforms, we decide not to include them in the \texttt{SEOBNRv4HM} model. The results of the maximum of the unfaithfulness due to the numerical errors suggest that in order to use NR waveforms to build an EOBNR model having maximum unfaithfulness against NR smaller than $1\%$ it would be necessary to have more accurate higher-order modes from NR simulations.

\section{Effective-one-body multipolar waveforms for nonprecessing binary black holes}
\label{sec:eob_formalism}

In this section we describe the main ingredients used to build the multipolar spinning, nonprecessing 
\texttt{SEOBNRv4HM} waveform model. We start briefly describing the dynamics in Sec.~\ref{subsec:dynamics}, and then 
focus on the structure of the gravitational modes in Sec. \ref{subsec:waveform}.

In the EOB formalism the real dynamics of two bodies with masses $m_{1,2}$ and
spins $\mathbf{S_{1,2}}$ is mapped into the effective dynamics of a
test particle with mass $\mu$ and spin $\mathbf{S_*}$ moving in a
deformed Kerr metric with mass $M = m_1 + m_2$ and spin
$\bm{S_{\textrm{Kerr}}}$ (for details see Ref.~\cite{Barausse:2011ys}). As discussed above, here we limit to nonprecessing 
spins $\mathbf{S_{1,2}}$ and introduce the dimensionless spin parameters $\chi_{1,2}$
defined as $\mathbf{S_\textrm{i}} = \chi_\textrm{i} m_\textrm{i}^2
\mathbf{\hat{L}}$, with $-1 \leq \chi_i \leq 1$.

\subsection{Effective-one-body dynamics}
\label{subsec:dynamics}

The EOB conservative orbital dynamics is obtained from the resummed EOB Hamiltonian through the 
energy mapping~\cite{Buonanno:1998gg}
\begin{equation}
H_{\textrm{EOB}} = M\sqrt{1+ 2\nu\left(\frac{H_{\textrm{eff}}}{\mu} -1 \right)} -M,
\end{equation}
where $\mu=m_1m_2/(m_1+m_2)$ is the reduced mass of the BBH and $\nu = \mu/M$ is the
symmetric mass ratio. When spins are nonprecessing the motion is constrained to a plane. 
Thus, the dynamical variables entering the Hamiltionian are the orbital phase $\phi$~\footnote{Abusing notation, we indicate the orbital phase with $\phi$, which we use to denote the azimuthal angle describing the sky location of the source in the detector frame. It will be clear from the context which of the two angles we refer to.},  
the radial separation $r$ (normalized to M) and their conjugate momenta $p_\phi$ and $p_r$
(normalized to $\mu$). The explicit form of $H_{\textrm{eff}}$ that we adopt here 
was derived in Refs.~\cite{Barausse:2009xi,Barausse:2011ys}, based on the
linear-in-spin Hamiltonian for spinning test particles of
Ref.~\cite{Barausse:2009aa}. The radial potential entering the 00-component of the EOB 
deformed metric, which also enters the effective Hamiltonian 
$H_{\textrm{eff}}$, is explicitly given in Eqs. (2.2) and (2.3) in Ref.~\cite{Bohe:2016gbl}. 
The Hamiltonian $H_{\textrm{eff}}$ depends also on the calibration parameters ($K$, $d_{\textrm{SO}}$, $d_{\textrm{SS}}$
$\Delta_{\textrm{peak}}^{22}$), which were determined in Ref.~\cite{Bohe:2016gbl} by requiring agreement against 
a large set of NR simulations (see Eqs. (4.12)--(4.15) therein). Here, we adopt the same values for these calibration parameters.

The dissipative dynamics in the EOB formalism is described by the radiation-reaction force given in 
Eq. (2.9) in Ref.~\cite{Bohe:2016gbl}. We notice that in this paper we do not change the dissipative and conservative 
dynamics of the \texttt{SEOBNRv4} model, and that the \texttt{SEOBNRv4HM} waveform models share the same two-body dynamics of 
\texttt{SEOBNRv4}. Here, we improve the accuracy of the gravitational modes with $(\ell, m) \neq (2,2)$, and use them in 
the gravitational waveform, but we do not employ these more accurate version of the modes in the radiation-reaction force. 
Furthermore, we note that the gravitational modes with $(\ell, m) \neq (2,2)$
are present in the radiation-reaction force, but they do not include the NQCs corrections (see Eq.~\eqref{eq:NQC_corrections}). As discussed also in Ref.~\cite{Pan:2011gk}, the latter modify the amplitude of the already subdominant higher-order modes (see Fig.~\ref{fig:rel_amp_hm_0spin}) by $\sim 10\%$ close to merger, where the effect of the radiation reaction is not very important for the plunging dynamics.

\subsection{Effective-one-body  gravitational modes} 
\label{subsec:waveform}

As usual in the EOB formalism~\cite{Buonanno:2000ef}, the gravitational modes entering Eq.~(\ref{eq:spHarmDec}) are composed of 
two main parts: inspiral \& plunge, and merger \& ringdown. We can write the generic mode as:
\begin{align}
\label{eq:complete_mode}
h_{\ell m}(t) = \begin{cases}
h_{\ell m}^{\mathrm{insp-plunge}}(t), &t < t_{\textrm{match}}^{\ell m}\\
h_{\ell m}^{\mathrm{merger-RD}}(t), &t > t_{\textrm{match}}^{\ell m},\\
\end{cases}
\end{align}
where $t_{\textrm{match}}^{\ell m}$ is defined as:
\begin{align}
t_{\textrm{match}}^{\ell m} =\begin{cases} 
t_{\textrm{peak}}^{22}, &  (\ell, m) =(2,2),(3,3),(2,1),(4,4)\\
t_{\textrm{peak}}^{22} - 10 M, & (\ell, m) = (5,5),\\
\end{cases}
\label{eq:matchtime}
\end{align}
with $t_{\textrm{peak}}^{22}$ being the peak of the amplitude of the
$(2,2)$ mode. By construction the amplitude and phase
of $h_{\ell m}(t)$ are $C^1$ at $t = t_{\textrm{match}}^{\ell m}$.  In
the following we shall discuss in more detail how
these two parts of the gravitational modes are built and why we choose 
a different matching point for the mode $(5,5)$. We note again that the mode
$(2,2)$ in the \texttt{SEOBNRv4HM} model is the same as in the 
\texttt{SEOBNRv4} model, and for this reason below we focus on the higher-order modes 
$(3,3),(2,1),(4,4),(5,5)$.

\subsection{Effective-one-body  waveform modes: inspiral-plunge} 
\label{subsec:waveform_inspiral}

The inspiral-plunge EOB modes are expressed in the following multiplicative form:
\begin{equation}
\label{eq:hlm_NQC}
h_{\ell m}^{\textrm{insp-plunge}} = h_{\ell m}^\textrm{F} N_{\ell m},
\end{equation}
where $h_{\ell m}^\textrm{F}$ is the factorized form of the PN GW modes~\cite{Arun:2008kb,Buonanno:2012rv} for quasi-circular orbits, aimed  
at capturing strong-field effects, as discussed in the test-mass limit~\cite{Damour:2007xr,Damour:2008gu,Pan:2010hz}. 
The factor $N_{\ell m}$ in Eq.~(\ref{eq:hlm_NQC}) is the nonquasi-circular (NQC) term, which includes possible radial  
effects that are no longer negligible during the late inspiral and 
plunge, and that are not captured by the rest of the waveform. More explicitly, the factorized term reads: 
\begin{equation}
\label{eq:hlm_factorized}
h_{\ell m}^{\textrm{F}} = h_{\ell m}^{(N, \epsilon)}\, \hat{S}^{(\epsilon)}_{\textrm{eff}}\, T_{\ell m}\,f_{\ell m}\, e^{i \delta_{\ell m}}\,,
\end{equation}
where $\epsilon$ is the parity of the multipolar waveform, defined as
\begin{equation}
\epsilon = \begin{cases}
0, \,\, \ell + m\,\, \text{is even} \\
1, \,\, \ell + m\,\, \text{is odd}.
\end{cases}
\end{equation}
The Newtonian term $h_{\ell m}^{(N,\epsilon)}$ reads: 
\begin{equation}
\label{eq:Newtonian}
h_{\ell m}^{(N,\epsilon)} = \frac{M \nu}{\textrm{D}_L}\, n_{\ell m}^{(\epsilon)} \,\ c_{\ell + \epsilon}(\nu)\,\, V_{\phi}^{\ell}\,\, Y^{\ell - \epsilon, -m}\left(\frac{\pi}{2},\phi\right),
\end{equation}
where $\textrm{D}_L$ is the distance from the source, $Y^{\ell m}(\theta,\phi)$ are the scalar spherical harmonics and the expression of the functions $n_{\ell m}^{(\epsilon)}$ and $c_{\ell + \epsilon}(\nu)$ are given in Appendix \ref{app:modes}. 
The function $V_{\phi}^{\ell}$ is defined as:
\begin{equation}
V^{\ell}_{\phi} \equiv v_{\phi}^{(\ell +  \epsilon)} \equiv M\,\, \Omega \,\, r_{\Omega},
\end{equation}
where
\begin{equation}
r_{\Omega} = \left[ \frac{\partial H_{\textrm{EOB}}(r, \phi, p_r = 0, p_\phi)}{\partial p_{\phi}}\right]^{-\frac{2}{3}},
\end{equation}
$\Omega = d\phi/dt$ being the angular frequency. We also define $v_\Omega = (M \, \Omega)^{1/3}$. 
 The term $\hat{S}^{(\epsilon)}_{\textrm{eff}}$ in Eq.~\eqref{eq:hlm_factorized} is an effective source term:
\begin{equation}
\hat{S}^{(\epsilon)}_{\textrm{eff}} = \begin{cases}
H_{\textrm{eff}}(r,p_{r_*},p_\phi),\,\,\epsilon = 0\\
L_{\textrm{eff}} = p_\phi \, (M \, \Omega)^{\frac{1}{3}},\,\,\,\, \epsilon = 1.\\
\end{cases}
\end{equation}
The function $T_{\ell m}$ in \eqref{eq:hlm_factorized} is a resummation of the leading-order logarithms of tail effects:
\begin{align}
T_{\ell m} &= \frac{\Gamma (\ell +1 -2 \ i H_{\textrm{EOB}} \Omega)}{\Gamma (\ell +1)} \exp[\pi \ m \ \Omega \ H_{\textrm{EOB}}] \nonumber\\
& \times \exp[2 \ i \ m \ \Omega \ H_{\textrm{EOB}} \ \log(2 \ m \ \Omega \ r_0)], 
\end{align}
where $r_0 = 2M/\sqrt{e}$. 

The functions $f_{\ell m}$ and $e^{\delta_{\ell m}}$ in
  Eq.~\eqref{eq:hlm_factorized} contain terms such
that when expanding in PN order $h_{\ell m}^\mathrm{F}$ one recovers $h_{\ell m}^{\mathrm{PN}}$ (i.e., the PN expansion of
the $(\ell, m)$ mode up to the PN order at which $h_{\ell
  m}^{\mathrm{PN}}$ is known today). In the \texttt{SEOBNRv4HM} model the expression for $f_{\ell m}$ and $\delta_{\ell m}$ are mostly taken from the \texttt{SEOBNRv4} model \cite{Bohe:2016gbl} with the addition of some newly computed PN terms (for more details and explicit
expressions of $f_{\ell m}$ and $\delta_{\ell m}$ see Appendix
\ref{app:modes}). For the modes $(2,1)$ and $(5,5)$, $f_{\ell m}$ includes also the calibration term $c_{\ell m} \
v_\Omega^{\beta_{\ell m}}$, where $\beta_{\ell m}$ denotes the first-order term 
at which the PN series of $_{\ell m}$ is not known today with its complete dependence on mass ratio and spins (see Eqs.~\eqref{eq:f_21}--\eqref{eq:f_55}). The calibration parameter $c_{\ell m}$ is evaluated
to satisfy the condition: 
\begin{align}
\left|h_{\ell m}^{F}(t_{\textrm{match}}^{\ell m})\right| &\equiv \left|h_{\ell m}^{(N, \epsilon)} \hat{S}^{(\epsilon)}_{\textrm{eff}} T_{\ell m} e^{\mathrm{i} \delta_{\ell m}} f_{\ell m}(c_{\ell m})\right| \bigg|_{t = t_{\textrm{match}}^{\ell m}}\,, \nonumber \\
&= \left|h_{\ell m}^\textrm{NR}(t_{\textrm{match}}^{\ell m})\right|, \qquad \mathrm{for}\,\, (\ell, m) = (2,1), \
(5,5),
\label{eq:cal_par}
\end{align}
where $\left|h_{\ell m}^\textrm{NR}(t_{\textrm{match}}^{\ell m})\right|$ is the amplitude of the NR modes at the matching point $t_{\textrm{match}}^{\ell m}$. The latter are given as fitting formulae for every point of the
parameter space $(\nu,\chi_1,\chi_2)$ in Appendix \ref{app:NQCfits}. We need to include the calibration parameter $c_{\ell m}$ for the modes $(\ell,m) = (2,1),(5,5)$ for reasons that we explain below in Sec.~\ref{subsec:waveform_zeros}.

Finally, the term $N_{\ell m}$ in Eq. \eqref{eq:hlm_NQC} is the NQC correction:
\begin{align}
\label{eq:NQC_corrections}
N_{\ell m} &= \left[1+ \frac{p_{r^*}^2}{(r\ \Omega)^2}\left(a_1^{h_{\ell m}} + \frac{a_2^{h_{\ell m}}}{r} + \frac{a_3^{h_{\ell m}}}{r^{3/2}} \right)\right] \nonumber \\ 
& \times \exp\left[i \left(b_1^{h_{\ell m}}\frac{p_{r^*}}{r \ \Omega} + b_2^{h_{\ell m}}\frac{p_{r^*}^3}{r \ \Omega} \right) \right], 
\end{align}
which is used to reproduce the shape of the NR modes close to the matching point $t_{\ell m}^{\textrm{match}}$. As done in the past~\cite{Taracchini:2013rva,Bohe:2016gbl}, the 5 constants $(a_1^{h_{\ell m}}, a_2^{h_{\ell m}}, a_3^{h_{\ell m}}, b_1^{h_{\ell m}}, b_2^{h_{\ell m}})$ are fixed by requiring that:
\begin{itemize}
\item The amplitude of the EOB modes is the same as that of the NR modes at the matching point $t_{\textrm{match}}^{\ell m}$:
\begin{equation}
\label{eq:NQC_condition_1}
\left| h_{\ell m}^{\textrm{insp-plunge}}(t_{\textrm{match}}^{\ell m}) \right| = \left|h_{\ell m}^\textrm{NR}(t_{\textrm{match}}^{\ell m})\right|;
\end{equation}
We notice that this condition is different from that in Eq.~\eqref{eq:cal_par} because it affects $h_{\ell m}^{\textrm{insp-plunge}}(t_{\textrm{match}}^{\ell m})$ and not $h_{\ell m}^\textrm{F}(t_{\textrm{match}}^{\ell m})$. Because of the calibration parameter in Eq.\eqref{eq:cal_par}, for the modes (2,1) and (5,5) this condition becomes simply $|N_{\ell m}| = 1$.
\item The first derivative of the amplitude of the EOB modes is the same as that of the NR modes at the matching point $t_{\textrm{match}}^{\ell m}$:
\begin{equation}
\label{eq:NQC_condition_2}
\left. \frac{d\left| h_{\ell m}^{\textrm{insp-plunge}}(t) \right|}{dt} \right|_{t =t_{\textrm{match}}^{\ell m}} = 
\left. \frac{d\left| h_{\ell m}^{\textrm{NR}}(t) \right|}{dt} \right|_{t =t_{\textrm{match}}^{\ell m}};
\end{equation}
\item The second derivative of the amplitude of the EOB modes is the same as that of the NR modes at the matching point $t_{\textrm{match}}^{\ell m}$:
\begin{equation}
\label{eq:NQC_condition_3}
\left. \frac{d^2\left| h_{\ell m}^{\textrm{insp-plunge}}(t) \right|}{dt^2} \right|_{t = t_{\textrm{match}}^{\ell m}} 
= \left. \frac{d^2\left| h_{\ell m}^{\textrm{NR}}(t) \right|}{dt^2} \right|_{t = t_{\textrm{match}}^{\ell m}};
\end{equation}
\item The frequency of the EOB modes is the same as that of the NR modes at the matching point $t_{\textrm{match}}^{\ell m}$:
\begin{equation}
\label{eq:NQC_condition_4}
\omega_{\ell m}^{\textrm{insp-plunge}}(t_{\textrm{match}}^{\ell m}) = \omega_{\ell m}^{\textrm{NR}}(t_{\textrm{match}}^{\ell m});
\end{equation}
\item The first derivative of the frequency of the EOB modes is the same as that of the NR modes at the matching point $t_{\textrm{match}}^{\ell m}$:
\begin{equation}
\label{eq:NQC_condition_5}
\left .\frac{d {\omega}_{\ell m}^{\textrm{insp-plunge}}(t)}{dt}\right|_{t = t_{\textrm{match}}^{\ell m}}  =
\left . \frac{d {\omega}_{\ell m}^{\textrm{NR}}(t)}{dt}\right|_{t = t_{\textrm{match}}^{\ell m}},
\end{equation}
\end{itemize}
where the RHS of Eqs. \eqref{eq:NQC_condition_1}--\eqref{eq:NQC_condition_5} (usually called ``input values''), are given as fitting formulae for every point of the
parameter space $(\nu,\chi_1,\chi_2)$ in Appendix \ref{app:NQCfits}. These fits are produced using the NR 
catalog and BH-perturbation-theory waveforms, as described in Appendix~\ref{sec:NRcatalog}. 

As we discuss in Appendices~\ref{app:NQCfits} and ~\ref{app:ringdownfits}, we find that for several binary configurations in the NR catalog, the numerical error is quite large for the mode $(5,5)$ close to merger. To minimize the impact of the numerical error on the fits of the input values,  
we are obliged to choose the  matching point for this mode earlier than for other modes, as indicated 
in Eq.~(\ref{eq:matchtime}).

\subsection{Minima in $(2,1)$, $(5,5)$-modes' amplitude and $c_{\ell m}$'s calibration parameters}
\label{subsec:waveform_zeros}

\begin{figure}[h]
\centering
\includegraphics[width=0.5\textwidth]{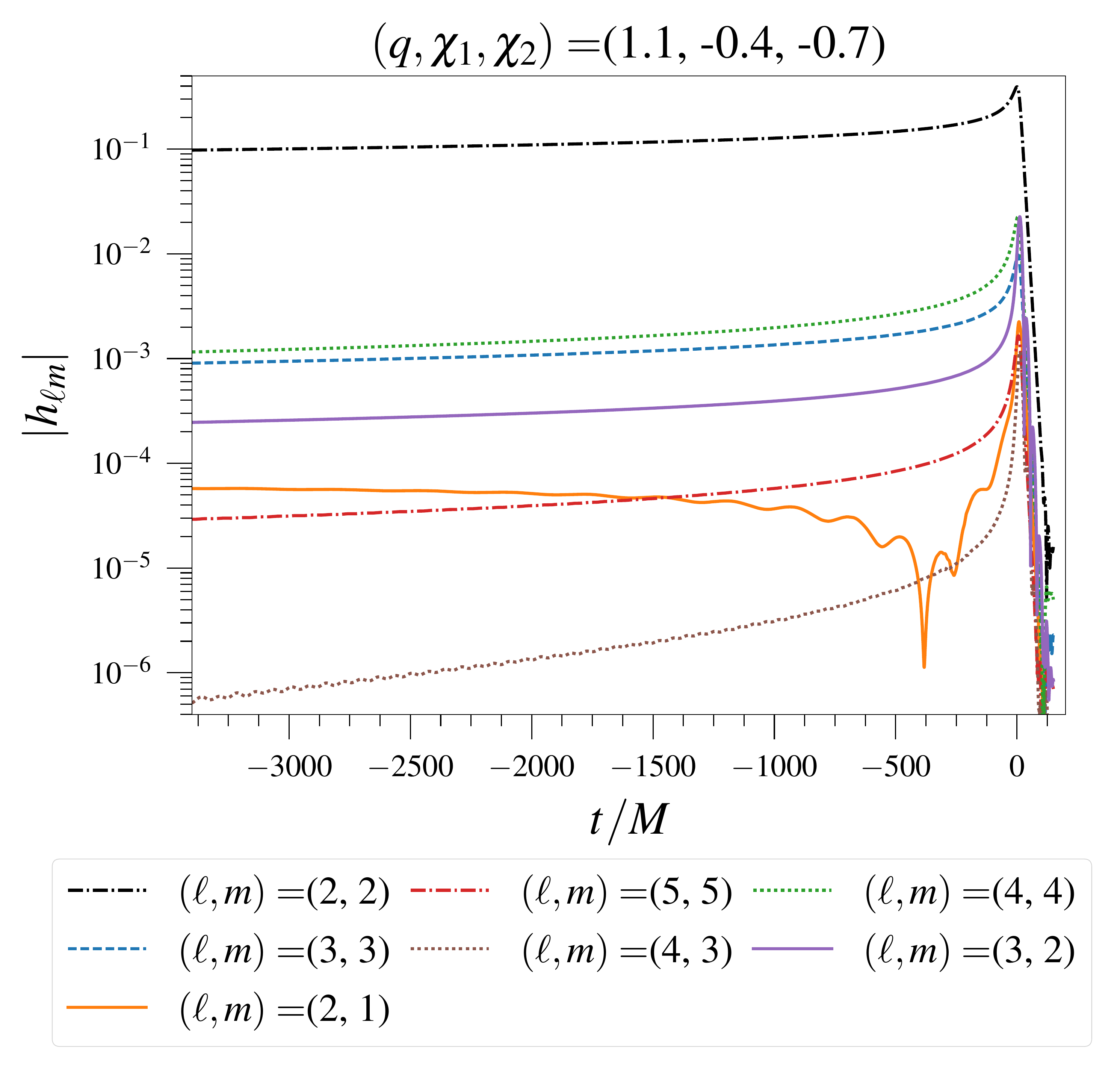}
\caption{Amplitude of the $(2,2),(2,1),(3,3),(4,4),(5,5)$,$(3,2),(4,3)$ modes versus time for the NR simulation 
\texttt{SXS:BBH:1377} with parameters $q = 1.1, \,\chi_1 = -0.4,\,\chi_2 = -0.7$. We produce such simulation to check if the 
analytical prediction that the $(2,1)$-mode's amplitude would have a non-monotonic behaviour toward merger holds. 
We choose as origin of time the peak of the $(2,2)$ mode.}
\label{fig:zero}
\end{figure}

We want now to come back to the motivation of introducing the $c_{\ell m}$'s calibration parameters in 
  Eq.~\eqref{eq:cal_par} for the modes $(2,1)$ and $(5,5)$. We note that those parameters are determined 
and included in the waveform  before applying the NQC conditions (\ref{eq:NQC_condition_1})--(\ref{eq:NQC_condition_5}). 
We introduce the $c_{\ell m}$'s to ``cure'' the behaviour of the modes $(2,1),(5,5)$ close to the matching point for a particular 
region of the parameter space. Indeed, we find that the factorized expression of the amplitude $\left|h_{\ell m}^\textrm{F}(t)\right|$ starts to  
decrease toward plunge and merger, approaching minimum values close to zero for $t \sim
  t_{\ell m}^{\textrm{match}}$ when the binary parameters have $q \sim 1$ and large $|\chi_A| = |(\chi_1-\chi_2)|/2$. 
Although the term $f_{\ell m}$ in Eq.~(\ref{eq:hlm_factorized}) is responsible of the zeros in the amplitude, we find 
that this unexpected behaviour is also present in the PN-expanded form of the mode, and persist in other 
mode resummations, like those suggested in Ref.~\cite{Pan:2010hz} (see Eq. 2 therein) and in Refs.~\cite{Nagar:2016ayt,Messina:2018ghh}.

Quite interestingly, in the case of the $(5,5)$ mode, we do not find such a non-monotonic behaviour toward merger 
in the NR simulations at our disposal, but we do find it for the $(2,1)$ mode in the same region of parameter 
space predicted by the analytical computation. In particular, we notice minima toward merger in \texttt{SXS:BBH:0612} with $(q = 1.6,\,
  \chi_1 = 0.5,\, \chi_2 = -0.5)$, \texttt{SXS:SXS:BBH:0614} $(q =
  2,\, \chi_1 = 0.75,\, \chi_2 = -0.5)$, \texttt{SXS:BBH:0254} $(q =
  2,\, \chi_1 = 0.6,\, \chi_2 = -0.6)$. We also produce a new NR simulation 
\texttt{SXS:BBH:1377} with $q = 1.1, \,\chi_1 = -0.4,\,\chi_2 = -0.7$ to check the 
presence of a minimum in the amplitude mode. Figure~\ref{fig:zero} shows indeed the presence of such a mimimum in 
the $(2,1)$ mode amplitude for  \texttt{SXS:BBH:1377}.

The minima (or zeros) of the $(2,1),(5,5)$ modes can sometime 
occur at times $t \sim t_{\ell m}^{\textrm{match}}$, that is close to the times where 
we impose the NQC conditions \eqref{eq:NQC_condition_1}--\eqref{eq:NQC_condition_5}. 
When that happens, the enforcement of such conditions yield a waveform which contains 
unwanted features\footnote{Since $|h_{\ell m}^{\mathrm{insp-plunge}}(t_{\mathrm{match}}^{\ell m})| \sim 0$, imposing the condition in Eq.~\eqref{eq:NQC_condition_1} with $\left|h_{\ell m}^\textrm{NR}(t_{\textrm{match}}^{\ell m})\right| \neq 0$ forces the function $|N_{\ell m}|$, hence the amplitude $|h_{\ell m}^{\mathrm{insp-plunge}}(t)|$, to assume unphysically large values for $t < t_{\mathrm{match}}^{\ell m}$.}. Considering that for the mode $(5,5)$ the mimima are absent in the NR simulations, 
thus they are likely an artefact of the analytical waveform, and that for the mode $(2,1)$ 
the minima are present only in the region of parameter space where the 
$(2,1)$ mode is much smaller than the other modes (i.e., when $q \sim 1$ and $|\chi_A| = |(\chi_1-\chi_2)|/2$ is large, 
see also Fig.~\ref{fig:zero}), we decide to remove the minima from the $(2,1)$ and $(5,5)$ EOB modes. We achieve 
this by introducing the calibration parameter $c_{\ell m}$, which enforces the condition that the EOB amplitude 
at $t_{\ell m}^{\textrm{match}}$ is equal to the NR amplitude (see Eq.~(\ref{eq:cal_par})). Note that the 
latter is imposed before the NQC conditions and removes the minima only when they appear for $t \sim t_{\mathrm{match}}^{\ell m}$. Modeling the minima in the $(2,1)$ modes could be considered 
in the future, when more accurate waveforms would be needed at higher SNRs.
  
Henceforth, we attempt to describe why the analytical modes (both in the PN and factorized form) present minima or zeros 
for the $(2,1)$ and $(5,5)$ cases when $q \sim 1$ and $|\chi_A| = |(\chi_1-\chi_2)|/2$ is large. 
Readers who might not be interested in this technical discussion, could skip the rest of this section 
and move to Sec.~\ref{subsec:waveform_merger}.

As discussed in Sec.~\ref{sec:motivations}, because of binary symmetry under rotation 
($\varphi_0 \rightarrow \varphi_0 + \pi$) the modes with odd $m$ vanish 
for equal-mass and equal-spins configurations. Thus, the nonspinning terms in those modes are proportional 
to $\delta m = (m_1 - m_2)/M$ while the spinning terms are an antisymmetric combination of $\delta m$, $\chi_A$ and $\chi_S =
(\chi_1+\chi_2)/2$ (e.g, $\chi_A$, $\chi_S \delta m$, $\chi_A^2 \delta m$), see for example Eqs.(38a)--(38i) in Ref.~\cite{Pan:2010hz}. In the limit $q\sim 1$ all 
the nonspinning and spinning terms proportional to $\delta m$ are suppressed, and the leading
spinning terms are proportional to $\chi_A$. For large values of $\chi_A$ and small values of $\delta m$ (very
unequal spins, almost equal mass) a cancellation between the leading-order spin 
correction and the dominant nonspinning PN term (which despite being of lower PN order 
is supressed by $\delta m$) can occur at some given frequency. The higher the difference in PN orders
between these two leading spinning and nonspinning contributions, the
higher the frequency at which the cancellation happens. For the
$(2,1)$ mode, there is only a half PN order difference between these terms (see Eq.~(38b) in Ref.~\cite{Pan:2010hz}), so the
cancellation arises at sufficiently low frequencies where this PN
analysis based on two leading terms can be reliable, and, indeed, we do
observe these minima in the NR simulations. In  Table~\ref{tab:zeros} we 
list the configurations in our NR catalog where the minimum happens and
its orbital frequency as measured in the NR simulation~\footnote{We estimate the orbital
  frequency in the NR simulation as half of the gravitational frequency of the $(2,2)$ mode.} 
and as predicted by PN modeling at 3PN order~\cite{Blanchet:2013haa,Buonanno:2012rv,Marsatetal2017}. As expected, the lower the
frequency, the more accurate the PN prediction. We note that the last row
shows results of a NR simulation that we specifically produce to confirm the presence of
the minimum in the mode (see also Fig.~\ref{fig:zero}). We note that for the binary's 
configuration listed in the first row of Table~\ref{tab:zeros},
the NR simulation shows a high-frequency minimum, which is not
reproduced by PN calculations, confirming that this analysis becomes less reliable
in the high-frequency regime.

Lastly, as already pointed out above, for the $(5,5)$
mode we do not observe any minimum in the NR simulations at our disposal. The most
likely explanation is that the cancellation of the leading terms
happens at frequencies high enough that the higher-order PN corrections
would change the result (i.e., they completely remove the minimum or push it
at frequency higher than the merger frequency).

\begin{center}
\begin{table}[h]
 \begin{tabular}{|c|c| c| c| c| c|} 
 \hline
 NR name & $q$ & $\chi_1$ & $\chi_2$ & $M \Omega_{0}^{\mathrm{NR}}$ & $M \Omega_{0}^{\mathrm{PN}}$ \\ [0.5ex] 
 \hline
 \texttt{SXS:BBH:0254}&2 & 0.6 & -0.6 & 0.17 & n/a \\ 
 \hline
 \texttt{SXS:BBH:0614}&2 & 0.75 & -0.5 & 0.082 & 0.057 \\
 \hline
 \texttt{SXS:BBH:0612}&1.6 & 0.5 & -0.5 & 0.068 & 0.047 \\
 \hline
 \texttt{SXS:BBH:1377}&1.1 & -0.4 & -0.7 & 0.033 & 0.029 \\
\hline
\end{tabular}
\caption{For each NR simulation, binary's parameters and values of the orbial 
frequencies $M\Omega_{0}^{\mathrm{NR}}$ and $M\Omega_{0}^{\mathrm{PN}}$ 
at which the minimum of the $(2,1)$ mode occurs.}
\label{tab:zeros}
\end{table}
\end{center}

\subsection{Effective-one-body  waveform modes: merger-ringdown} 
\label{subsec:waveform_merger}

We build the merger-ringdown EOB waveforms following Refs.~\cite{Baker:2008mj,Damour:2014yha,Nagar:2016iwa,Bohe:2016gbl}, notably 
the implementation in Ref.~\cite{Bohe:2016gbl}. The merger-ringdown mode reads:
\begin{equation}
\label{eq:merger-RD_wave}
h_{\ell m}^{\textrm{merger-RD}}(t) = \nu \ \tilde{A}_{\ell m}(t)\ e^{i \tilde{\phi}_{\ell m}(t)} \ e^{-i \sigma_{\ell m 0}(t-t_{\textrm{match}}^{\ell m})},
\end{equation}
where $\sigma_{\ell m 0}$ is the (complex) frequency of the
  least-damped QNM of the final BH. We denote $\sigma_{\ell m}^\textrm{R} \equiv \Im (\sigma_{\ell m0}) < 0$ and 
$\sigma_{\ell m}^\textrm{I} \equiv -\Re (\sigma_{\ell m0})$. For each mode $(\ell,m)$, we employ the  
  frequency values tabulated in Refs.~\cite{Berti:2005ys,Berti:2009kk} as
  functions of the BH's mass and spin. We compute the remnant-BH's mass using the same fitting 
  formula in Ref.~\cite{Taracchini:2013rva}, which is based on the
  phenomenological formula in Ref.~\cite{Barausse:2012qz}, but
  we replace its equal-mass limit (see Eq. (11) in Ref.~\cite{Barausse:2012qz}) with the fit
  in Ref.~\cite{Hemberger:2013hsa} (see Eq. (9) of Ref.~\cite{Hemberger:2013hsa}). The remnant-BH's spin 
is computed using the spin formula in Ref.~\cite{Hofmann:2016yih} (see Eq. (7) in Ref.~\cite{Hofmann:2016yih}).

For the two functions $\tilde{A}_{\ell m}(t)$ and $\tilde{\phi}_{\ell m}(t)$, we use the ans\"atze~\cite{Bohe:2016gbl}:
\begin{equation}
\label{eq:ansatz_amp}
\tilde{A}_{\ell m}(t) = c_{1,c}^{\ell m} \tanh[c_{1,f}^{\ell m}\ (t-t_{\textrm{match}}^{\ell m}) \ +\ c_{2,f}^{\ell m}] \ + \ c_{2,c}^{\ell m},
\end{equation}
\begin{equation}
\label{eq:ansatz_phase}
\tilde{\phi}_{\ell m}(t) = \phi_{\textrm{match}}^{\ell m} - d_{1,c}^{\ell m} \log\left[\frac{1+d_{2,f}^{\ell m} e^{-d_{1,f}^{\ell m}(t-t_{\textrm{match}}^{\ell m})}}{1+d_{2,f}^{\ell m}}\right],
\end{equation}
where $ \phi_{\textrm{match}}^{\ell m}$ is the phase of the
inspiral-plunge mode $(\ell, m)$ at $t = t_{\textrm{match}}^{\ell m}$.
The coefficients $d_{1,c}^{\ell m}$ and $c_{i,c}^{\ell m}$ \footnote{The subscript ``c'' means ``constrained'' 
while ``f'' stands for ``free''.} with $i = 1,2$
are fixed by imposing that the functions $\tilde{A}_{\ell m}(t)$ and $\tilde{\phi}_{\ell m}(t)$ 
in Eq. \eqref{eq:complete_mode} are of class $C^1$ at $t = t_{\textrm{match}}^{\ell m}$. Those 
constraints allow us to express $c_{i,c}^{\ell m}$ in terms of $c_{1,f}^{\ell
    m},\ c_{2,f}^{\ell m},\ \sigma^\textrm{R}_{\ell m},\ |h_{\ell
    m}^{\textrm{insp-plunge}}(t_{\textrm{match}}^{\ell
    m})|,\ \partial_t|h_{\ell
    m}^{\textrm{insp-plunge}}(t_{\textrm{match}}^{\ell m})|$ as
\begin{align}    
c_{1,c}^{\ell m} &= \frac{1}{c_{1,f}^{\ell
    m} \nu} \big[ \partial_t|h_{\ell
    m}^{\textrm{insp-plunge}}(t_{\textrm{match}}^{\ell m})| \nonumber \\
    &- \sigma^\textrm{R}_{\ell m} |h_{\ell
    m}^{\textrm{insp-plunge}}(t_{\textrm{match}}^{\ell
    m})|\big] \cosh^2{(c_{2,f}^{\ell m})}, \\
c_{2,c}^{\ell m} &= -\frac{ |h_{\ell
    m}^{\textrm{insp-plunge}}(t_{\textrm{match}}^{\ell
    m})|}{\nu} + \frac{1}{c_{1,f}^{\ell
    m} \nu} \big[ \partial_t|h_{\ell
    m}^{\textrm{insp-plunge}}(t_{\textrm{match}}^{\ell m})|  \nonumber \\
    &- \sigma^\textrm{R}_{\ell m} |h_{\ell
    m}^{\textrm{insp-plunge}}(t_{\textrm{match}}^{\ell
    m})|\big] \cosh{(c_{2,f}^{\ell m})}\sinh{(c_{2,f}^{\ell m})}, \\ \nonumber   
\end{align}
and 
$d_{1,c}^{\ell m}$ in terms of $d_{1,f}^{\ell m},\ d_{2,f}^{\ell m}, \sigma^\textrm{I}_{\ell
      m},\ \omega_{\ell
      m}^{\textrm{insp-plunge}}(t_{\textrm{match}}^{\ell m})$ as
\begin{align}    
d_{1,c}^{\ell m} &= \left[\omega_{\ell m}^{\textrm{insp-plunge}}(t_{\textrm{match}}^{\ell m}) -  \sigma^\textrm{I}_{\ell
      m}\right]\frac{1+ d_{2,f}^{\ell m}}{d_{1,f}^{\ell m}d_{2,f}^{\ell m}}.
\end{align}
Let us emphasize again that the values of $|h_{\ell
    m}^{\textrm{insp-plunge}}(t_{\textrm{match}}^{\ell
    m})|,\ \partial_t|h_{\ell
    m}^{\textrm{insp-plunge}}(t_{\textrm{match}}^{\ell m})|$ and $\omega_{\ell
      m}^{\textrm{insp-plunge}}(t_{\textrm{match}}^{\ell m})$ are fixed by the NQCs conditions in Eqs.~\eqref{eq:NQC_condition_1}~\eqref{eq:NQC_condition_2}~\eqref{eq:NQC_condition_4} to be the same as the NR values $\left|h_{\ell m}^\textrm{NR}(t_{\textrm{match}}^{\ell m})\right|, \partial_t|h_{\ell m}^\textrm{NR}(t_{\textrm{match}}^{\ell m})|$ and $\omega_{\ell m}^{\textrm{insp-plunge}}(t_{\textrm{match}}^{\ell m})$ which are given in Appendix~\ref{app:NQCfits} as function of $\nu$ and a combination of the spins $\chi_1$ and $\chi_2$.     
Thus, we are left 
with only two free parameters in the amplitude $c_{i,f}^{\ell m}$ and 
in the phase $d_{i,f}^{\ell m}$. To obtain those 
parameters we first extract them applying a least-square fit in each point of the
parameter space $(\nu,\chi_1,\chi_2)$ for which we have NR and 
Teukolsky-equation--based waveforms. Then, we interpolate those values in the rest 
of the parameter space using polynomial fits in $\nu$ and a combination of $\chi_1$ and $\chi_2$, as given 
explicitly in Appendix~\ref{app:ringdownfits}.

Regarding the accuracy of our merger-ringdown model, for the modes (2,1) and (3,3) the average fractional difference in the amplitude between the model and the NR waveform is of the
  order of percent, while the average phase difference is $\lesssim 0.1$ radians. For the modes (4,4) and (5,5) we are unable to 
  determine a similar average error, because those modes are 
  affected by numerical error at merger and during ringdown, as we discuss in Appendix \ref{app:ringdownfits}. 
  We find that the average fractional difference in the amplitude (phase) between the
  model and the NR simulation can be in some cases on the order of $10\%$ ($\lesssim 0.3$ rad), but this can be comparable to the difference between NR waveforms at different 
extraction radius (see Fig.~\ref{fig:noisyNR} in Appendix~\ref{app:ringdownfits}). We notice that although the
  errors in those modes are not as small as those of the modes (2,1) and (3,3), they
  are still acceptable considering the relatively small amplitude of
  the modes (4,4) and (5,5) with respect to the (2,1) and (3,3).

In summary, given a binary configuration $(m_1, m_2, \chi_1,\chi_2)$, 
the merger-ringdown model that we have developed is uniquely
  determined by the following parameters $(m_1,m_2,\chi_1,\chi_2,t_{\textrm{match}}^{\ell m},
  \phi_{\textrm{match}}^{\ell m}, \sigma^\textrm{I}_{\ell m},
  \sigma^\textrm{R}_{\ell m})$, the latter being a function of the
  remnant-BH's mass and spin determined by the NR fits. It is possible to use
  this merger-ringdown model as a stand-alone model (i.e., independently from the 
inspiral-plunge part), if we also provide equations relating $\phi_{\textrm{match}}^{\ell m}$ (i.e., the phase of the mode $(\ell, m)$ at $t_{\textrm{match}}^{\ell m}$) with $\phi_{\textrm{match}}^{2 2}$.
Indeed even if a
  global time and phase shift is possible, the relations between the phases of different modes are fixed.
  The latter are given as a fit for every point of the parameter space $(\nu,
  \chi_1,\chi_2)$ in Appendix~\ref{app:fitphasediff}.  We note that in this stand-alone
  merger-ringdown model, one can also treat $\sigma^\textrm{I}_{\ell m}$
  and $\sigma^\textrm{R}_{\ell m}$ as free parameters (i.e., we do not 
  compute them from Refs.~\cite{Berti:2005ys,Berti:2009kk}). In this case the 
  merger-ringdown model is a function of
  $(m_1,m_2,\chi_1,\chi_2,t_{\textrm{match}}^{\ell m},
  \phi_{\textrm{match}}^{\ell m}, \sigma^\textrm{I}_{\ell m},
  \sigma^\textrm{R}_{\ell m}, M_{\textrm{final}})$ where
  $M_{\textrm{final}}$ is the remnant-BH's, which is used only
  to rescale $\sigma^\textrm{I}_{\ell m}$ and $\sigma^\textrm{R}_{\ell
    m}$.

\begin{figure*}
    \centering
    \begin{minipage}{0.5\textwidth}
        \centering
        \includegraphics[width=1.0\textwidth]{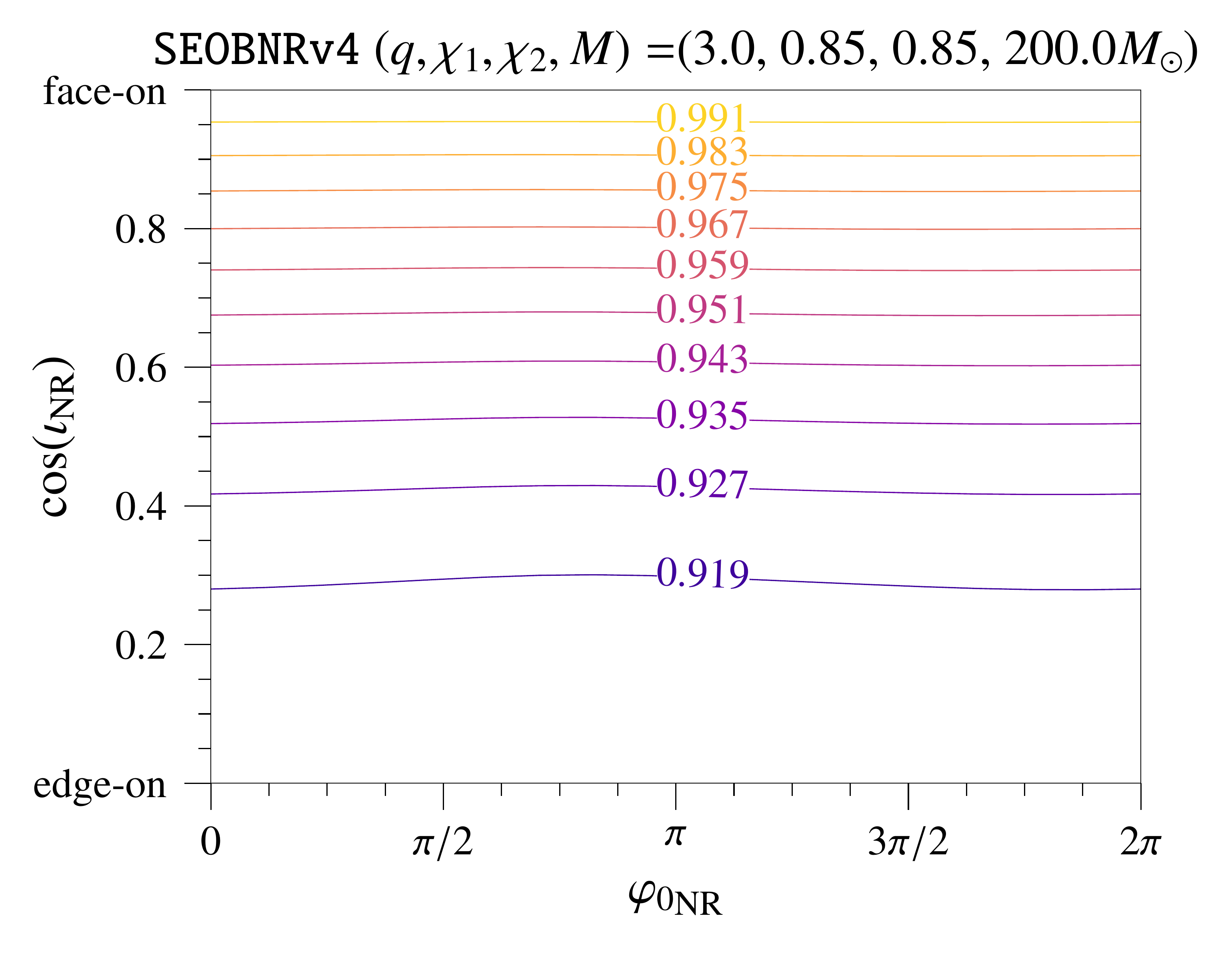} % first figure itself
    \end{minipage}\hfill
    \begin{minipage}{0.5\textwidth}
        \centering
        \includegraphics[width=1.0\textwidth]{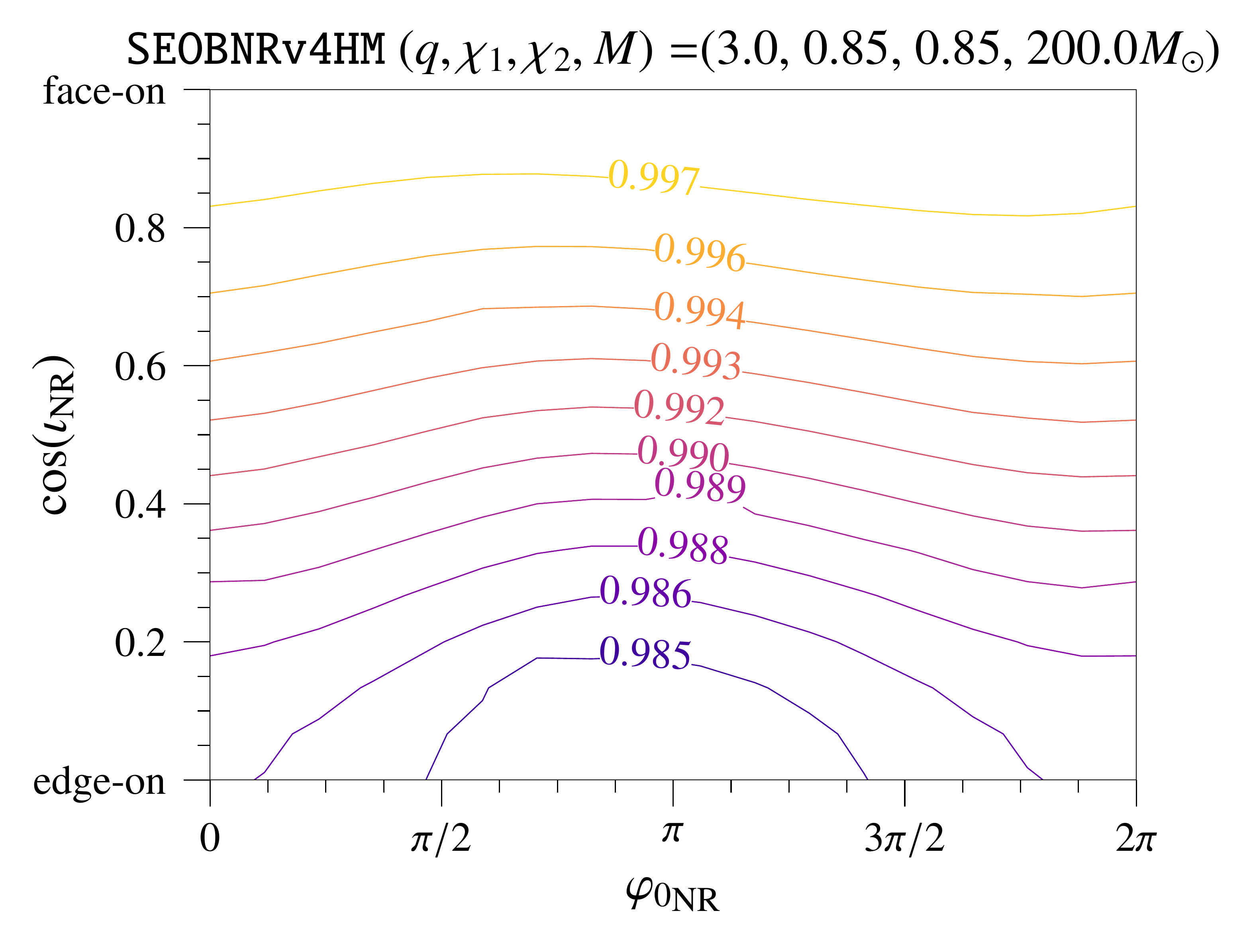} % second figure itself
    \end{minipage}
\caption{Faithfulness $\mathcal{F}(\cos(\iota_{\textrm{NR}}),{\varphi_0}_{\textrm{NR}},\kappa_{\textrm{NR}} = 0)$ for the configuration $(q = 3,\, M = 200 M_\odot,\, \chi_1 = 0.85, \, \chi_2 = 0.85)$: NR $(\ell \leq 5, \, m \neq 0)$ vs \texttt{SEOBNRv4} (left panel), NR $(\ell \leq 5, \, m \neq 0)$ vs \texttt{SEOBNRv4HM} (right panel). We plot the faithfulness for a fixed $\kappa_{\textrm{NR}}$ because we have noted that $\mathcal{F}(\iota_{\textrm{NR}},{\varphi_0}_{\textrm{NR}},\kappa_{\textrm{NR}})$ is mildly dependent on this variable.}
\label{fig:q3v4vsHMthetaphi}
\end{figure*}

\section{Performance of the multipolar effective-one-body waveform model}
\label{sec:comparison}

We study the accuracy of the multipolar waveform model
\texttt{SEOBNRv4HM} by computing its faithfulness against waveforms
in the NR catalog at our disposal.  In Secs.~\ref{comparison:NR1} and
\ref{comparison:NR2}, we perform a detailed comparison against three
NR simulations, notably a moderate--mass-ratio configuration,
\texttt{SXS:BBH:0293} $(q = 3,\, \chi_1 = 0.85,\, \chi_2 = 0.85)$, and
two high--mass-ratio configurations, \texttt{SXS:BBH:0065} $(q = 8,\,
\chi_1 = 0.5,\, \chi_2 = 0)$ and \texttt{ET:AEI:0004} $(q = 8,\,
\chi_1 = 0.85,\, \chi_2 = 0.85)$. We also compare the results above
with those obtained when the (2,2)--waveform-model \texttt{SEOBNRv4}
is employed. Finally, in
Sec.~\ref{comparison:NRcatalog} we summarize the agreement of
the \texttt{SEOBRNv4HM} model against the entire NR catalog composed of 157
simulations.

\subsection{Moderate mass ratio: \texttt{SXS:BBH:0293}}
\label{comparison:NR1}

In the left panel of Fig.~\ref{fig:q3v4vsHMthetaphi} we show a contour
plot of the faithfulness
$\mathcal{F}(\cos(\iota_{\textrm{NR}}),{\varphi_0}_{\textrm{NR}},\kappa_{\textrm{NR}})\big|_{\kappa_{\textrm{NR}}
  = 0}$ between the NR waveform \texttt{SXS:BBH:0293} with modes
($\ell \leq 5, \,m \neq 0$), and the waveform generated with
\texttt{SEOBNRv4}, for a total mass of $M = 200 M_\odot$. In order to
reduce the dimensionality of the plot, we fix the value of
$\kappa_{\textrm{NR}}$. However, we find that the dependence of the faithfulness on this
variable is mild. We can see that the 
faithfulness depends mainly on the inclination angle
$\iota_{\textrm{NR}}$ and degrades when we move from a face-on
$\{\mathcal{F}(\cos(\iota_{\textrm{NR}}) = 0) \sim 99\%\}$ to an
edge-on orientation $\{\mathcal{F}(\cos(\iota_{\textrm{NR}}) = 1)
\sim 92\%\}$. This situation is different if we include the
higher-order modes in the model (i.e, $(3,3),(2,1),(4,4),(5,5)$), 
as can be seen in the right panel of Fig.~\ref{fig:q3v4vsHMthetaphi} 
where we use the \texttt{SEOBNRv4HM} waveform model.
In this case the faithfulness degrades much less if we go from a
face-on ($\mathcal{F} \sim 99.7\%$) to an edge-on ($\mathcal{F} \sim
98.5\%$) orientation. The small residual degradation is due to the
fact that the dominant mode is still better modeled than the
higher-order modes and for this reason for a face-on orientation
(where the signal is dominated by the dominant mode) the faithfulness
is larger than for an edge-on orientation where the higher-order modes
contribute the most. Another contribution to the residual degradation
in an edge-on orientation stems from the fact that in the
\texttt{SEOBNRv4HM} model we still miss some subdominant higher-order
modes, which instead we have included in the NR waveform.

\begin{figure}[htb]
\centering
\includegraphics[width=0.5\textwidth]{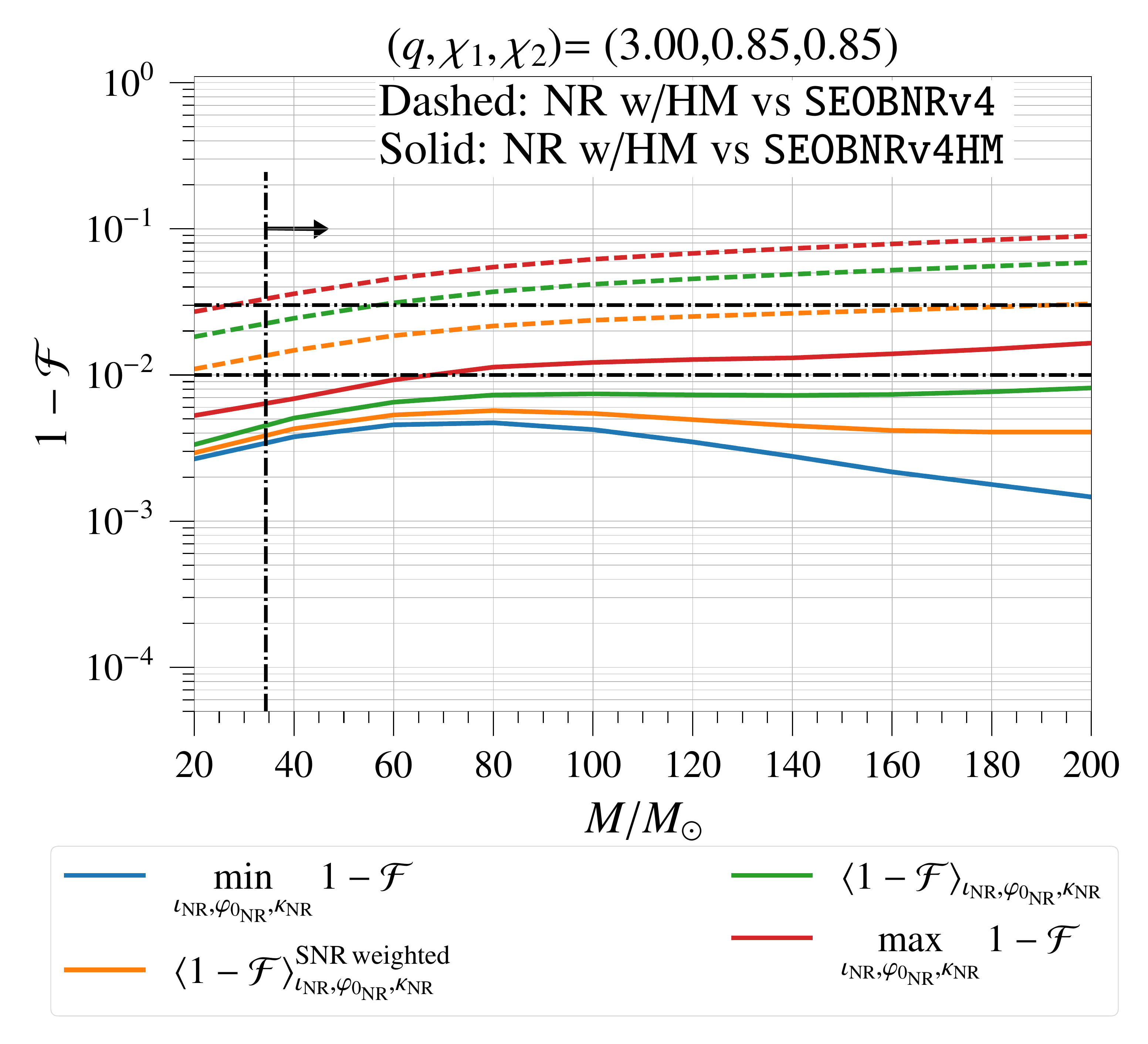}
\caption{Unfaithfulness $(1-\mathcal{F})$ for the
  configuration $(q = 3,\, \chi_1 = \chi_2 = 0.85)$ in the mass range
  $20 M_\odot \leq M \leq 200 M_\odot$. Dashed (plain) curves refer to results for
  \texttt{SEOBNRv4} (\texttt{SEOBNRv4HM}). The minima of the unfaithfulness for the two
  models (blue curves), lie on top of each other because they are
  reached for a face-on orientation, where the higher modes
  contribution is zero. The unfaithfulness averaged over the three
  angles
  $\iota_{\textrm{NR}},{\varphi_0}_{\textrm{NR}},\kappa_{\textrm{NR}}$ are
  obtained assuming an isotropic distribution for the source
  orientation, homogeneous distribution in GW polarization and isotropic distribution in sky position (green curves and orange curves for the average weighted
  with the SNR). The minimum of the unfaithfulness (red curves) in
  practice correspond to an edge-on orientation, minimized over the
  other two angles. The vertical dotted-dashed black line is the
  smallest mass at which the $(2,1)$ mode is entirerly in
  the Advanced LIGO band. The $(\ell, m')$ mode is entirerly in the Advanced LIGO band
  starting from a mass $m'$ times the mass associated with the $(2,1)$ mode.  
 The horizontal dotted-dashed black lines
  represent the values of $1\%$ and $3\%$ unfaithfulness.}
\label{fig:unfaith_mass_q3chi085}
\end{figure}

As done in Sec. \ref{sec:faithfulness} we summarize the results of the
faithfulness calculation in Fig.~\ref{fig:unfaith_mass_q3chi085}, where 
we show the minimum and maximum of the unfaithfulness over the
NR orientations, GW polarization and sky position, respectively indicated as $\min_{\iota_{\mathrm{NR}},{\varphi_0}_{\mathrm{NR}},\kappa_{\mathrm{NR}}}
(1 -\mathcal{F})$ (blue) and
$\max_{\iota_{\mathrm{NR}},{\varphi_0}_{\mathrm{NR}},\kappa_{\mathrm{NR}}}
(1 -\mathcal{F})$ (red); the average of the unfaithfulness
over these three angles $
\langle1-\mathcal{F}\rangle_{\iota_{\mathrm{NR}},{\varphi_0}_{\mathrm{NR}},\kappa_{\mathrm{NR}}}$
(green), and the average of the unfaithfulness weighted with the
cube of the SNR: $\langle1-
\mathcal{F}\rangle_{\iota_{\mathrm{NR}},{\varphi_0}_{\mathrm{NR}},\kappa_{\mathrm{NR}}}^{\mathrm{SNRweighted}}$
(orange). All the averages are computed assuming an isotropic distribution 
for the source orientation, homogeneous distribution in GW polarization and isotropic distribution in sky position.  
All these quantities are shown as a function of the total mass of the system. In the plots the plain curves are the
results of the unfaithfulness between the NR and \texttt{SEOBNRv4HM} waveforms, 
while dashed curves are the results of the unfaithfulness between NR
and \texttt{SEOBNRv4} waveforms. In this case, the maximum and the averaged
values of the unfaithfulness for the \texttt{SEOBNRv4} model are one order of
magnitude larger than the ones with the \texttt{SEOBNRv4HM} model. The minimum of
the unfaithfulness is the same for both models (blue curves lying on
top of each other) because it is reached for a face-on orientation, where
the contribution of the higher-order modes used for \texttt{SEOBNRv4HM} is zero. Indeed the -2 spin-weighted 
spherical harmonics associated to these higher-order modes 
go to zero for face-on orientations. We note also that in \texttt{SEOBNRv4}, as expected, the disagreement grows strongly with the
total mass of the system, because higher-order modes are more 
important toward merger and ringdown.

\subsection{High mass ratios: \texttt{SXS:BBH:0065} and \texttt{ET:AEI:0004}}
\label{comparison:NR2}

\begin{figure*}
    \centering
    \begin{minipage}{0.5\textwidth}
        \centering
        \includegraphics[width=1.0\textwidth]{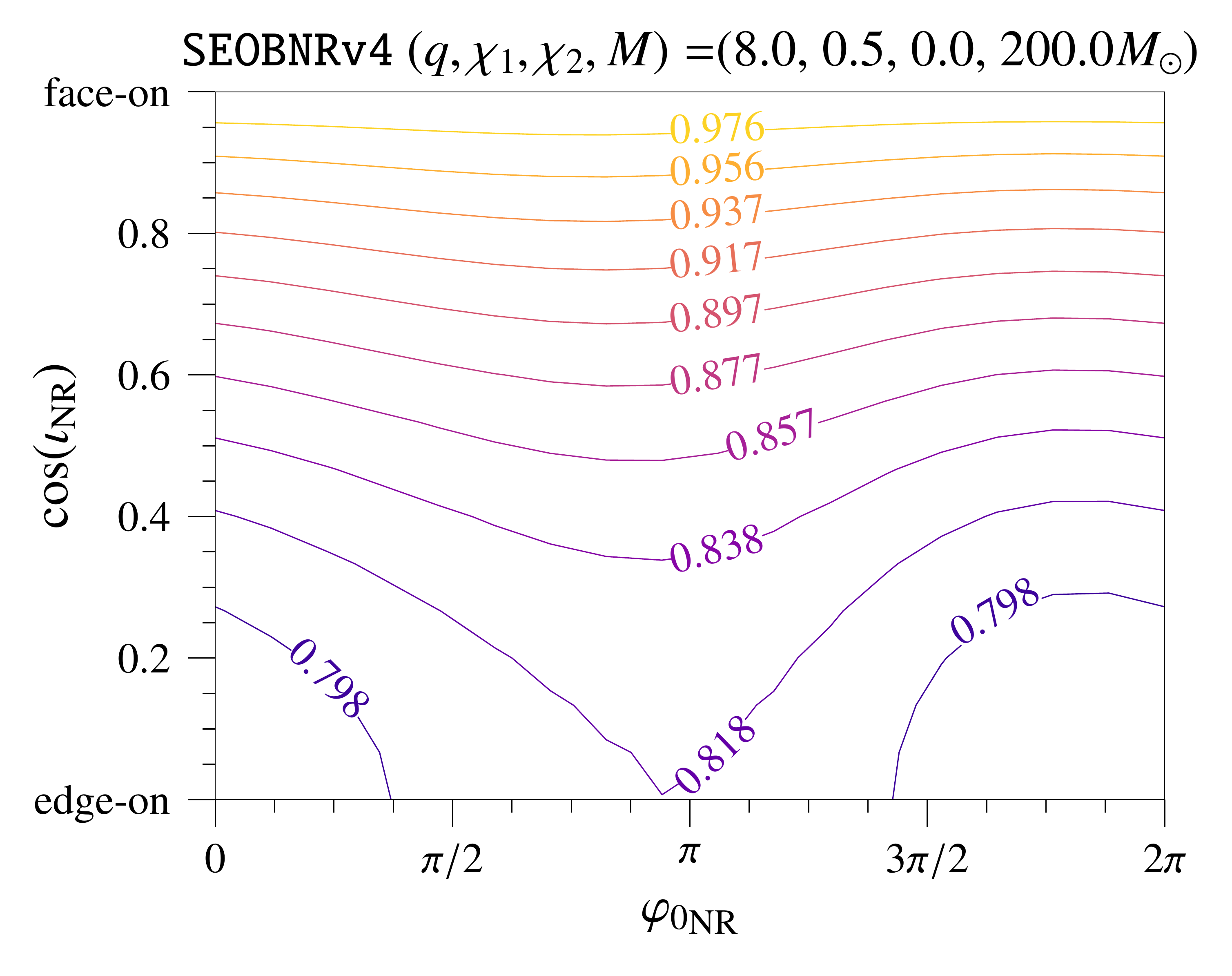} % first figure itself
    \end{minipage}\hfill
    \begin{minipage}{0.5\textwidth}
        \centering
        \includegraphics[width=1.0\textwidth]{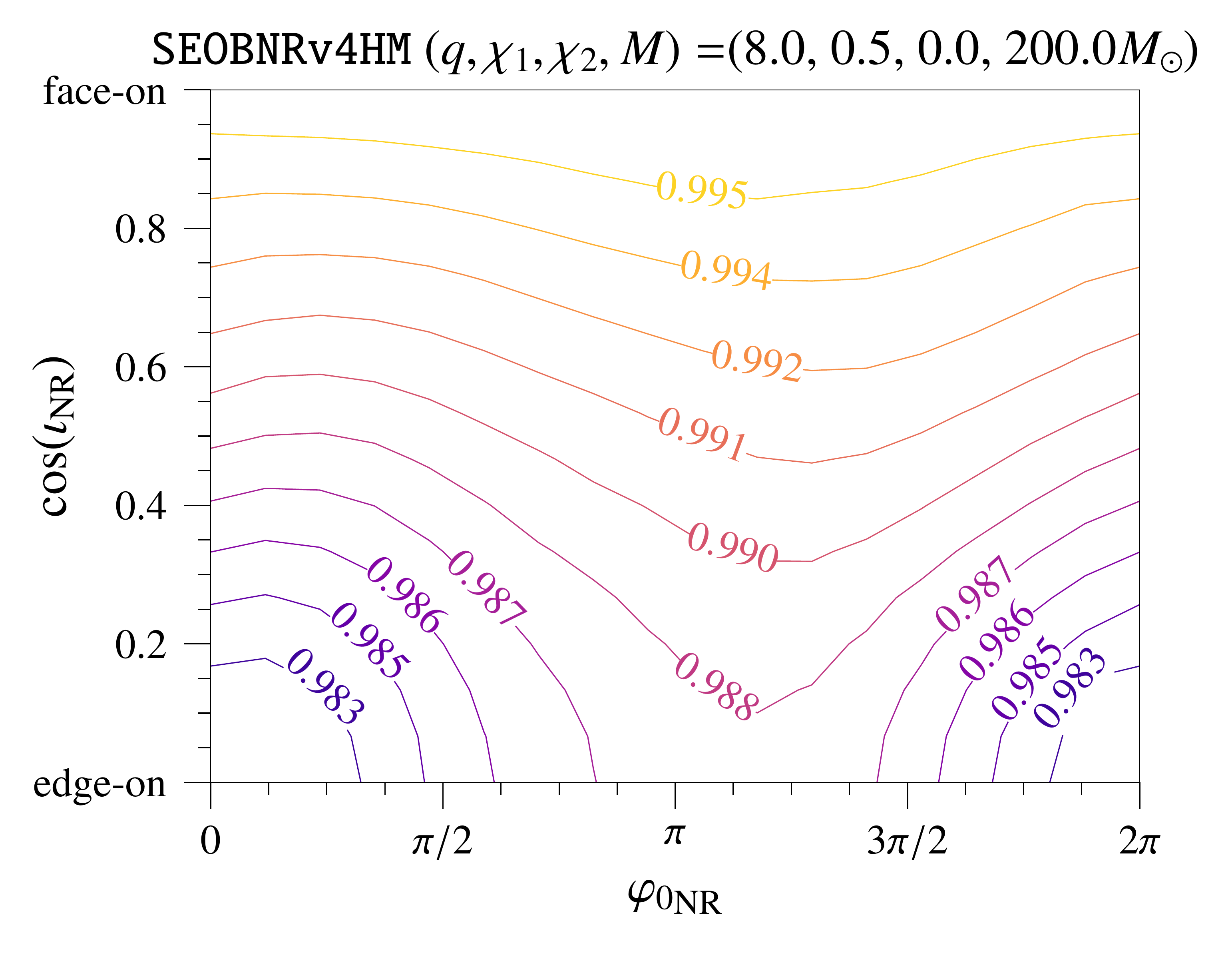} % second figure itself
    \end{minipage}
\caption{Faithfulness $\mathcal{F}(\cos(\iota_{\textrm{NR}}),{\varphi_0}_{\textrm{NR}},\kappa_{\textrm{NR}} = 0)$ for the configuration $(q = 8,\, M = 200 M_\odot,\, \chi_1 = 0.5, \, \chi_2 = 0)$: NR $(\ell \leq 5,\, m \neq 0)$ vs \texttt{SEOBNRv4} (left panel), NR $(\ell \leq 5, \, m \neq 0)$ vs \texttt{SEOBNRv4HM} (right panel).}
\label{fig:q8v4vsHMthetaphi}
\end{figure*}

\begin{figure*}
    \centering
    \begin{minipage}{0.5\textwidth}
        \centering
        \includegraphics[width=1.0\textwidth]{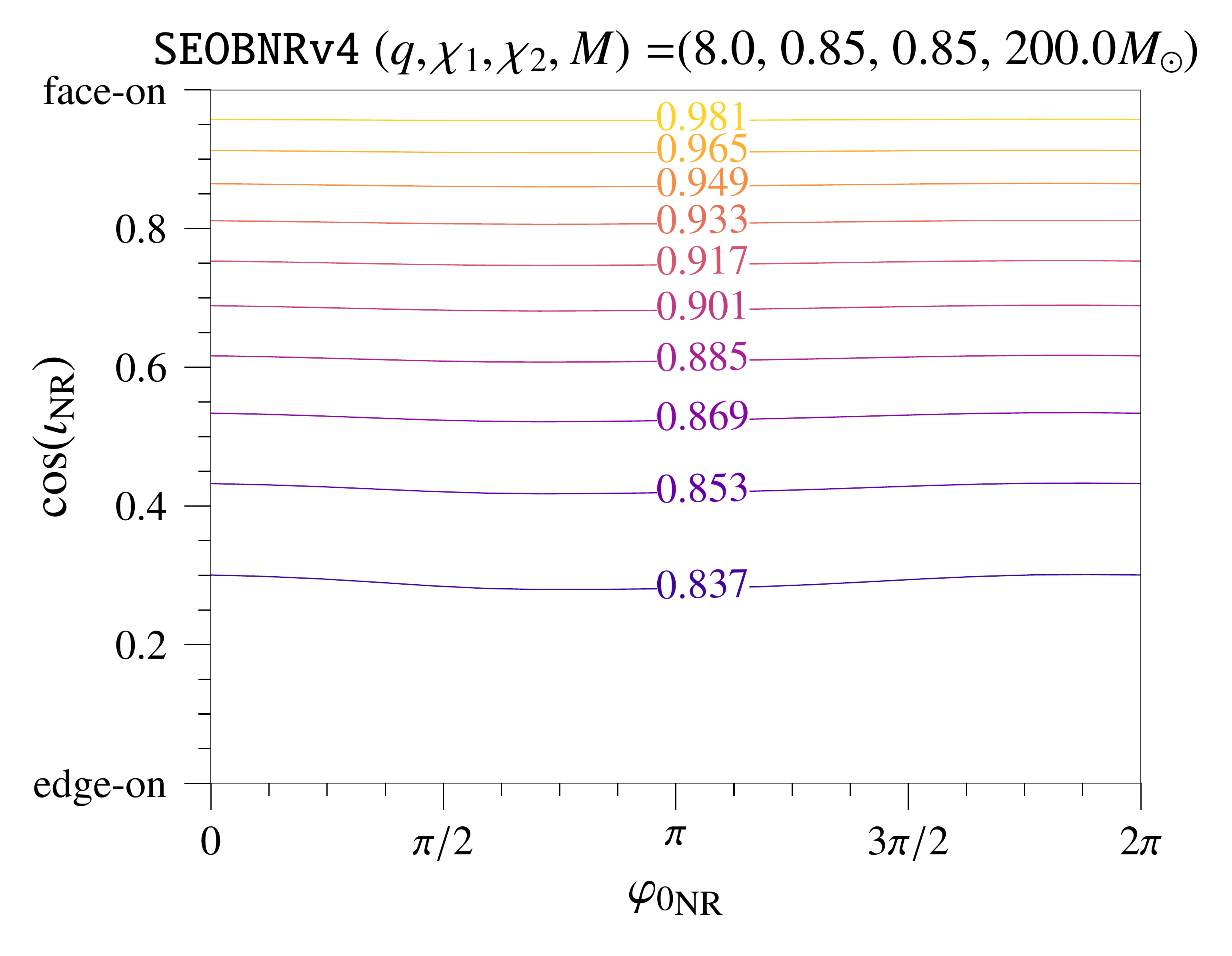} % first figure itself
    \end{minipage}\hfill
    \begin{minipage}{0.5\textwidth}
        \centering
        \includegraphics[width=1.0\textwidth]{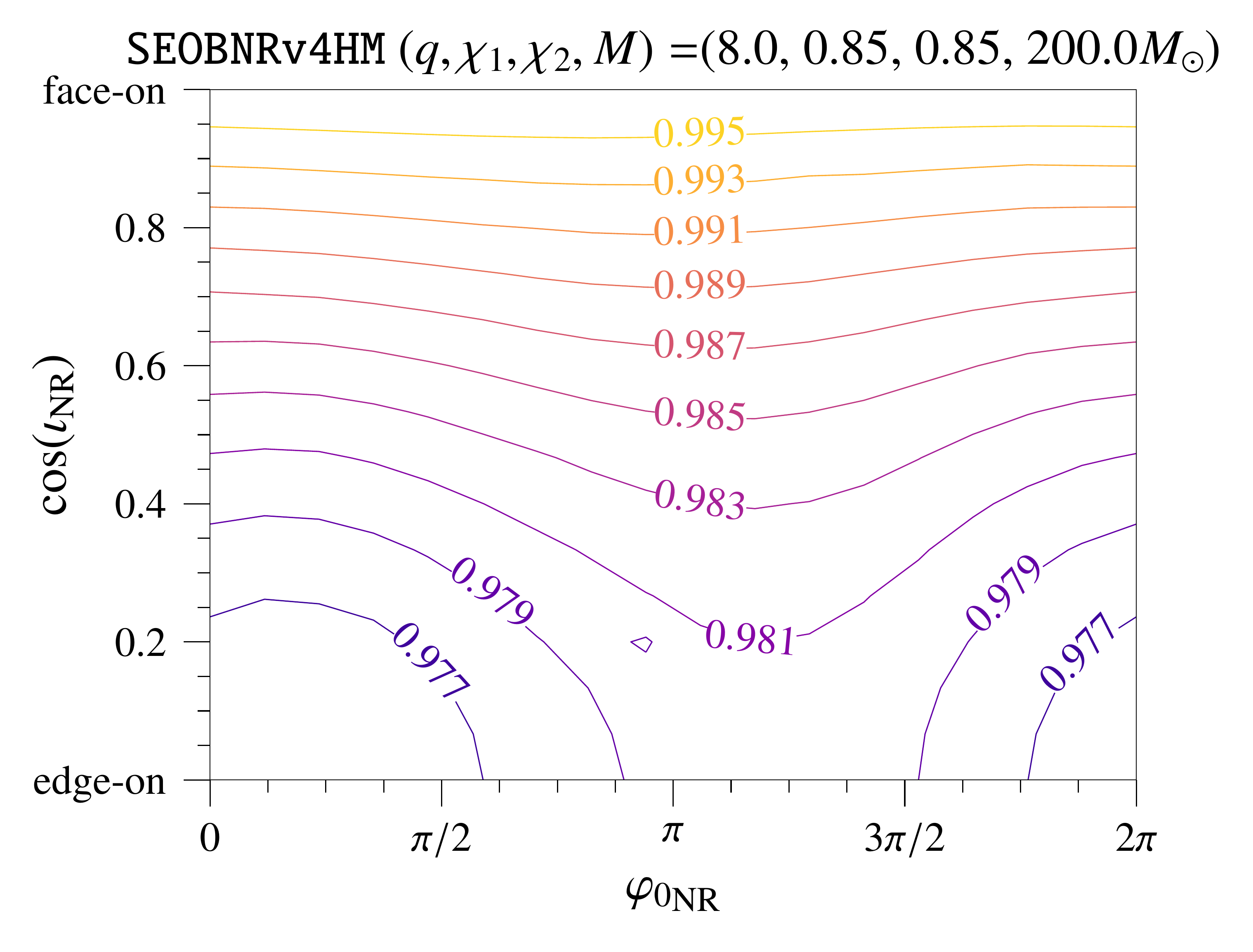} % second figure itself
    \end{minipage}
\caption{Faithfulness $\mathcal{F}(\cos(\iota_{\textrm{NR}}),{\varphi_0}_{\textrm{NR}},\kappa_{\textrm{NR}} = 0)$ for the configuration $(q = 8,\, M = 200 M_\odot,\, \chi_1 = 0.85, \, \chi_2 = 0.85)$: NR $(\ell \leq 5,\, m \neq 0)$ vs \texttt{SEOBNRv4} (left panel), NR $(\ell \leq 5,\, m \neq 0)$ vs \texttt{SEOBNRv4HM} (right panel).}
\label{fig:q8sv4vsHMthetaphi}
\end{figure*}

More striking conclusions about the improvement of the waveform model due to the inclusion of higher-order modes can be drawn looking at the comparison with the two NR simulations \texttt{SXS:BBH:0065} and \texttt{ET:AEI:0004}, 
for which higher-order modes are expected to be more important,  
because of the higher mass ratio. For the first
configuration $(q = 8,\, M = 200 M_\odot,\, \chi_1 = 0.5, \, \chi_2 =
0)$ we see in Fig.~\ref{fig:q8v4vsHMthetaphi} that the
faithfulness between the NR $(\ell \leq 5, \, m\neq 0)$ and the
\texttt{SEOBNRv4} waveforms (left panel) degrades much faster than
before as a function of the inclination angle $\iota_{\textrm{NR}}$,
reaching $\mathcal{F} \lesssim 90\%$ already for values of
$\cos(\iota_{\textrm{NR}}) \sim 0.7$ ($\iota_{\textrm{NR}} \sim 45
\degree)$, being very large for the edge-on inclination
$\mathcal{F} \sim 80\%$. Similarly to what happens for the example discussed 
in Sec.~\ref{comparison:NR1}, the situation is much better if we include in the 
model the higher modes, as can be seen in Fig.~\ref{fig:q8v4vsHMthetaphi} (right panel). 
Now, the degradation as a function of $\iota_{\textrm{NR}}$ is much weaker
and for edge-on orientations the faithfulness reaches values close to
$\mathcal{F} \sim 98 \%$. Similar conclusions can be drawn by looking at Fig.~\ref{fig:q8sv4vsHMthetaphi}, whch refers 
to the simulation \texttt{ET:AEI:0004} $(q = 8,\, M = 200 M_\odot,\, \chi_1 = 0.85, \,
\chi_2 = 0.85)$. The only relevant difference with respect to the
aforementioned case is that in this case the faithfulness of
the \texttt{SEOBNRv4HM} waveform is a little bit smaller and it goes down to
$\mathcal{F} \sim 97.7\%$ in the edge-on orientations. At a fixed binary orientation, the faithfulness of the (2,2)--waveform-model \texttt{SEOBNRv4} against the NR waveform for the configuration $(q = 8,\, M = 200 M_\odot,\, \chi_1 = 0.85 = \chi_2 =
0.85)$ is always larger than that for the configuration  $(q = 8,\, M = 200 M_\odot,\, \chi_1 = 0.5, \, \chi_2 =
0)$. This can be explained considering that, as discussed in Sec.~\ref{sec:motivations}, for a fixed mass ratio the $(2,1)$ mode is increasingly suppressed when the spin of the heavier BH grows, while the other higher-order modes are mostly constant as a function of the spins. Since in the first case $\chi_1$, that is the spin of the heavier BH, is larger than in the second case, the $(2,1)$ mode is more suppressed in the first case than in the second one. For this reason the faithfulness with the \texttt{SEOBNRv4} model, including only the dominant mode, is higher for the first configuration.

\begin{figure*}
    \centering
    \begin{minipage}{0.5\textwidth}
        \centering
        \includegraphics[width=1.0\textwidth]{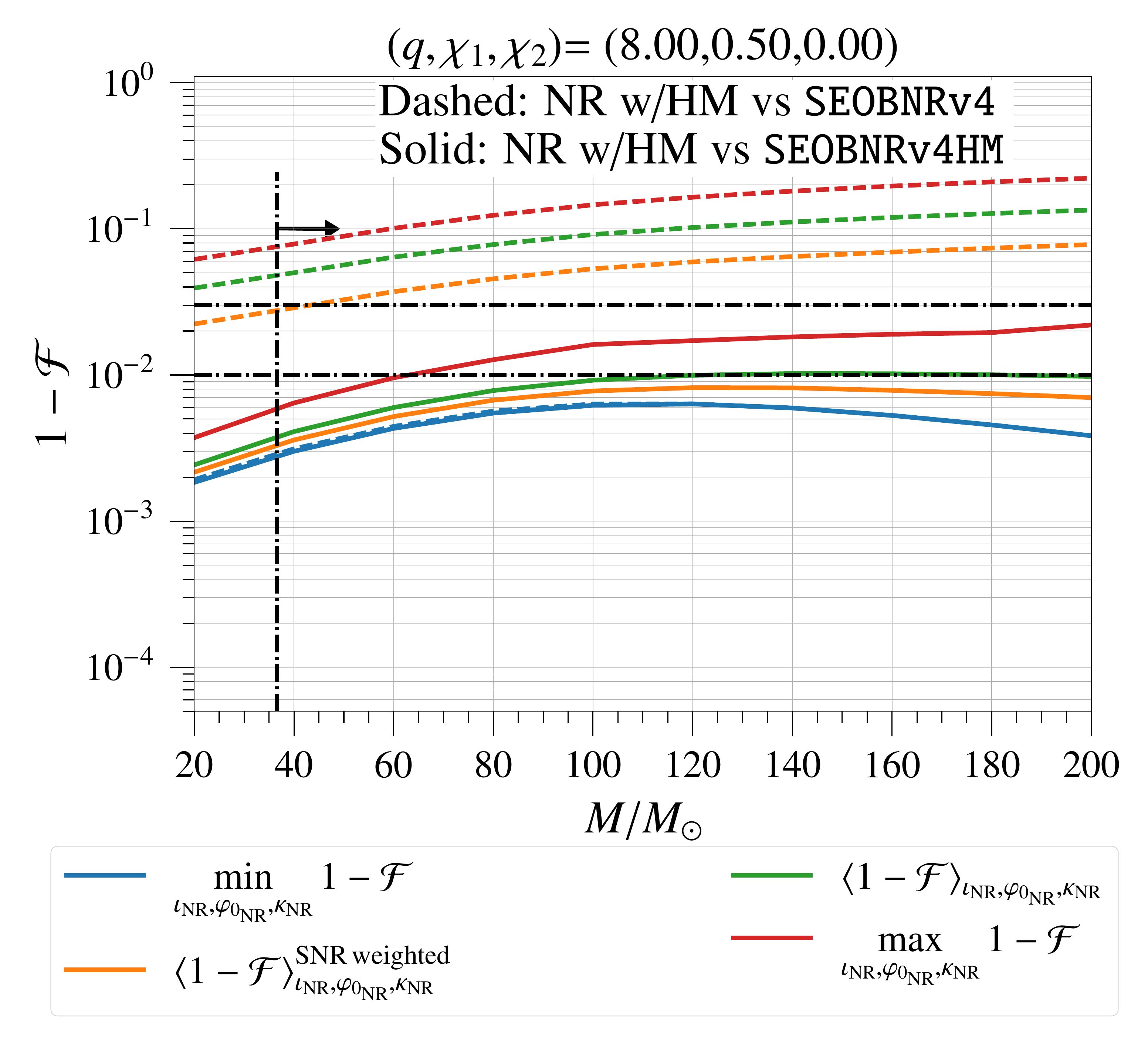} % first figure itself
    \end{minipage}\hfill
    \begin{minipage}{0.5\textwidth}
        \centering
        \includegraphics[width=1.0\textwidth]{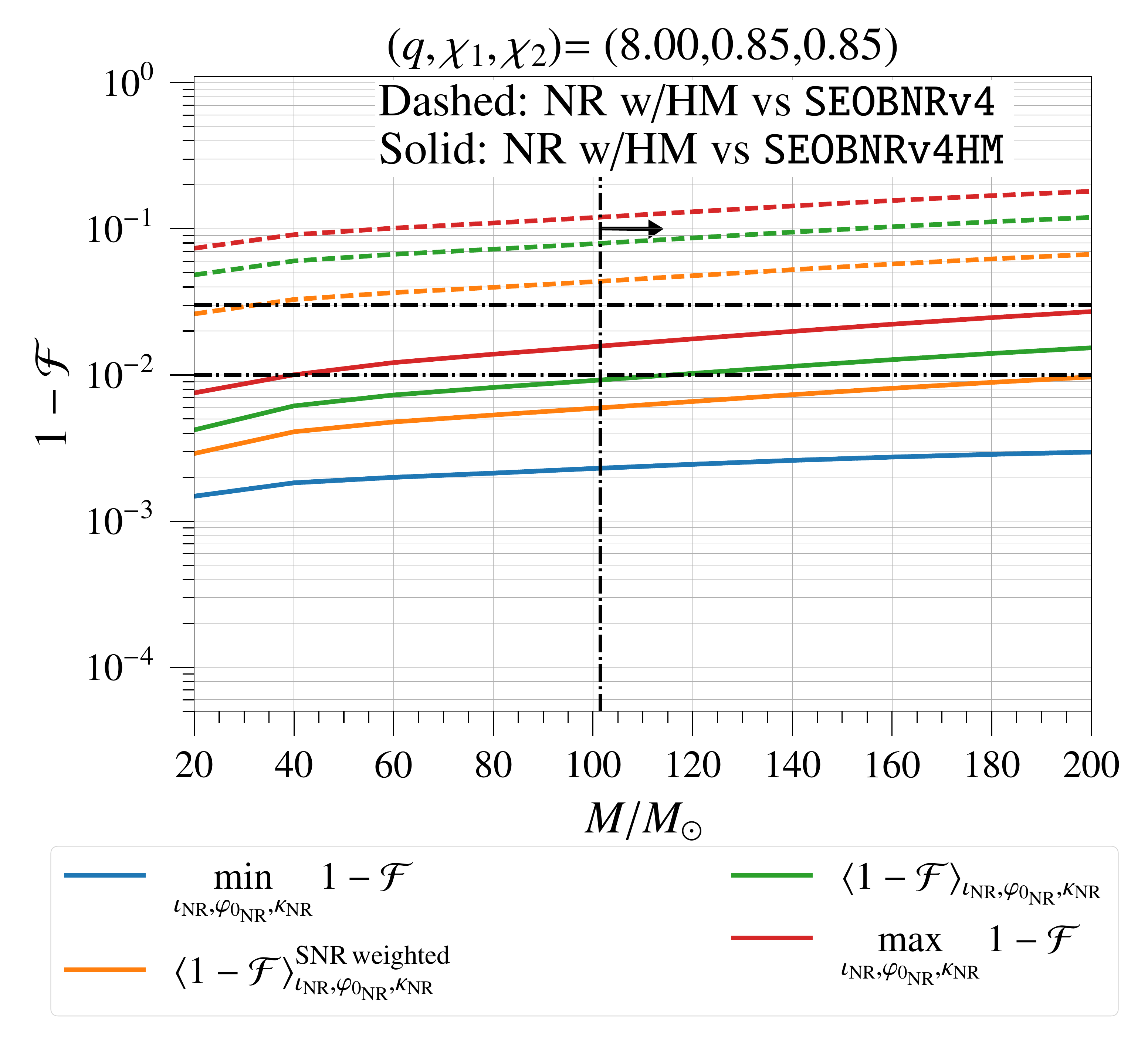} % second figure itself
    \end{minipage}
\caption{Unfaithfulness $(1-\mathcal{F})$ in the mass range $20 M_\odot \leq M \leq 200 M_\odot$ for the configuration $(q = 8,\, \chi_1 = 0.5,\, \chi_2 = 0)$ (left panel) and $(q = 8,\, \chi_1 = 0.85,\, \chi_2 = 0.85)$ (right panel). Plotted data as in Fig.~\ref{fig:unfaith_mass_q3chi085}}
\label{unfaith_mass_q8}
\end{figure*}

As for the previous configuration, in Fig.~\ref{unfaith_mass_q8}, we
show the summary of the faithfulness results as maximum, minimum and
averages of the unfaithfulness, respectively for \texttt{SXS:BBH:0065}
(left panel) and \texttt{ET:AEI:0004} (right panel). For these binary
configurations, even if the maxima of the unfaithfulness have larger
values with respect to the case discussed in the previous section (
$\sim 2\%$ for \texttt{SXS:BBH:0065} and $\sim 2.7\%$ for
\texttt{ET:AEI:0004} at a total mass of $M = 200M_\odot$), we still
have acceptable values of the unfaithfulness averaged over the
orientations, sky position and polarizations: respectively $\sim 1\%$
and $\sim1.6\%$ for a total mass of $M = 200M_\odot$. This is a big
improvement with respect to the \texttt{SEOBNRv4} model, which gives
averaged values of the unfaithfulness larger than $10\%$ for both
configurations and the same total mass.  For the configuration
  with $q = 8,\, \chi_1 = 0.85 = \chi_2 = 0.85$, the unfaithfulness
  against the NR simulation was also computed for the multipolar
  waveform model developed in Ref.~\cite{London:2017bcn}, and found to 
  be around $\sim 5\%$ for $\iota_\mathrm{s} =
  \pi/2$, when averaging over the angles $\kappa_\mathrm{s}$ and
  ${\varphi_0}_s$ for a total mass $M = 100 M_\odot$. In our model the
  maximum of the unfaithfulness
  (i.e., $\max_{\iota_{\mathrm{s}},{\varphi_0}_{\mathrm{s}},\kappa_{\mathrm{s}}}(1
  -\mathcal{F})$) over the angles
  $\iota_{\mathrm{s}},{\varphi_0}_{\mathrm{s}}$ and
  $\kappa_{\mathrm{s}}$ is around $1.5\%$ at $M = 100 M_\odot$. The
  reason for the better accuracy of \texttt{SEOBNRv4HM} model with respect
  to the waveform model in Ref.~\cite{London:2017bcn} for this ``extreme''
  binary configuration might be due to the fact that the simple scaling argument used
  there to build the higher-order modes is not very accurate for
  high-mass ratio and high-spin binary systems. We leave to the future a 
direct, comprehensive comparison between the two waveform models.

As discussed in Sec.~\ref{sec:faithfulness}, an important quantity to assess the 
improvement that \texttt{SEOBNRv4HM} could yield for detecting BBHs 
is the average unfaithfulness weighted with the cube of the
SNR. For this quantity our model yields values of $\sim 0.7\%$ for
\texttt{SXS:BBH:0065} and $\sim 1\%$ for \texttt{ET:AEI:0004} at a
total mass of $M = 200M_\odot$ compared to values around $\sim 7\%$
returned by the \texttt{SEOBNRv4} model.

\subsection{Comparison with entire numerical-relativity catalog}
\label{comparison:NRcatalog}

\begin{figure*}
    \centering
    \begin{minipage}{0.5\textwidth}
        \centering
        \includegraphics[width=1.0\textwidth]{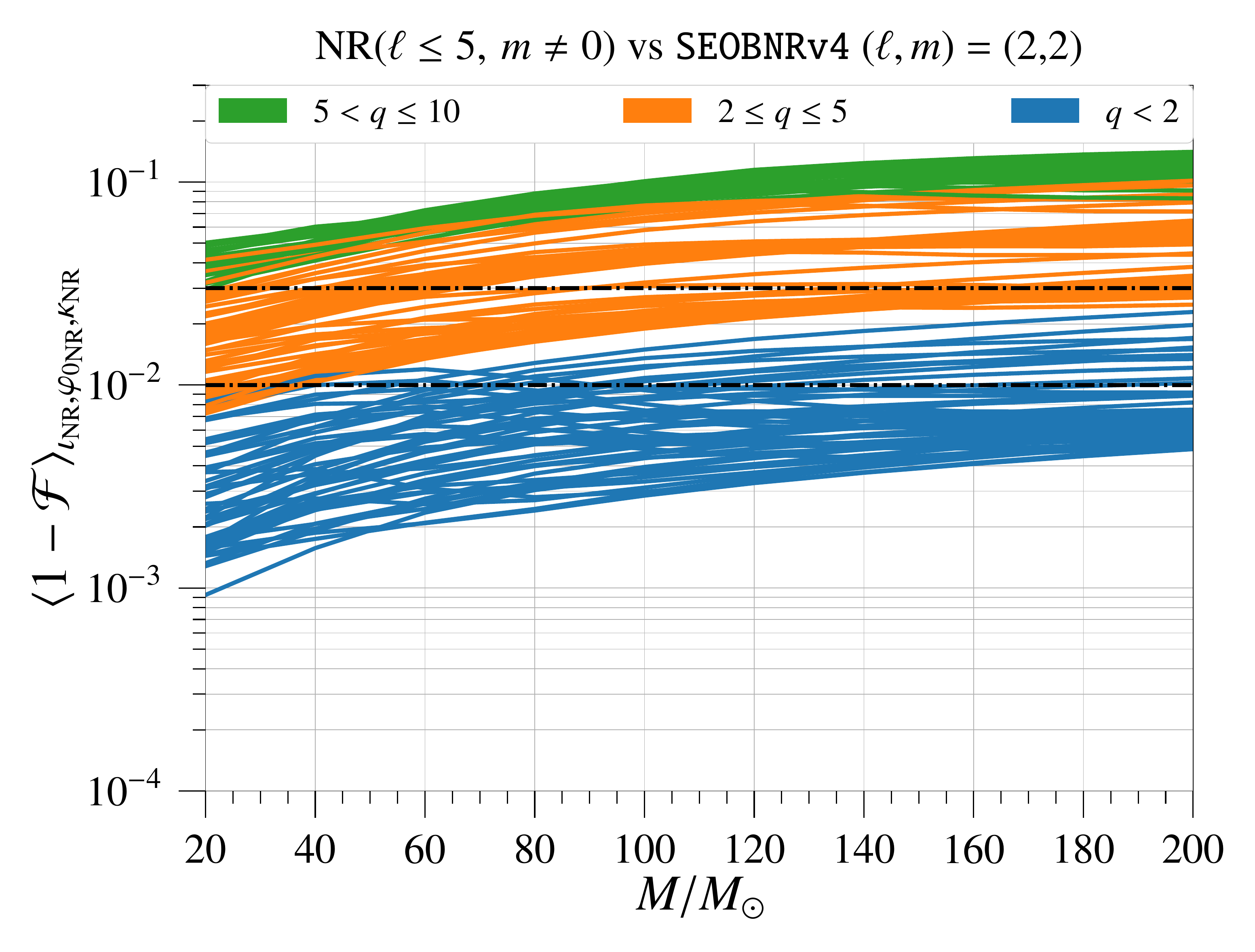} % first figure itself
    \end{minipage}\hfill
    \begin{minipage}{0.5\textwidth}
        \centering
        \includegraphics[width=1.0\textwidth]{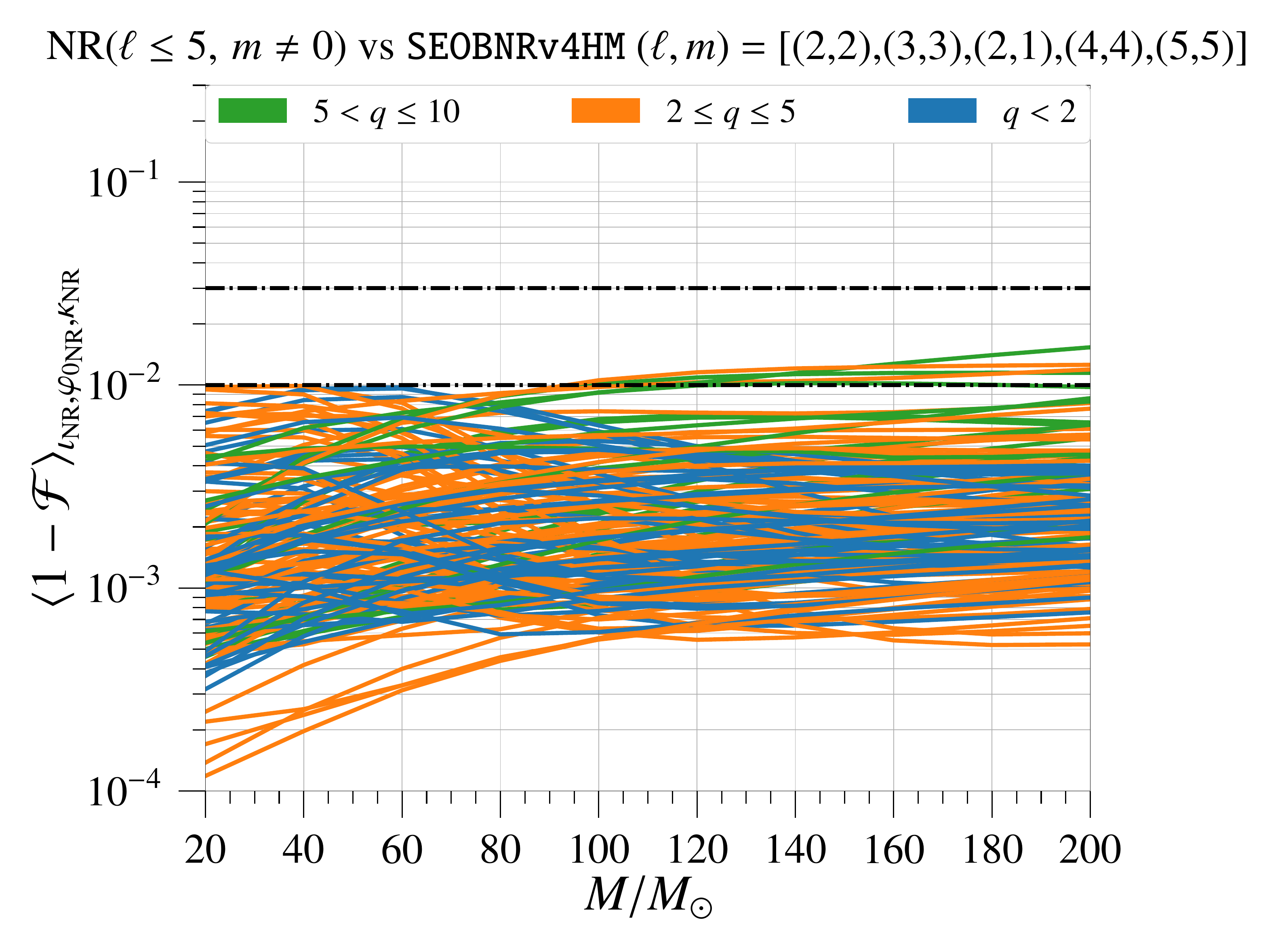} % second figure itself
    \end{minipage}
\caption{Unfaithfulness $(1-\mathcal{F})$ averaged over the three angles $(\iota_{\textrm{NR}},{\varphi_0}_{\textrm{NR}},\kappa_{\textrm{NR}})$ as a function of the total mass, in the range $20 M_\odot \leq M \leq 200 M_\odot$. Left panel NR $(\ell \leq 5, \, m \neq 0)$ vs \texttt{SEOBNRv4}, right panel NR $(\ell \leq 5,\, m \neq 0)$ vs \texttt{SEOBNRv4HM}. The horizontal dotted-dashed black lines represent the values of $1\%$ and $3\%$ unfaithfulness.}
\label{fig:skyaverageall}
\end{figure*}

Having studied in detail some particular configurations, we can now examine how the model works over the entire NR waveform catalog at our 
disposal. In Fig.~\ref{fig:skyaverageall} we plot the angle-averaged 
unfaithfulness as a function of the total mass of the system,
computed between the NR waveforms with modes $(\ell
\leq 5, m \neq 0)$ and the \texttt{SEOBNRv4} model (left panel), 
\texttt{SEOBNRv4HM} model (right panel). Comparing the two panels, we can see that \texttt{SEOBNRv4HM} 
yields unfaithfulnesses one order of magnitude
smaller than those of the \texttt{SEOBNRv4} model.  In the plots
different colors correspond to different ranges of mass ratios, and
from the left panel it is visible that in the case of the 
\texttt{SEOBNRv4} model, there is a clear hierarchy for which configurations
with higher mass ratios have also larger unfaithfulness. This effect
is removed in the \texttt{SEOBNRv4HM} model, as visible in the right panel of the
same figure. In general for all of NR simulations the averaged
unfaithfulness against \texttt{SEOBNRv4HM} is always smaller than $1 \%$ in the
mass range $20 M_\odot \leq M \leq 200 M_\odot$ with the exception of
few simulations for which the unfaithfulness reaches values $\leq 1.5\%$ for a total mass of $M = 200 M_\odot$: \texttt{SXS:BBH:0202} $(q = 7,\, \chi_1 = 0.6,\,
\chi_2 = 0)$, \texttt{ET:AEI:0004} $(q = 8,\, \chi_1 = 0.85,\,
\chi_2 = 0.85)$, \texttt{ET:AEI:0001} $(q = 5,\, \chi_1 = 0.8,\,
\chi_2 = 0)$ and \texttt{SXS:BBH:0061} $(q = 5,\, \chi_1 = 0.5,\,
\chi_2 = 0)$. These are the configurations in the NR catalog having the most extreme values of mass ratio and spins.
The results of this analysis does not change considerably if
we include in the NR waveforms only the modes used in the \texttt{SEOBNRv4HM} 
model, because, when looking at averaged unfaithfulness, the error is dominated by the imperfect modeling of the 
$(2,1), (3,3), (4,4),(5, 5)$ modes, and not by neglecting other 
subdominant higher modes, as discussed in Sec.~\ref{sec:faithfulness}.

The comparison between the unfaithfulness averaged
over the three angles $(\iota_{\textrm{NR}},{\varphi_0}_{\textrm{NR}},\kappa_{\textrm{NR}})$ and
weighted by the cube of the SNR of two waveform models against NR waveforms displays similar features, with the only difference of having overall smaller values of the unfaithfulness 
(always $\leq 1\%$ for the \texttt{SEOBNRv4HM} model). This happens because weighting with the SNR 
favours orientations closer to face-on for which the best
modeled $(2,2)$ mode is dominant.

\begin{figure*}
    \centering
    \begin{minipage}{0.5\textwidth}
        \centering
        \includegraphics[width=1.0\textwidth]{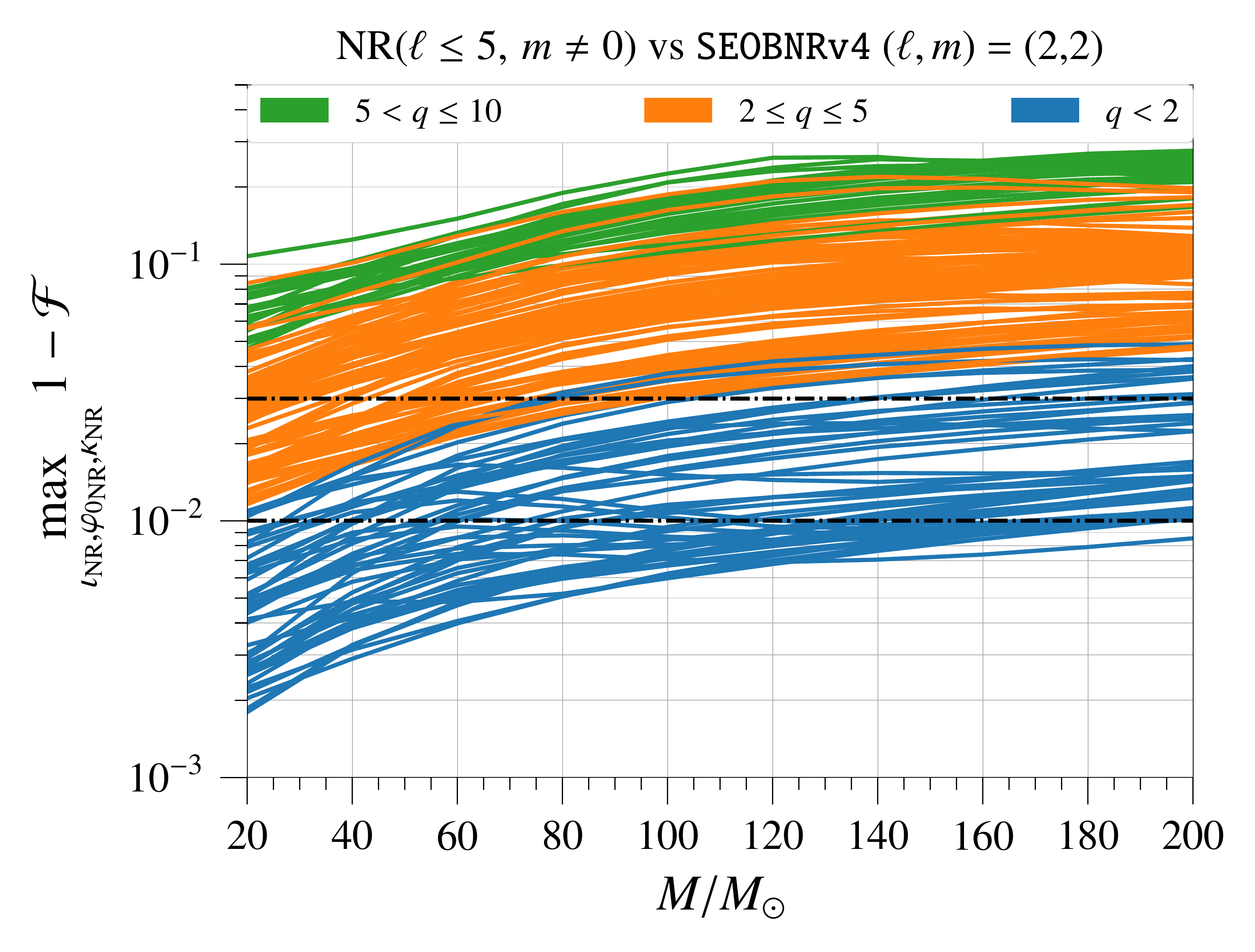} % first figure itself
    \end{minipage}\hfill
    \begin{minipage}{0.5\textwidth}
        \centering
        \includegraphics[width=1.0\textwidth]{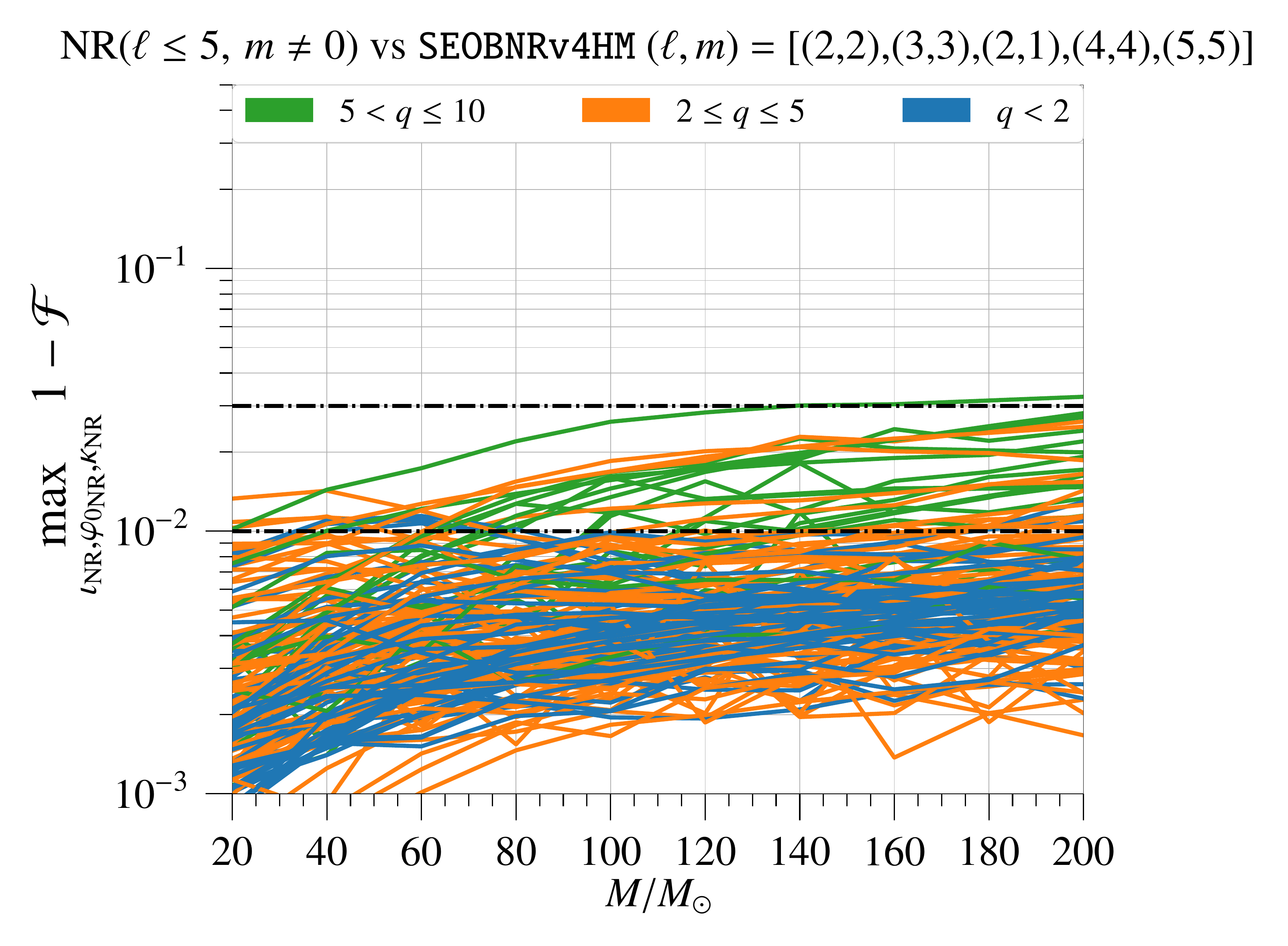} % second figure itself
    \end{minipage}
\caption{Maximum of unfaithfulness $(1-\mathcal{F})$ over the three angles $(\iota_{\textrm{NR}},{\varphi_0}_{\textrm{NR}},\kappa_{\textrm{NR}})$ as a function of the total mass, in the range $20 M_\odot \leq M \leq 200 M_\odot$. Left panel NR $(\ell \leq 5, \, m \neq 0)$ vs \texttt{SEOBNRv4}, right panel NR $(\ell \leq 5,\, m \neq 0)$ vs \texttt{SEOBNRv4HM}. The horizontal dotted-dashed black lines represent the values of $1\%$ and $3\%$ unfaithfulness. The jaggedness of the curves in the plot (right panel) is caused by the numerical noise present in the NR 
higher-order modes, which are not very well resolved. We find that this feature is not present when these noisy modes are removed from the calculation of the faithfulness.}
\label{fig:skyworstall}
\end{figure*}

Finally, in the right panel of Fig.~\ref{fig:skyworstall} we show the maximum of the
unfaithfulness over the three angles $(\iota_{\textrm{NR}},{\varphi_0}_{\textrm{NR}},\kappa_{\textrm{NR}})$  
between the \texttt{SEOBNRv4HM} model and the NR waveforms with 
the modes $(\ell \leq 5, \, m\neq 0)$. In the left panel of the same figure we show
the same comparison but this time using the \texttt{SEOBNRv4} model. 
Here we see that the \texttt{SEOBNRv4HM} waveforms have unfaithfulness 
smaller than $3\%$ in the mass range considered for all the NR simulations 
with the exception of one case, namely \texttt{SXS:BBH:0621} $(q = 7,\, \chi_1 = -0.8,\, \chi_2 = 0)$ for which the unfaithfulness at $M = 200 M_\odot$ is $(1-\mathcal{F}) \sim 3.1\%$. 

In general, over the NR simulations of our catalog, the maximum of the
unfaithfulness is always smaller than $1\%$ in the total mass range
$20 M_\odot \leq M \leq 200 M_\odot$ for nonspinning configurations up
to mass ratio $q = 8$. Nonspinning cases with $q \geq 8$ and
configurations with high spins and mass ratios $q \geq 5$ have maximum
unfaithfulness in the range $1\% \leq (1-\mathcal{F}) \leq 3\%$. For
the former the unfaithfulness decreases to values smaller than $1\%$
when the comparison is done including only the modes $(2,2),(2,1),(3,3),(4,4),(5,5)$ 
in the NR waveforms (i.e., excluding smaller higher-order modes like $(3,2),(4,3)$). This is
not true for high-spin, high--mass-ratio configurations where the
unfaithfulness due to a nonperfect modeling dominates over that due
to neglecting smaller higher-order modes.  It is important to stress
that, as discussed in Sec.~\ref{sec:faithfulness}, the maximum
unfaithfulness due to the numerical error in the NR waveforms of our
catalog is in the range $[0.1 \%, 1\%]$. This means that 
when comparing the NR waveforms with the
\texttt{SEOBNRv4HM} model a fraction of
the maximum unfaithfulness as large as $1\%$ could be due to numerical error. Given that
maximum unfaithfulness are reached for edge-on configurations where
the higher-order modes are more relevant, NR waveforms with better
resolved higher-order modes would be needed in order to attempt to
build a model with maximum unfaithfulness smaller than $1\%$.

\section{Conclusions}
\label{sec:concl}

We have worked within the spinning EOB framework and have built a multipolar waveform
model for BBHs with nonprecessing spins that includes the higher-order modes
$(\ell, m) = (2,1),(3,3),(4,4),(5,5)$, besides the dominant $(2,2)$
mode. In order to improve the agreement with the NR results we
included recently computed PN corrections~\cite{Marsatetal2017,Faye:2014fra,Fujita:2012cm} 
in the resummed GW modes, and also used nonperturbative informations from
NR waveforms in the NQCs corrections of the higher-order modes, and in the 
calibration parameters $c_{\ell m}$'s (the latter only for the modes $(2,1),(5,5)$).  
We also extended to higher-order modes the phenomenological ansatz for 
the merger-ringdown signal that was originally proposed in 
Refs.~\cite{Baker:2008mj,Damour:2014yha,Nagar:2016iwa,Bohe:2016gbl} for the 
dominant $(2,2)$ mode.
 
We have found that the unfaithfulness averaged over orientations, polarizations and
sky positions between the \texttt{SEOBNR4HM} model and NR waveforms of the 
catalog at our disposal, is always smaller than $1\%$ with the exception of four configurations
for which the unfaithfulness is smaller than $1.5\%$. Moreover, the unfaithfulness are 
one order of magnitude smaller than those obtained with the \texttt{SEOBNRv4} model~\cite{Bohe:2016gbl}, 
which only contains the $(2,2)$ mode. The maximum unfaithfulness over orientations, polarizations and sky 
positions between \texttt{SEOBNR4HM} and NR waveforms is always smaller than $3\%$ with the exception of one configuration
for which the faithfulness is smaller than $3.1\%$. Also for the maximum unfaithfulness the results are 
one order of magnitude smaller than those obtained with the \texttt{SEOBNRv4} model~\cite{Bohe:2016gbl}. 
We have also found that, in the nonspinning limit, the \texttt{SEOBNRv4HM} model returns values of the unfaithfulness 
smaller than its (nonspinning) predecessor waveform model, that is \texttt{EOBNRv2HM}~\cite{Pan:2011gk} 
(see Appendix~\ref{sec:EOBNRv2HM}).

Other studies are needed to fully assess the accuracy of \texttt{SEOBNRv4HM} for GW astronomy. In
particular it will be important to understand if unfaithfulnesses below $1\%$ can affect the recovery of 
binary parameters, and if so which parameters will be mainly biased, for which SNR and in which region 
of the parameter space. In particular, we expect that the multipolar \texttt{SEOBNRv4HM} model will be more precise 
than the \texttt{SEOBNRv4} model for recovering the binary's inclination angle and the distance from the source. 
Indeed, those parameters are degenerate with each other when only the $(2,2)$ mode is present, and the inclusion 
of higher-order modes can help in disentagle them (e.g., see Ref.~\cite{OShaughnessy:2014shr}).  We postpone this kind of studies to the future because for
computational reasons, we would need to develop a reduced-order-model (ROM)~\cite{Purrer:2015tud} version of the 
\texttt{SEOBNRv4HM} model. Another important test for the future would be the comparison between 
\texttt{SEOBNRv4HM} model and other multipolar, inspiral-merger-ringdown in the literature, such as the \texttt{IMRPhenom} 
models proposed in Refs.~\cite{Mehta:2017jpq, London:2017bcn}. It will be relevant to compare those models especially 
outside the range of binary configurations where the NR waveforms are available, in order to identify if there are 
regions where the two models predict significantly different waveforms.

We also expect that the multipolar spinning, nonprecessing waveform model developed here 
will be a more accurate model to carry out parameterized tests of General Relativity~\cite{TheLIGOScientific:2016src} 
when BBHs with high mass-ratio, high total mass and in a non face-on orientation will be detected. Furthermore, the 
\texttt{SEOBNRv4HM} model can be employed to search for more than one gravitational quasi-normal mode in the ringdown 
portion of the signal, coherently with multiple detections~\cite{Dreyer:2003bv,Berti:2005ys,Meidam:2014jpa,Yang:2017zxs}. 
In fact, those studies can also be performed with our multipolar, stand-alone merger-ringdown model.

The \texttt{SEOBNRv4HM} waveform model employs the same conservative and dissipative dynamics of the
\texttt{SEOBNRv4} model, which was calibrated to NR simulations by requiring very good 
agreement with the NR $(2,2)$ GW mode. Further improvements of the 
\texttt{SEOBNRv4} waveform model could be achieved in the future by recalibrating the two-body dynamics.  
Such calibration would require the production of a new set of NR waveforms (with more accurate higher-order modes) 
in the region of high mass-ratios, say $q\geq 4$, and high spins, say $\chi_{1,2} \geq 0.6$ where few NR
simulations are currently available and where the disagreement between current analytical inspiral-merger-ringdown 
waveforms is the worst (e.g., see Figs.~ 5 and 6 in Ref.~\cite{Bohe:2016gbl}). Those NR waveforms would need to be sufficiently 
long to make the calibration procedure sufficiently robust (see Sec.VI, and Fig.~7 and 8 in Ref.~\cite{Bohe:2016gbl}).

In the near future our priority is to include the next largest modes in the \texttt{SEOBNRHM} model, notably the 
$(3,2),(4,3)$ modes. The work would need to take into account the mixing between spherical-harmonic and spheroidal harmonics during the merger-ringdown 
stage, as observed in Refs.~\cite{Buonanno:2006ui,Kelly:2012nd}, and investigated more recently in Refs.~\cite{Berti:2014fga,London:2014cma}. 
Insights might need to be gained also from merger-ringdown waveforms in the test-particle limit~\cite{Taracchini:2013wfa,Harms:2015ixa,Harms:2016ctx}. However, to develop a more accurate multipolar model, one would also need to reduce the numerical error in NR waveforms 
around merger and during ringdown, in particular for the modes (4,4) and (5,5). 
Another important and timely application of this work, is its extension to the spinning, precessing case, 
thus improving, the current \texttt{SEOBNRv3} model~\cite{Pan:2013rra,Babak:2016tgq,Abbott:2016izl}, which only contains the $(2,2)$ and $(2,1)$ modes.

\FloatBarrier
\section*{Acknowledgments}

It is a pleasure to thank Juan Calderon Bustillo, Ian Harry, Sylvain Marsat, Harald Pfeiffer, and Noah Sennett for helpful discussions.
Computational work for this manuscript was carried out on the computer clusters {\tt Vulcan} 
and {\tt Minerva} at the Max Planck Institute for Gravitational Physics in Potsdam. 

\appendix

\section{Explicit expressions of higher-order {\it factorized} modes}
\label{app:modes}

Here we list expressions needed to build the $h_{\ell m}$'s of the \texttt{SEOBNRv4HM} model.

The functions $n_{\ell m}^{(\epsilon)}$ and $c_{\ell + \epsilon}(\nu)$ used in Eq.~\eqref{eq:Newtonian} are defined as (see Ref.~\cite{Taracchini:2012ig}):
\begin{align}
n_{\ell m}^{(0)} &= (i m)^\ell \frac{8\pi}{(2\ell + 1)!!} \sqrt{\frac{(\ell + 1)(\ell + 2)}{\ell (\ell -1)}},\\
n_{\ell m}^{(0)} &= -(i m)^\ell \frac{16\pi i}{(2\ell +1)!!}\sqrt{\frac{(2\ell + 1)(\ell + 2)(\ell^2 - m^2)}{(2\ell - 1)(\ell + 1)\ell (\ell -1)}},
\end{align}
and
\begin{equation}
\label{eq:clm}
c_{\ell + \epsilon}(\nu) = \left(\frac{1}{2} - \frac{1}{2}\sqrt{1-4\nu}\right)^{\ell + \epsilon -1} + (-1)^{\ell + \epsilon}\left(\frac{1}{2} + \frac{1}{2}\sqrt{1-4\nu} \right)^{\ell + \epsilon -1}.
\end{equation}
We define also the function
\begin{equation}
\text{eulerlog}\left(m,v_{\Omega }\right)\equiv \gamma + \log(2mv_{\Omega}),
\end{equation}
which is used in the expression of the factorized modes. Here $\gamma$ is the Euler constant.

The quantity $f_{\ell m}$ in Eq. (\ref{eq:hlm_factorized}) is:
\begin{equation}
f_{\ell m} = \begin{cases}
\rho_{\ell m}^\ell, \qquad \qquad \quad \,\,\, \ell \,\, \text{is even}, \\
(\rho_{\ell m}^{\mathrm{NS}})^\ell + f_{\ell m}^{\mathrm{S}}, \qquad \ell \,\, \text{is odd}.
\end{cases}
\end{equation}
The functions $\rho_{\ell m}$, $\rho_{\ell m}^{\mathrm{NS}}$, $ f_{\ell
  m}^{\mathrm{S}}$ are defined below; the superscript ``NS'' stands for nonspinning,  
and the superscript ``S'' indicates spinning. Below, we also list the phase terms $\delta_{\ell
  m}$. 

The quantities $f_{\ell m}$ and $\delta_{\ell m}$ for the \texttt{SEOBNRv4HM} model are mostly taken from the \texttt{SEOBNRv4} 
model in Ref. \cite{Bohe:2016gbl} with the additions of several new terms:
%\begin{widetext}
\begin{itemize}
\item 3PN nonspinning terms in $\rho_{33}^{\mathrm{NS}}$ from Ref. \cite{Faye:2014fra};
\item 5PN test-mass, nonspinning terms in $\rho_{33}^{\mathrm{NS}}$ from Ref. \cite{Fujita:2012cm};
\item 5PN test-mass, nonspinning terms in $\rho_{21}^{\mathrm{NS}}$ from Ref. \cite{Fujita:2012cm};
\item 2PN and 2.5PN spinning terms in $\rho_{44}$ from Ref.~\cite{Marsatetal2017}; 
\item 3PN, 4PN and 5PN test-mass, nonspinning terms in $\rho_{55}^{\mathrm{NS}}$ from Ref. \cite{Fujita:2012cm};
\item 2PN, 2.5PN and 3PN spinning terms in $f_{33}^{\mathrm{S}}$ from Ref.~\cite{Marsatetal2017};
\item 2PN, 2.5PN and 3PN spinning terms in $f_{21}^{\mathrm{S}}$ from Ref.~\cite{Marsatetal2017}; 
\item 1.5PN and 2PN spinning terms in $f_{55}^{\mathrm{S}}$ from Ref.~\cite{Marsatetal2017};
\item  3PN and 4.5PN test-mass, nonspinning terms in $\delta_{55}$ from Ref. \cite{Fujita:2012cm}.
\end{itemize}
%\end{widetext}
Furthermore, we find that resummations of the $f_{\ell m}$ function
for the $(3,3), (2,1),(4,4),(5,5)$ modes of the kind proposed in
Refs.~\cite{Nagar:2016ayt,Messina:2018ghh} (see Eq. (47) and Eq. (48)
in the latter) do not always improve the agreement with the NR
waveforms of our catalog. For this reason we decide not to implement
those resummations when building the \texttt{SEOBNRv4HM}
model. It is worth to mention that whereas in our model the resummed
  expressions are computed as a function of $v_\Omega =
  (M\Omega)^{1/3}$, in Refs.~\cite{Nagar:2016ayt,Messina:2018ghh} they
  are expressed as a function of $v_\phi$ defined in Eq.~(69) of
  Ref.~\cite{Damour:2014sva}. While the two variables are very similar
  at low frequency, they can  differ toward merger where the aforementioned resummation 
may be more effective.
\begin{widetext}
\begin{align}
\rho_{33}^{\mathrm{NS}} =  &1+\left(-\frac{7}{6}+\frac{2 \nu }{3}\right) v_{\Omega
   }^2+\left(-\frac{6719}{3960}-\frac{1861 \nu }{990}+\frac{149 \nu ^2}{330}\right)
   v_{\Omega
   }^4 \nonumber \\ 
   &+\left[\frac{3203101567}{227026800}+\left(-\frac{129509}{25740}+\frac{41 \pi
   ^2}{192}\right) \nu -\frac{274621 \nu ^2}{154440}+\frac{12011 \nu
   ^3}{46332}-\frac{26}{7} \text{eulerlog}\left(3,v_{\Omega }\right)\right] v_{\Omega
   }^6 \nonumber \\
   &+\left(-\frac{57566572157}{8562153600}+\frac{13}{3}
   \text{eulerlog}\left(3,v_{\Omega }\right)\right) v_{\Omega
   }^8+\left(-\frac{903823148417327}{30566888352000}+\frac{87347
   \text{eulerlog}\left(3,v_{\Omega }\right)}{13860}\right) v_{\Omega }^{10}\,, \\
 \rho_{21}^{\mathrm{NS}} =&1+\left(-\frac{59}{56}+\frac{23 \nu }{84}\right) v_{\Omega
   }^2+\left(-\frac{47009}{56448}-\frac{10993 \nu }{14112}+\frac{617 \nu
   ^2}{4704}\right) v_{\Omega }^4+\left(\frac{7613184941}{2607897600}-\frac{107}{105}
   \text{eulerlog}\left(1,v_{\Omega }\right)\right) v_{\Omega
   }^6 \nonumber \\
   &+\left(-\frac{1168617463883}{911303737344}+\frac{6313
   \text{eulerlog}\left(1,v_{\Omega }\right)}{5880}\right) v_{\Omega
   }^8+\frac{\left(-63735873771463+14061362165760 \text{eulerlog}\left(1,v_{\Omega
   }\right)\right) v_{\Omega }^{10}}{16569158860800}\,, \\
  \rho_{44} =& 1+\left(\frac{1614-5870 \nu +2625 \nu ^2}{1320 (-1+3 \nu )}\right) v_{\Omega }^2+ \Bigg[\left(\frac{2}{3 (-1+3 \nu )}-\frac{41 \nu }{15 (-1+3 \nu )}+\frac{14 \nu ^2}{5 (-1+3 \nu )}\right) \chi
   _S \nonumber \\
   &+\delta m \left(\frac{2}{3 (-1+3 \nu )}-\frac{13 \nu }{5 (-1+3 \nu )}\right) \chi _A\Bigg] v_{\Omega }^3 +\Bigg[-\frac{14210377}{8808800 (1-3 \nu )^2}+\frac{32485357 \nu
   }{4404400 (1-3 \nu )^2}-\frac{1401149 \nu ^2}{1415700 (1-3 \nu )^2} \nonumber \\
   &-\frac{801565 \nu ^3}{37752 (1-3 \nu )^2}+\frac{3976393 \nu ^4}{1006720 (1-3 \nu )^2}+\frac{\chi _A^2}{2}-2 \nu  \chi _A^2+\delta m \chi _A \chi _S+\frac{\chi _S^2}{2}\Bigg] v_{\Omega }^4 \nonumber \\
   &+ \Bigg[\left(-\frac{69}{55
   (1-3 \nu )^2}+\frac{16571 \nu }{1650 (1-3 \nu )^2}-\frac{2673 \nu ^2}{100 (1-3 \nu )^2}+\frac{8539 \nu ^3}{440 (1-3 \nu )^2}+\frac{591 \nu ^4}{44 (1-3 \nu )^2}\right) \chi _S\nonumber \\
   &+\delta m
   \left(-\frac{69}{55 (1-3 \nu )^2}+\frac{10679 \nu }{1650 (1-3 \nu )^2}-\frac{1933 \nu ^2}{220 (1-3 \nu )^2}+\frac{597 \nu ^3}{440 (1-3 \nu )^2}\right) \chi
   _A\Bigg] v_{\Omega }^5 \nonumber \\
   &+\left(\frac{16600939332793}{1098809712000}-\frac{12568 \text{eulerlog}\left(4,v_{\Omega }\right)}{3465}\right) v_{\Omega
   }^6 +\left(-\frac{172066910136202271}{19426955708160000}+\frac{845198 \text{eulerlog}\left(4,v_{\Omega }\right)}{190575}\right) v_{\Omega
   }^8\nonumber \\
   &+\left(-\frac{17154485653213713419357}{568432724020761600000}+\frac{22324502267 \text{eulerlog}\left(4,v_{\Omega }\right)}{3815311500}\right) v_{\Omega }^{10}\,, \\
   \rho_{55}^{\mathrm{NS}} =& 1+\left(\frac{487}{390 (-1+2 \nu )}-\frac{649 \nu }{195 (-1+2 \nu )}+\frac{256 \nu ^2}{195 (-1+2 \nu )}\right) v_{\Omega }^2-\frac{3353747 v_{\Omega
   }^4}{2129400} \nonumber \\
   &+\left(\frac{190606537999247}{11957879934000}-\frac{1546}{429} \text{eulerlog}\left(5,v_{\Omega }\right)\right) v_{\Omega
   }^6+\left(-\frac{1213641959949291437}{118143853747920000}+\frac{376451 \text{eulerlog}\left(5,v_{\Omega }\right)}{83655}\right) v_{\Omega
   }^8\nonumber \\
   &+\left(-\frac{150082616449726042201261}{4837990810977324000000}+\frac{2592446431 \text{eulerlog}\left(5,v_{\Omega }\right)}{456756300}\right) v_{\Omega }^{10}\,, \\
f_{33}^{\mathrm{S}} = &\left[\left(-2+\frac{19 \nu }{2}\right) \frac{\chi _A}{\delta m}+\left(-2+\frac{5 \nu }{2}\right) \chi _S\right] v_{\Omega }^3 + \left[\left(\frac{3}{2}-6 \nu\right)\chi
   _A^2+(3-12 \nu ) \frac{\chi _A}{\delta m} \chi _S+\frac{3 \chi
   _S^2}{2}\right] v_{\Omega }^4 \nonumber \\
   &+
   \left[\left(\frac{2}{3}-\frac{593 \nu }{60}+\frac{407 \nu ^2}{30}\right)\frac{\chi
   _A}{\delta m}+\left(\frac{2}{3}+\frac{11 \nu }{20}+\frac{241 \nu
   ^2}{30}\right) \chi _S\right]v_{\Omega }^5 \nonumber \\
   &+ \left[\left(-\frac{7}{4} + \frac{11\nu}{2} -12\nu^2 \right) \chi _A^2+\left(-\frac{7}{2}-\nu +44 \nu
   ^2\right)\frac{\chi _A}{\delta m} \chi _S+\left(-\frac{7}{4}-\frac{27 \nu }{2}+6
   \nu ^2\right) \chi _S^2\right]v_{\Omega }^6 \nonumber \\
   &+ \left[\left(-\frac{81}{20}+\frac{7339 \nu }{540}\right)
   \frac{\chi _A}{\delta m}+\left(-\frac{81}{20}+\frac{593 \nu }{108}\right) \chi
   _S\right] i (H_{\mathrm{EOB}} \Omega) ^2\,, \\
   \label{eq:f_21}
   f_{21}^{\mathrm{S}} =& \left(-\frac{3}{2} \chi _S-\frac{3 \chi _A}{2 \delta m}\right)v_{\Omega } + \left[\left(\frac{61}{12}+\frac{79 \nu }{84}\right) \chi
   _S+\left(\frac{61}{12}+\frac{131 \nu }{84}\right)\frac{\chi _A}{\delta m}\right]v_{\Omega }^3+ \left[(-3-2 \nu ) \chi _A^2+\left(-3+\frac{\nu }{2}\right) \chi
   _S^2+\left(-6+\frac{21 \nu }{2}\right)\chi _S \frac{\chi _A }{\delta m}\right]v_{\Omega }^4\nonumber \\
   &+ \bigg\{\left(\frac{3}{4 \delta m}-\frac{3 \nu }{\delta m}\right) \chi
   _A^3+\left[-\frac{81}{16}+\frac{1709 \nu }{1008}+\frac{613 \nu ^2}{1008}+\left(\frac{9}{4}-3 \nu \right) \chi _A^2\right] \chi _S + \nonumber \\
   &+\frac{3 \chi _S^3}{4}+\left[-\frac{81}{16}-\frac{703
   \nu ^2}{112}+\frac{8797 \nu }{1008}+\left(\frac{9}{4}-6 \nu \right) \chi _S^2\right]\frac{\chi _A }{\delta m}\bigg\}v_{\Omega }^5 \nonumber \\
   &+ \left[\left(\frac{4163}{252}-\frac{9287 \nu }{1008}-\frac{85 \nu
   ^2}{112}\right) \chi _A^2+\left(\frac{4163}{252}-\frac{2633 \nu }{1008}+\frac{461 \nu ^2}{1008}\right) \chi _S^2+\left(\frac{4163}{126}-\frac{1636 \nu }{21}+\frac{1088 \nu ^2}{63}\right)\chi _S
   \frac{\chi _A }{\delta m}\right]v_{\Omega }^6+\mathbf{c_{21}} v_{\Omega }^7\,,\\
   \label{eq:f_55}
   f_{55}^{\mathrm{S}} = & \left[\left(-\frac{70 \nu }{3 (-1+2 \nu )}+\frac{110 \nu ^2}{3 (-1+2 \nu )}+\frac{10}{3 (-1+2 \nu )}\right) \frac{\chi _A}{\delta m}+\left(\frac{10}{3
   (-1+2 \nu )}-\frac{10 \nu }{-1+2 \nu }+\frac{10 \nu ^2}{-1+2 \nu }\right) \chi _S\right]v_{\Omega }^3\nonumber \\
&+ \left[\frac{5}{2}\delta m^2 \chi _A^2+5 \delta m \chi _A \chi _S+\frac{5
   \chi _S^2}{2}\right]v_{\Omega }^4+\mathbf{c_{55}} v_{\Omega }^5\,.
\end{align}
\begin{align}
\delta_{33} =& \frac{13}{10}(H_{\mathrm{EOB}}\Omega) + \frac{39\pi}{7}(H_{\mathrm{EOB}}\Omega)^2 + \left(-\frac{227827}{3000} + \frac{78\pi^2}{7}\right)(H_{\mathrm{EOB}}\Omega)^3 - \frac{80897\nu}{2430} v_{\Omega}^5\,, \\
\delta_{21} =& \frac{2}{3} (\Omega  H_{\mathrm{EOB}})+\frac{107}{105} \pi  (\Omega H_{\mathrm{EOB}})^2+\left(-\frac{272}{81}+\frac{214 \pi ^2}{315}\right) \Omega ^3
   H_{\text{EOB}}^3 -\frac{493}{42} \nu  v_{\Omega}^{5}\,, \\
\delta_{44} =& \frac{(112+219 \nu )}{120 (1-3 \nu )}(\Omega  H_{\mathrm{EOB}})+\frac{25136 \pi}{3465}(\Omega  H_{\mathrm{EOB}})^2 +\left(\frac{201088}{10395}\pi^2 - \frac{55144}{375} \right) (\Omega  H_{\mathrm{EOB}})^3\,, \\
\delta_{55} =& \frac{(96875+857528 \nu )}{131250 (1-2 \nu )}(\Omega  H_{\mathrm{EOB}}) + \frac{3865\pi}{429}(\Omega  H_{\mathrm{EOB}})^2 + \frac{-7686949127 + 954500400\pi^2}{31783752}(\Omega  H_{\mathrm{EOB}})^3\,.
\end{align}
\end{widetext}
We notice that $f_{33}^S$ is a complex quantity because it contains an  
imaginary term recently computed in PN theory~\cite{Marsatetal2017}
\begin{equation}
i\delta_{33}^\mathrm{S} \equiv \left[\left(-\frac{81}{20}+\frac{7339 \nu }{540}\right)
   \frac{\chi _A}{\delta m}+\left(-\frac{81}{20}+\frac{593 \nu }{108}\right) \chi
   _S\right] i (H_{\mathrm{EOB}} \Omega) ^2,
\end{equation}
where with the superscript ``S'' we indicate the spin
  dependence. The term proportional to $\chi_A/\delta m$ seems to diverge when $\delta m \rightarrow 0$, but this divergence is apparent 
 because, as it happens for all the functions $f_{\ell
    m}^{\mathrm{S}}$, it is removed by the factor $\delta
  m$ that appears in the function $c_{\ell +\epsilon}(\nu)$ (see
  Eq.\eqref{eq:clm}) at Newtonian order (see
  Eq.\eqref{eq:Newtonian}). If one includes the term
  $\delta_{33}^\mathrm{S}$ in the resummation with the complex
  exponential, one obtains the expression $e^{i(\delta_{33}+
    \delta_{33}^\mathrm{S})}$ which is not well-behaved in the limit
  $\delta m \rightarrow 0$.  For this reason we do not include this
  new PN term in the resummation $f_{3 3}e^{i(\delta_{33}+
    \delta_{33}^\mathrm{S})}$, but, instead, we compute the latter quantity
  excluding this term (i.e., $f_{3 3}e^{i\delta_{33}}$) and we then add
  the new complex term to the real amplitude $f_{3 3}$. We can do so because 
  $e^{i\delta_{3 3}} i\delta_{33}^\mathrm{S} = i\delta_{33}^\mathrm{S}
  + \mathcal{O}(\Omega^3)$, where the latter is a PN correction at higher order with respect to the order at which 
we currently know PN terms. 

We remember also that the modes $(2,1),(5,5)$ contain the calibration 
parameters $c_{21}$ and $c_{55}$ computed imposing the condition in Eq.~\eqref{eq:cal_par}.

\section{Fits of nonquasi-circular input values}
\label{app:NQCfits}

We build the fits of the nonquasi-circular (NQC) input values using NR
waveforms with the highest level of resolution available and the
extrapolation order $N = 2$. Depending on the
  mode, the fits use a different number of NR waveforms, because for
  some binary configurations the large numerical error prevents us to
  use some NR modes. For each mode, in order to choose which NR
  simulations to use for the fits, we first remove all the NR
  simulations showing clearly unphysical features (e.g., strong
  oscillations in the post-merger stage that are not consistent among waveforms 
at different resolution and extrapolation order). For the modes (3,3) and (2,1) all the NR
  waveforms pass this selection, while for the modes (4,4) and (5,5)
  we remove respectively 10 and 42 NR simulations. 
For each NQC input value (i.e., amplitude and its first and second derivative, and
  frequency and its first derivative) 
we weight the value extracted by
  a given NR simulation with the inverse of the NR error. The latter
  is estimated as $\sqrt{(\delta^{\mathrm{NQC}}_{\mathrm{res}})^2 +
    (\delta^{\mathrm{NQC}}_{\mathrm{extr}}})^2$, where
  $\delta^{\mathrm{NQC}}_{\mathrm{res}}$ is the difference between the
  NQC input values extracted from the NR waveform with the same
  extrapolation order ($N = 2$) and different resolutions (i.e., the
  highest and second highest
  resolution). The quantity $\delta^{\mathrm{NQC}}_{\mathrm{extr}}$ is instead the
  difference between the NQC input values extracted from the NR
  waveform with the same resolution level (the highest) and different
  extrapolation order (i.e., $N = 2$ and $N = 3$).

We find it convenient to define a few variables that enter the fits below:
\begin{align}
\chi_{33} &= \chi_S\delta m + \chi_A\,,\\
\chi_{21A} &= \frac{\chi_S}{1-1.3\nu}\delta m + \chi_A\,,\\
\chi_{44A} &= (1-5\nu)\chi_S+\chi_A\delta_m\,,\\
\chi_{21D} &= \frac{\chi_S}{1-2\nu}\delta m + \chi_A\,,\\
\chi_{44D} &= (1-7\nu)\chi_S+\chi_A\delta_m\,, \\
\chi&= \chi_S+\chi_A\frac{\delta m}{1-2\nu}.\\ \nonumber
\end{align}
We notice that the variables $\chi_{33}, \,\chi_{21A}\,,\chi_{21D}$ are by definition
zero in the equal-mass, equal-spin limit. They are used for the fits
of the amplitude (and its derivative) to guarantee that in this limit 
the modes with  $m$ odd vanish, since they have to satisfy the 
symmetry under rotation $\varphi_0 \rightarrow \varphi_0 + \pi$.
\subsection{Amplitude's fits}

\begin{widetext}
\begin{align}
\frac{|h_{3 3}^{\textrm{NR}}(t^{3 3}_{\textrm{match}})|}{\nu} = &|(0.101092 + 0.470410 \nu+1.073546 \nu ^2) \chi
   _{33} \nonumber \\
   &+\delta m (0.563658 - 0.054609 \nu+2.309370 \nu
   ^2+0.029813 \chi _{33}^2 - 0.0968810 \nu \ \chi_{33}^2)|\,, \\
\frac{|h_{2 1}^{\textrm{NR}}(t^{2 1}_{\textrm{match}})|}{\nu} = &|\delta m (-0.428179 +0.113789 \nu -0.773677 \nu ^2  -0.0101951 \chi_{21A}+0.0470041
   \chi_{21A}^2 -0.0932613 \chi_{21A}^2 \nu) \nonumber \\
   &+\chi_{21A} (0.292567 -0.197103 \nu)+ \delta m 0.0168769 \chi_{21A}^3|\,, \\
\frac{|h_{4 4}^{\textrm{NR}}(t^{4 4}_{\textrm{match}})|}{\nu} = &0.264658\, +0.0675842 \chi _{44 A}+0.029251 \chi _{44 A}^2+ (-0.565825-0.866746 \chi _{44 A}+0.00523419 \chi _{44 A}^2)\nu \nonumber \\
&+ (-2.50083+6.88077 \chi _{44 A}-1.02347 \chi _{44
   A}^2)\nu ^2+ (7.69745\, -16.5515 \chi _{44 A})\nu ^3\,, \\
 \frac{|h_{5 5}^{\textrm{NR}}(t^{5 5}_{\textrm{match}})|}{\nu}  =&\delta m (0.0953727\, -0.128585 \nu )+\delta m(0.0309164\, -0.0997875 \nu ) \chi _{33}+\delta m (0.0437835\, -0.212609 \nu ) \chi _{33}^2 \nonumber \\
 &+0.00503392 |\delta m
   (-0.16815+8.54945 \nu )+\chi _{33}|\,.
\end{align}
\subsection{Amplitude--first-derivative's fits}

\begin{align}
\frac{1}{\nu}\left. \frac{d\left| h_{33}^{\textrm{NR}}(t) \right|}{dt} \right|_{t = t^{3 3}_{\textrm{match}}} = &\delta m (-0.00309944+0.0100765 \nu) \chi_{33}^2\nonumber \\
&+
   0.00163096\sqrt{\delta m^2 (8.81166\, +104.478 \nu)+\delta m
   (-5.35204+49.6862 \nu) \chi_{33}+ \chi_{33}^2}\,, \\
   \frac{1}{\nu}\left. \frac{d\left| h_{21}^{\textrm{NR}}(t) \right|}{dt} \right|_{t = t^{2 1}_{\textrm{match}}} = & \delta m (0.00714753\, -0.0356440 \nu ) +0.00801714  |
   -\delta m (0.787561\, +1.61127 \nu +11.30606 \nu ^2)+\chi _{21
   D}| \nonumber \\
   & +\delta m (-0.00877851+0.0305467 \nu ) \chi _{21 D}\,, \\
\frac{1}{\nu}\left. \frac{d\left| h_{44}^{\textrm{NR}}(t) \right|}{dt} \right|_{t = t^{4 4}_{\textrm{match}}} = &0.00434759\, -0.00146122 \chi _{44 D}-0.00242805 \chi _{44 D}^2+ (0.0233207\,
   -0.0224068 \chi _{44 D}+0.0114271 \chi _{44 D}^2)\nu\nonumber \\
   &+
   (-0.460545+0.433527 \chi _{44 D})\nu ^2+ (1.27963\, -1.24001 \chi
   _{44 D})\nu ^3\,, \\
   \frac{1}{\nu}\left. \frac{d\left| h_{55}^{\textrm{NR}}(t) \right|}{dt} \right|_{t = t^{5 5}_{\textrm{match}}} = &\delta m (-0.0083898+0.0467835 \nu )+\delta m (-0.00136056+0.00430271
   \nu ) \chi _{33} \nonumber \\
   &+\delta m (-0.00114121+0.00185904 \nu ) \chi
   _{55}^2+0.000294422 |\delta m (37.1113\, -157.799 \nu )+\chi
   _{55}|\,.
\end{align}
\subsection{Amplitude--second-derivative's fits}
\begin{align}
\frac{1}{\nu}\left. \frac{d^2\left| h_{33}^{\textrm{NR}}(t) \right|}{dt^2} \right|_{t = t^{3 3}_{\textrm{match}}} = &\delta m (0.000960569\, -0.000190807 \nu)
   \chi_{33}\nonumber \\
   &- 0.000156238 | \delta m(4.67666 + 79.2019 \nu -1097.41 \nu^2 + 6512.96 \nu^3 -13263.4 \nu^4) + \chi_{33}|\,, \\
   \frac{1}{\nu}\left. \frac{d^2\left| h_{21}^{\textrm{NR}}(t) \right|}{dt^2} \right|_{t = t^{2 1}_{\textrm{match}}} = & 0.000371322 \delta m
    -| \delta m
   (-0.000365087-0.00305417 \nu ) +\delta m (-0.000630623 
   -0.000868048 \nu \nonumber \\
   &+0.0223062 \nu ^2) \chi _{21 D}^2 +0.000340243  \chi _{21
   D}^3+0.000283985  \delta m \, \chi _{21 D}|\,, \\
      \frac{1}{\nu}\left. \frac{d^2\left| h_{44}^{\textrm{NR}}(t) \right|}{dt^2} \right|_{t = t^{4 4}_{\textrm{match}}} = & -0.000301723+0.000321595 \chi +  (0.00628305\, +0.00115988 \chi)\nu\nonumber \\
   &+ (-0.0814352-0.0138195 \chi)\nu ^2+ (0.226849\, +0.0327575 \chi)\nu ^3\,, \\
       \frac{1}{\nu}\left. \frac{d^2\left| h_{55}^{\textrm{NR}}(t) \right|}{dt^2} \right|_{t = t^{5 5}_{\textrm{match}}}  = &\delta m (0.000127272\, +0.000321167 \nu )+\delta m
   (-0.0000662168+0.000328855 \nu ) \chi _{33} \nonumber \\
   &+(-0.0000582462+0.000139443 \nu ) \chi
   _{33}^2.
\end{align}
\subsection{Frequency and frequency-derivative fits}
\begin{align}
\omega_{33}^{\textrm{NR}}(t^{3 3}_{\textrm{match}}) = &0.397395\, +0.164193 \chi+0.163553 \chi^2+0.0614016 \chi^3+
   (0.699506\, -0.362674 \chi-0.977547 \chi^2)\nu\nonumber \\
   &+
   (-0.345533+0.319523 \chi+1.93342 \chi^2)\nu^2\,, \\
   \omega_{21}^{\textrm{NR}}(t^{2 1}_{\textrm{match}}) = & 0.174319\, +0.0535087 \chi +0.0302288 \chi ^2+  (0.193894\, -0.184602 \chi
   -0.112222 \chi ^2)\nu\nonumber \\
   &+ (0.167006\, +0.218731 \chi )\nu ^2\,, \\
  \omega_{44}^{\textrm{NR}}(t^{4 4}_{\textrm{match}}) = & 0.538936\, +0.166352 \chi +0.207539 \chi ^2+0.152681 \chi ^3 \nonumber \\
  &+ \left(0.76174+0.00958786 \chi -1.3023 \chi
   ^2-0.556275 \chi ^3\right)\nu
   + \left(0.967515\,
   -0.220593 \chi +2.6781 \chi ^2\right)\nu ^2\nonumber \\
   &- 4.89538 \nu^3 \,,\\
   \omega_{55}^{\textrm{NR}}(t^{5 5}_{\textrm{match}})   = &0.643755\, +0.223155 \chi +0.295689 \chi ^2+0.173278 \chi ^3\nonumber \\
   &+ 
   \left(-0.470178-0.392901 \chi -2.26534 \chi ^2-0.5513 \chi ^3\right)\nu
   +\left(2.31148\, +0.882934 \chi +5.8176 \chi ^2\right)\nu ^2\,.
\end{align}
\begin{align}
\dot{\omega}_{33}^{\textrm{NR}}(t^{3 3}_{\textrm{match}}) = &0.0103372\, -0.00530678 \chi ^2-0.00508793 \chi ^3\nonumber \\
&+ \left(0.0277356\, +0.0188642
   \chi+0.0217545 \chi^2+0.0178548 \chi^3\right)\nu +(0.0180842\,
   -0.0820427 \chi)\nu^2, \\  
   \dot{\omega}_{21}^{\textrm{NR}}(t^{2 1}_{\textrm{match}}) = &0.00709874\, -0.00177519 \chi -0.00356273 \chi ^2-0.0019021 \chi ^3\nonumber \\&
   + 
   (0.0248168\, +0.00424406 \chi +0.0147181 \chi ^2)\nu +
   (-0.050429-0.0319965 \chi )\nu ^2\,, \\
   \dot{\omega}_{44}^{\textrm{NR}}(t^{4 4}_{\textrm{match}}) =&0.0139979\, -0.00511782 \chi -0.00738743 \chi ^2+ \left(0.0528489\, +0.016323 \chi
   +0.0253907 \chi ^2\right)\nu  \nonumber \\
   &+ (-0.0652999+0.0578289 \chi )\nu ^2\,,\\
   \dot{\omega}_{55}^{\textrm{NR}}(t^{5 5}_{\textrm{match}})   =&0.0176343\, -0.000249257 \chi -0.0092404 \chi ^2-0.00790783 \chi ^3\nonumber \\ 
   &+\left(-0.13660+0.0561378 \chi +0.164063 \chi ^2+0.0773623 \chi ^3\right)\nu
   +\left(0.987589\, -0.313921 \chi -0.592615 \chi ^2\right)\nu ^2 \nonumber \\
   &-1.694335\nu^3\,.
\end{align}
\end{widetext}
 
\section{Fits for amplitude and phase of merger-ringdown model}
\label{app:ringdownfits}

\begin{figure}[h]
  \centering
  \includegraphics[width=0.5\textwidth]{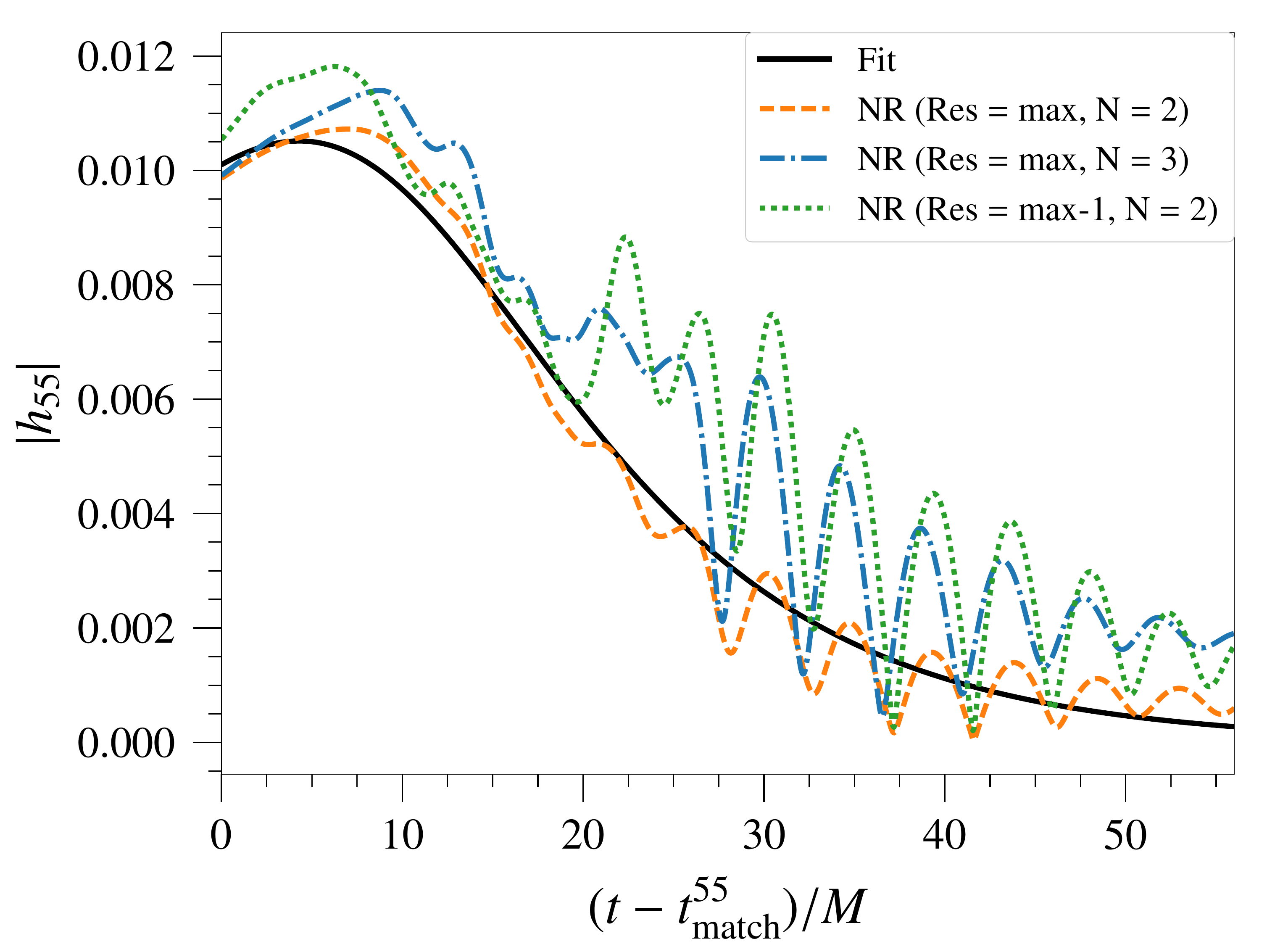}
\caption{Amplitudes of the (5,5) NR mode of the simulation \texttt{SXS:BBH:0065} $(q = 8, \chi_1 = 0.5, \chi_2 = 0)$ for extraction order $N = 2$ and highest resolution (dashed orange), extraction order $N = 3$ and highest resolution (dotted-dashed blue), extraction order $N = 2$ and second highest resolution (dotted green). In solid black we show the result of the fit of the merger-ringdown signal used in the \texttt{SEOBNRv4HM} model.}
\label{fig:noisyNR}
\end{figure}

For these fits we apply the same selection of the NR waveforms discussed for the fits of the input values for the NQC. In particular, in performing the fits for the amplitude (phase) of the merger-ringdown signal, we weigh the contribution of the values extracted from every NR waveform with the same weight used for the NQC input value of the amplitude (frequency). It should be noted that  in some cases, especially in the ringdown, the NR error in the (4,4) and (5,5) modes limits our ability to accurately model this part of the waveform (see Fig.~\ref{fig:noisyNR}).

\begin{align}
c_{1,f}^{33} = &0.0763873\, +0.254345 \nu -1.08927 \nu ^2-0.0309934 \chi\nonumber \\
& +0.251688 \nu  \chi -0.798091
   \nu ^2 \chi\,, \\
c_{2,f}^{33} = & -0.832529+2.76799 \nu -7.02815 \nu ^2-0.59888 \chi \nonumber \\
&+5.90437 \nu  \chi -18.2326 \nu ^2
   \chi\,, \\
c_{1,f}^{21} = &  0.0778033\, +0.24091 \nu -0.745633 \nu ^2-0.0507064 \chi\nonumber \\
& +0.385826 \nu  \chi -0.969553
   \nu ^2 \chi\,, \\
   c_{2,f}^{21} = & -1.24519+6.1342 \nu -14.6725 \nu ^2-1.19579 \chi \nonumber \\
   &+15.667 \nu  \chi -44.4198 \nu ^2 \chi\,, \\
   c_{1,f}^{44} = &-0.0639271+0.345195 \nu -1.76435 \nu ^2-0.0364617 \chi \nonumber \\
   &+1.27774 \nu  \chi -14.8253 \nu
   ^2 \chi +40.6714 \nu ^3 \chi\,, \\
   c_{2,f}^{44} = &0.781328\, -5.1869 \nu +14.0264 \nu ^2+0.809471 \chi\nonumber \\
   & -5.38343 \nu  \chi +0.105163 \nu ^2
   \chi +46.9784 \nu ^3 \chi\,,\\
c_{1,f}^{55} =&-0.0670461-0.247549 \nu +0.758804 \nu ^2+0.0219059 \chi \nonumber \\
   & -0.0943771 \nu  \chi +0.435777
   \nu ^2 \chi\,, \\
c_{2,f}^{55} =&1.67634\, -5.60456 \nu +16.7513 \nu ^2+0.49257 \chi\nonumber \\
   & -6.2091 \nu  \chi +16.7785 \nu ^2
   \chi\,.
\end{align}
\begin{align}
d_{1,f}^{33} = & 0.110853\, +0.99998 \nu -3.39833 \nu ^2+0.0189591 \chi \nonumber \\
& -0.72915 \nu  \chi +2.5192 \nu ^2
   \chi\,,\\
   d_{2,f}^{33} = & 2.78252\, -7.84474 \nu +27.181 \nu ^2+2.87968 \chi\nonumber \\
   & -34.767 \nu  \chi +127.139 \nu ^2
   \chi\,,\\
   d_{1,f}^{21} = &0.156014\, +0.0233469 \nu +0.153266 \nu ^2+0.1022 \chi\nonumber \\
   & -0.943531 \nu  \chi +1.79791 \nu
   ^2 \chi\,, \\
   d_{2,f}^{21} = &2.78863\, -0.814541 \nu +5.54934 \nu ^2+4.2929 \chi\nonumber \\
   & -15.938 \nu  \chi +12.6498 \nu ^2
   \chi\,, \\
   d_{1,f}^{44} = &0.11499\, +1.61265 \nu -6.2559 \nu ^2+0.00838952 \chi \nonumber \\
   & -0.806998 \nu  \chi +7.59565 \nu
   ^2 \chi -19.3237 \nu ^3 \chi\,, \\
   d_{2,f}^{44} = & 3.11182\, +15.8853 \nu -79.6493 \nu ^2+5.39934 \chi \nonumber \\
   & -87.9242 \nu  \chi +657.716 \nu ^2
   \chi -1555.3 \nu ^3 \chi\,,\\
    d_{1,f}^{55} = & 0.142876\, +0.256174 \nu -1.45316 \nu ^2-0.0411121 \chi\nonumber \\
    & +0.300264 \nu  \chi -0.91379 \nu
   ^2 \chi\,, \\
    d_{2,f}^{55} = & 2.95824\, -17.3332 \nu +50.7161 \nu ^2+3.09389 \chi\nonumber \\
    & -30.6289 \nu  \chi +88.3659 \nu ^2 \chi\,.
\end{align}

\section{Fits for the phase difference between higher-order modes and (2,2) mode at the matching point $t_{\textrm{match}}^{\ell m}$}
\label{app:fitphasediff}

The relations between $\phi_{\textrm{match}}^{\ell m}$ (i.e., the phase of the $(\ell, m)$ modes computed at $t_{\textrm{match}}^{\ell m}$) and $\phi_{\textrm{match}}^{2 2}$ are
\begin{align}
\Delta\phi_{\textrm{match}}^{3 3} &\equiv \phi_{\textrm{match}}^{3 3} - \frac{3}{2}(\phi_{\textrm{match}}^{2 2} - \pi)  \pmod{\pi}\,, \\
\Delta\phi_{\textrm{match}}^{2 1} &\equiv \phi_{\textrm{match}}^{2 1} - \frac{1}{2}(\phi_{\textrm{match}}^{2 2} - \pi)  \pmod{\pi}\,,\\
\Delta\phi_{\textrm{match}}^{4 4} &\equiv \phi_{\textrm{match}}^{4 4} - (2\phi_{\textrm{match}}^{2 2} - \pi) \pmod{2\pi}\,,\\
\Delta\phi_{\textrm{match}}^{5 5} &\equiv \phi_{\textrm{match}}^{5 5} - \frac{1}{2}(5\phi_{\textrm{match}}^{2 2} - \pi) \pmod{\pi}, 
\end{align}
where the RHS is the scaling of the phase at leading PN order, and the LHS is the deviation from the latter, 
computed at $t_{\textrm{match}}^{\ell m}$. The term $\Delta\phi_{\textrm{match}}^{\ell m}$ is extracted from 
each NR and Teukolsky--equation-based waveforms in our catalog and then fitted as a function of $(\nu, \chi)$. We find
\begin{align}
\Delta\phi_{\textrm{match}}^{3 3} =&3.20275\, -1.47295 \sqrt{\delta m}+1.21021 \delta m-0.203442 \chi \nonumber \\
& +\delta m^2 (-0.0284949-0.217949 \chi ) \chi \pmod{\pi}\,,\\
\Delta\phi_{\textrm{match}}^{2 1} = & 2.28855\, +0.200895 \delta m-0.0403123 \chi \nonumber \\
&+\delta m^2 \left(-0.0331133-0.0424056 \chi -0.0244154 \chi ^2\right) \nonumber\\
&\pmod{\pi}\,, \\
\Delta\phi_{\textrm{match}}^{4 4} = & 5.89306\, +\nu ^2 (-36.7321-21.9229 \chi ) \nonumber \\
&-0.499652 \chi -0.292006 \chi ^2\nonumber \\
&+\nu ^3 (160.102\, +67.0793 \chi )\nonumber \\
&+\nu  \left(2.48143\, +3.26618 \chi +1.38065 \chi ^2\right) \pmod{2\pi}\,, \\
\Delta\phi_{\textrm{match}}^{5 5} =& 3.61933\, -1.52671 \delta m-0.172907 \chi \nonumber \\
& +\delta m^2 \left(0.72564\, -0.44462 \chi -0.528597 \chi ^2\right)\nonumber \\
& \pmod{\pi} \,.
\end{align}
The error on the phase of each mode caused by the fit of $\Delta\phi_{\textrm{match}}^{\ell m}$ is on average of the order of 0.05 rad.

\section{Fits for time difference between modes' amplitude peaks}
\label{app:timeshiftfits}

As originally observed in Refs.~\cite{Barausse:2011kb,Pan:2011gk}, gravitational modes peak 
at different times ($t_{\mathrm{peak}}^{\ell m}$) with respect to 
the dominant $(2,2)$ mode. Using the NR catalog at our disposal, we fit 
the times shifts $\Delta t_{\ell m} \equiv t_{\mathrm{peak}}^{\ell m} -
t_{\mathrm{peak}}^{2 2}$ as function of $\nu$ and $\chi_{\mathrm{eff}} = (m_1 \chi_1 +
m_2 \chi_2)/M$. We find
\begin{align}
\Delta t_{33} =& 4.20646\, +4.215 \chi_{\mathrm{eff}}+2.12487 \chi_{\mathrm{eff}}^2 \nonumber \\
&+(-10.9615+5.20758 a) \nu +(53.3674\, -65.0849 a) \nu ^2\,, \\
\Delta t_{21} =& 12.892\, +1.14433 \chi_{\mathrm{eff}}+1.12146 \chi_{\mathrm{eff}}^2 \nonumber \\
&+\left(-61.1508-96.0301\chi_{\mathrm{eff}} -85.4386 \chi_{\mathrm{eff}}^2\right) \nu \nonumber \\
&+\left(144.497\, +366.374 \chi_{\mathrm{eff}}+322.06 \chi_{\mathrm{eff}}^2\right) \nu ^2\,, \\
\Delta t_{44} =&  7.49641\, +6.7245 \chi_{\mathrm{eff}} +3.11618  \chi_{\mathrm{eff}}^2\nonumber \\
&+\left(-48.5578-78.8077 \chi_{\mathrm{eff}} -92.1608  \chi_{\mathrm{eff}}^2\right) \nu \nonumber \\
& +\left(91.483\, +231.917 \chi_{\mathrm{eff}} +388.074  \chi_{\mathrm{eff}}^2\right) \nu ^2 \,,\\
\Delta t_{55} =& 10.031\, +5.80884 \chi_{\mathrm{eff}}+(-103.252-75.8935 \chi_{\mathrm{eff}}) \nu \nonumber \\
& +(366.57\, +282.552 \chi_{\mathrm{eff}}) \nu ^2\,.
\end{align}
\begin{figure}[h]
  \centering
  \includegraphics[width=0.5\textwidth]{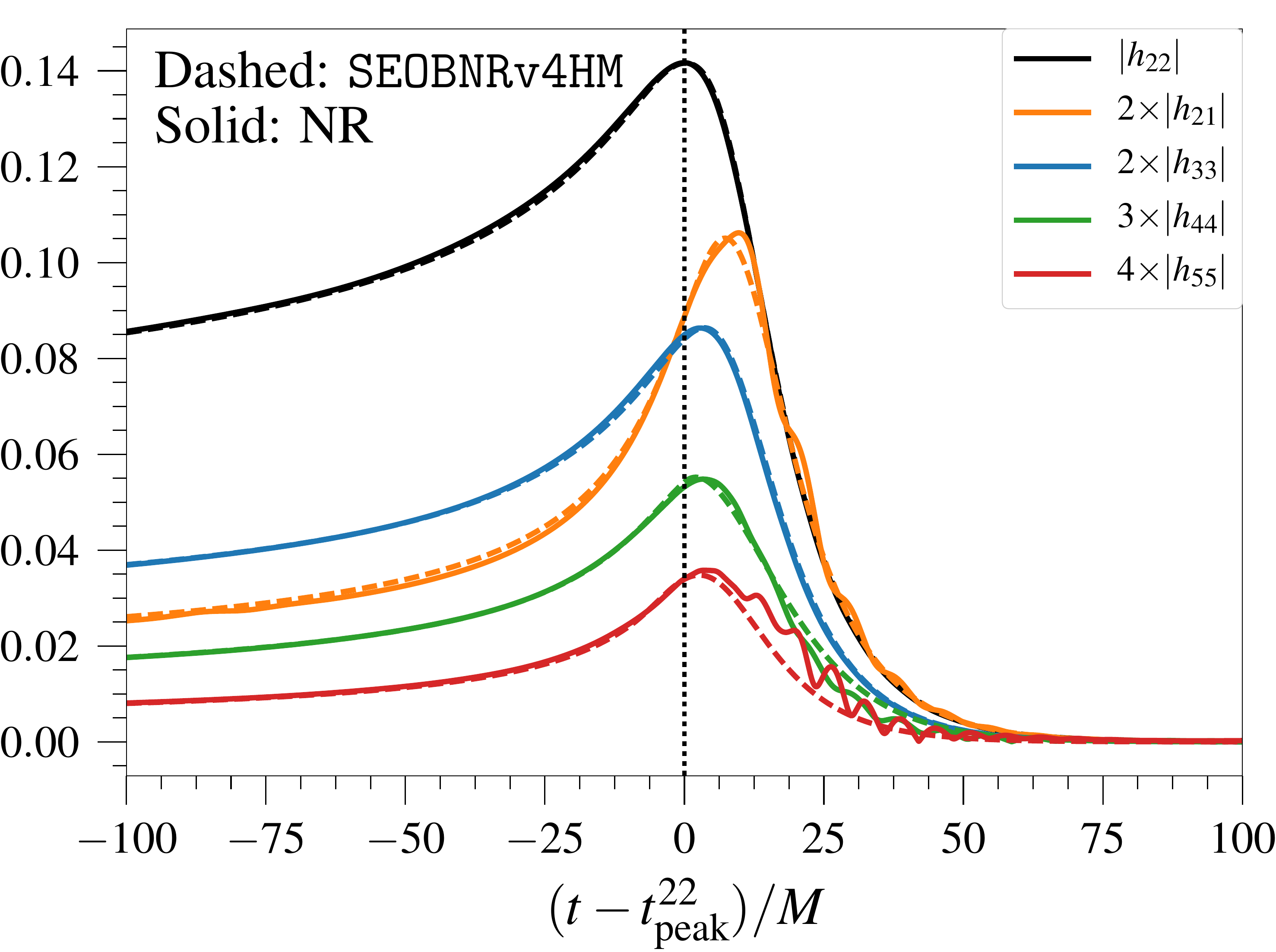}
\caption{Amplitudes of different modes for the \texttt{SEOBNRv4HM} (dashed) and NR (solid) waveforms 
with $(q = 8,\, \chi_1 = -0.5,\, \chi_2 = 0)$ (\texttt{SXS:BBH:0064}) versus time. The time origin corresponds to 
the $(2,2)$ mode's peak.}
\label{fig:timeshift}
\end{figure}

The above expressions could be employed in building phenomenological
  models for the ringdown signal when multipole modes are
  present~\cite{London:2014cma}. We notice that these fits are not
  used for building \texttt{SEOBNRv4HM} waveforms, whose merger-ringdown model 
is constructed through Eqs.~(\ref{eq:merger-RD_wave})--(\ref{eq:ansatz_phase}), 
starting from $t_{\mathrm{match}}^{\ell m}$ in Eq.~(\ref{eq:matchtime}). The 
merger-ringdown \texttt{SEOBNRv4HM} waveforms reproduce the time shifts $\Delta t_{\ell m}$ 
between the NR modes' amplitude peaks by construction, as it can be seen in Fig.~\ref{fig:timeshift} 
for a particular binary configuration.
  
We emphasize that while in the \texttt{EOBNRv2HM} model~\cite{Pan:2011gk} the 
merger-ringdown attachment was done at each modes' peak time, in
  \texttt{SEOBNRv4HM} we do it at the $(2,2)$ mode's peak for all modes 
except the $(5,5)$ mode. We make this change here because 
  typically $\Delta t_{\ell m} = t_{\mathrm{peak}}^{\ell m} - t_{\mathrm{peak}}^{2 2} > 0$, and 
at these late times we find that for some binary configurations either the
EOB dynamics becomes unreliable or the error in the NR waveforms is too large and prevents 
us to accurately extract the input values for the NQC conditions (i.e., Eqs. \eqref{eq:NQC_condition_1}
--\eqref{eq:NQC_condition_5}).

\section{Numerical-relativity catalog}
\label{sec:NRcatalog}

In the tables below we list the binary configurations of the NR
simulations used to build and test the \texttt{SEOBNRv4HM} waveform 
model. The NR waveforms were produced with the (pseudo) Spectral Einstein code (\texttt{SpEC}) 
of the Simulating eXtreme Spacetimes (\texttt{SXS}) project and the \texttt{Einstein Toolkit} (ET) 
code. In particular, we list the mass ratio $q$, the dimensionless
spins $\chi_{1,2}$, the eccentricity $e$, the initial frequency
$\omega_{22}$ of the dominant $(\ell, m) = (2,2)$ mode and the number
of orbits $N_{\mathrm{orb}}$ up to the waveform peak.

\begin{figure}[h]
  \centering
  \includegraphics[width=0.5\textwidth]{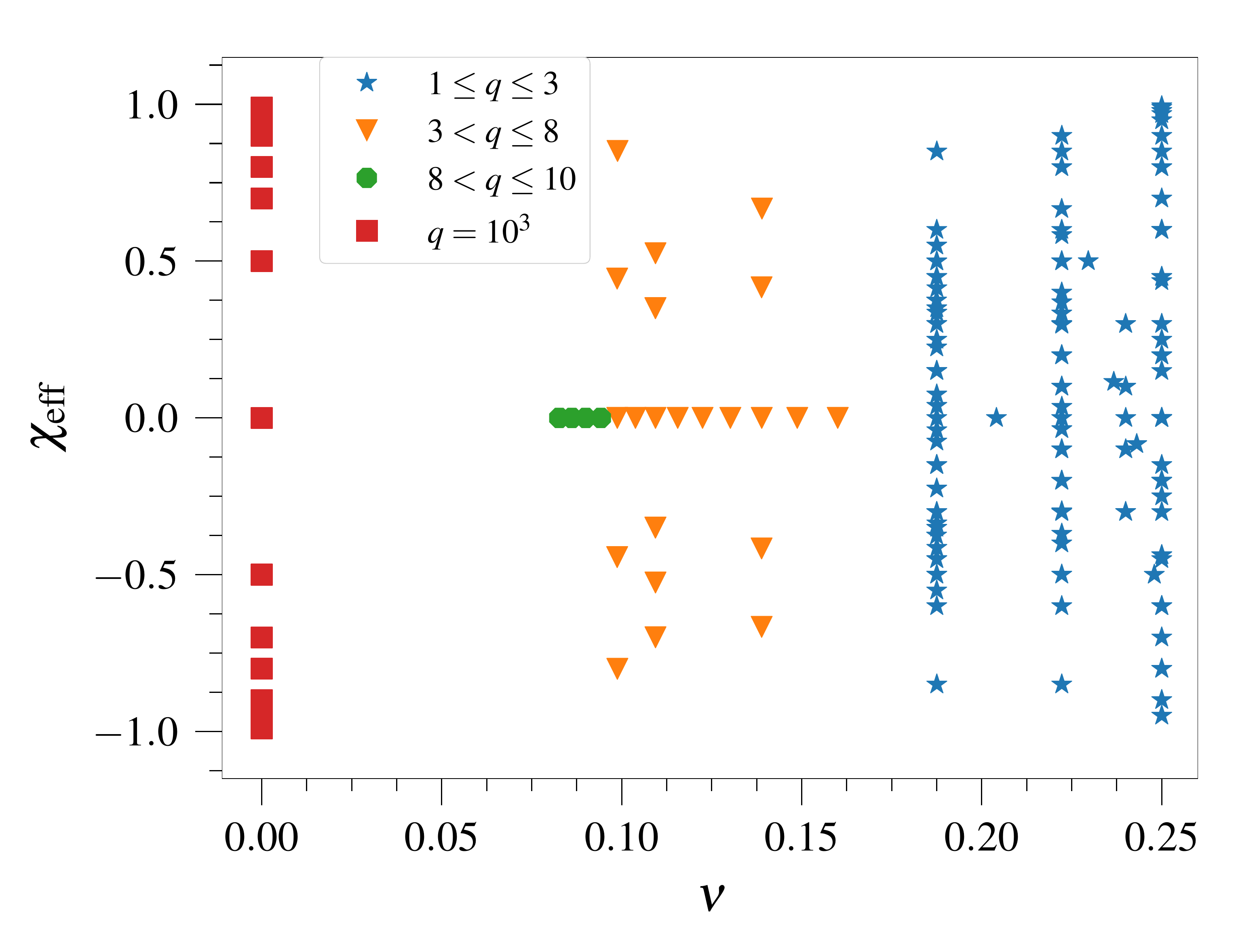}
\caption{2D projection of the 3D parameter space of the NR and BH-perturbation-theory waveforms used to build the \texttt{SEOBNRv4HM} model. 
The $x$-axis is $\nu$ and the $y$-axis is the effective spin $\chi_{\mathrm{eff}} = (\chi_1 m_1 + \chi_2 m_2)/M$. In the legend we 
highlight four different regions of coverage, as discussed in the text.}
\label{fig:NRcatalog}
\end{figure}

In Fig.~\ref{fig:NRcatalog} we show the coverage of NR and BH-perturbation-theory 
waveforms when projected on the binary's parameters $\nu$ and $\chi_{\rm eff}=(\chi_1 m_1 + \chi_2 m_2)/M$. 
We highlight four regions. In the first region $1 \leq q \leq 3$ there is a large number of configurations 
with both BHs carrying spin. The spins magnitude are as high as $\chi_{1,2}
= 0.99$ in the equal-mass limit, while they are limited to $\chi_{1,2}
= 0.85$ for $q = 3$. The second region is between $3 < q \leq 8$, and 
most of the simulations have spins only on the heavier BH. The values
of the spin of the heavier BH span in the region $-0.8 \leq \chi_1
\leq 0.85$. The third region is between $8 < q \leq 10$ and it includes
only nonspinning waveforms.  Finally, the fourth region covers 
13 waveforms computed solving the Teukolsky equation in the framework of BH
pertubation theory~\cite{Barausse:2011kb,Taracchini:2014zpa}. They
have $q = 10^{3}$ and dimensionless spins values in the range $-0.99
\leq \chi \leq 0.99$.

\newcommand{\numoriginal}{38}
\newcommand{\numpublicnew}{9}
\newcommand{\numchu}{94}

\subsection{SXS and ET waveform produced for testing \texttt{SEOBNRv4} (Ref.~\cite{Bohe:2016gbl})}
\label{sec:original}
\begingroup
%\squeezetable
%% \begin{table}[H]
%%\begin{tabular}{lllllll}
 \begin{longtable}{ccccccc}
  \doubleline

ID & $q$ & $\chi_1$ & $\chi_2$ & $e$ & $M\omega_{22}$ & $N_\mathrm{orb}$\\
\hline
SXS:BBH:0610 & $1.2$ & $-0.50$ & $-0.50$ & $7.4\times 10^{-5}$ & $0.01872$ & $12.1$ \\
SXS:BBH:0611 & $1.4$ & $-0.50$ & $+0.50$ & $6.0\times 10^{-4}$ & $0.02033$ & $12.5$ \\
SXS:BBH:0612 & $1.6$ & $+0.50$ & $-0.50$ & $3.7\times 10^{-4}$ & $0.02156$ & $12.8$ \\
SXS:BBH:0613 & $1.8$ & $+0.50$ & $+0.50$ & $1.8\times 10^{-4}$ & $0.02383$ & $13.1$ \\
SXS:BBH:0614 & $2.0$ & $+0.75$ & $-0.50$ & $6.7\times 10^{-4}$ & $0.02355$ & $13.1$ \\
SXS:BBH:0615 & $2.0$ & $+0.75$ & $+0.00$ & $7.0\times 10^{-4}$ & $0.02401$ & $13.3$ \\
SXS:BBH:0616 & $2.0$ & $+0.75$ & $+0.50$ & $8.0\times 10^{-4}$ & $0.02475$ & $13.3$ \\
SXS:BBH:0617 & $2.0$ & $+0.50$ & $+0.75$ & $7.8\times 10^{-4}$ & $0.02342$ & $13.1$ \\
SXS:BBH:0618 & $2.0$ & $+0.80$ & $+0.80$ & $5.9\times 10^{-4}$ & $0.02578$ & $13.4$ \\
SXS:BBH:0620 & $5.0$ & $-0.80$ & $+0.00$ & $3.4\times 10^{-3}$ & $0.02527$ & $8.2$ \\
SXS:BBH:0621 & $7.0$ & $-0.80$ & $+0.00$ & $3.2\times 10^{-3}$ & $0.02784$ & $7.1$ \\
SXS:BBH:0619 & $2.0$ & $+0.90$ & $+0.90$ & $2.9\times 10^{-4}$ & $0.02520$ & $13.5$ \\
ET:AEI:0001 & $5.0$ & $+0.80$ & $+0.00$ & $9.2\times 10^{-4}$ & $0.03077$ & $10.5$ \\
ET:AEI:0002 & $7.0$ & $+0.80$ & $+0.00$ & $6.1\times 10^{-4}$ & $0.03503$ & $10.4$ \\
ET:AEI:0004 & $8.0$ & $+0.85$ & $+0.85$ & $3.0\times 10^{-3}$ & $0.04368$ & $7.4$ \\

  \doubleline
%  \vspace{0.5cm}
  %% \caption{\numoriginal{} public SpEC waveforms used to calibrate the previous EOBNR model}
%\end{tabular}
\end{longtable}
\label{tab:original}
%% \caption{}
%\end{table}
\endgroup

\subsection{SXS waveforms from Ref.~\cite{Mroue:2013xna}}
\label{sec:original}
\begingroup
%\squeezetable
%% \begin{table}[H]
%%\begin{tabular}{lllllll}
 \begin{longtable}{ccccccc}
  \doubleline
ID & $q$ & $\chi_1$ & $\chi_2$ & $e$ & $M\omega_{22}$ & $N_\mathrm{orb}$\\
\hline
%% \endfirsthead
%% ID & $q$ & $\chi_1$ & $\chi_2$ & $e$ & $M\omega$ & $N_\mathrm{orb}$\\
%% \hline
%% \endhead
SXS:BBH:0004 & $1.0$ & $-0.50$ & $+0.00$ & $3.7\times 10^{-4}$ & $0.01151$ & $30.2$ \\
SXS:BBH:0005 & $1.0$ & $+0.50$ & $+0.00$ & $2.5\times 10^{-4}$ & $0.01227$ & $30.2$ \\
SXS:BBH:0007 & $1.5$ & $+0.00$ & $+0.00$ & $4.2\times 10^{-4}$ & $0.01229$ & $29.1$ \\
SXS:BBH:0013 & $1.5$ & $+0.50$ & $+0.00$ & $1.4\times 10^{-4}$ & $0.01444$ & $23.8$ \\
SXS:BBH:0016 & $1.5$ & $-0.50$ & $+0.00$ & $4.2\times 10^{-4}$ & $0.01149$ & $30.7$ \\
SXS:BBH:0019 & $1.5$ & $-0.50$ & $+0.50$ & $7.6\times 10^{-5}$ & $0.01460$ & $20.4$ \\
SXS:BBH:0025 & $1.5$ & $+0.50$ & $-0.50$ & $7.6\times 10^{-5}$ & $0.01456$ & $22.4$ \\
SXS:BBH:0030 & $3.0$ & $+0.00$ & $+0.00$ & $2.0\times 10^{-3}$ & $0.01775$ & $18.2$ \\
SXS:BBH:0036 & $3.0$ & $-0.50$ & $+0.00$ & $5.1\times 10^{-4}$ & $0.01226$ & $31.7$ \\
SXS:BBH:0045 & $3.0$ & $+0.50$ & $-0.50$ & $6.4\times 10^{-4}$ & $0.01748$ & $21.0$ \\
SXS:BBH:0046 & $3.0$ & $-0.50$ & $-0.50$ & $2.6\times 10^{-4}$ & $0.01771$ & $14.4$ \\
SXS:BBH:0047 & $3.0$ & $+0.50$ & $+0.50$ & $4.7\times 10^{-4}$ & $0.01743$ & $22.7$ \\
SXS:BBH:0056 & $5.0$ & $+0.00$ & $+0.00$ & $4.9\times 10^{-4}$ & $0.01589$ & $28.8$ \\
SXS:BBH:0060 & $5.0$ & $-0.50$ & $+0.00$ & $3.4\times 10^{-3}$ & $0.01608$ & $23.2$ \\
SXS:BBH:0061 & $5.0$ & $+0.50$ & $+0.00$ & $4.2\times 10^{-3}$ & $0.01578$ & $34.5$ \\
SXS:BBH:0063 & $8.0$ & $+0.00$ & $+0.00$ & $2.8\times 10^{-4}$ & $0.01938$ & $25.8$ \\
SXS:BBH:0064 & $8.0$ & $-0.50$ & $+0.00$ & $4.9\times 10^{-4}$ & $0.01968$ & $19.2$ \\
SXS:BBH:0065 & $8.0$ & $+0.50$ & $+0.00$ & $3.7\times 10^{-3}$ & $0.01887$ & $34.0$ \\
SXS:BBH:0148 & $1.0$ & $-0.44$ & $-0.44$ & $2.0\times 10^{-5}$ & $0.01634$ & $15.5$ \\
SXS:BBH:0149 & $1.0$ & $-0.20$ & $-0.20$ & $1.8\times 10^{-4}$ & $0.01614$ & $17.1$ \\
SXS:BBH:0150 & $1.0$ & $+0.20$ & $+0.20$ & $2.9\times 10^{-4}$ & $0.01591$ & $19.8$ \\
SXS:BBH:0151 & $1.0$ & $-0.60$ & $-0.60$ & $2.5\times 10^{-4}$ & $0.01575$ & $14.5$ \\
SXS:BBH:0152 & $1.0$ & $+0.60$ & $+0.60$ & $4.3\times 10^{-4}$ & $0.01553$ & $22.6$ \\
SXS:BBH:0153 & $1.0$ & $+0.85$ & $+0.85$ & $8.3\times 10^{-4}$ & $0.01539$ & $24.5$ \\
SXS:BBH:0154 & $1.0$ & $-0.80$ & $-0.80$ & $3.3\times 10^{-4}$ & $0.01605$ & $13.2$ \\
SXS:BBH:0155 & $1.0$ & $+0.80$ & $+0.80$ & $4.7\times 10^{-4}$ & $0.01543$ & $24.1$ \\
SXS:BBH:0156 & $1.0$ & $-0.95$ & $-0.95$ & $5.4\times 10^{-4}$ & $0.01643$ & $12.4$ \\
SXS:BBH:0157 & $1.0$ & $+0.95$ & $+0.95$ & $1.4\times 10^{-4}$ & $0.01535$ & $25.2$ \\
SXS:BBH:0158 & $1.0$ & $+0.97$ & $+0.97$ & $7.9\times 10^{-4}$ & $0.01565$ & $25.3$ \\
SXS:BBH:0159 & $1.0$ & $-0.90$ & $-0.90$ & $5.6\times 10^{-4}$ & $0.01588$ & $12.7$ \\
SXS:BBH:0160 & $1.0$ & $+0.90$ & $+0.90$ & $4.2\times 10^{-4}$ & $0.01538$ & $24.8$ \\
SXS:BBH:0166 & $6.0$ & $+0.00$ & $+0.00$ & $4.4\times 10^{-5}$ & $0.01940$ & $21.6$ \\
SXS:BBH:0167 & $4.0$ & $+0.00$ & $+0.00$ & $9.9\times 10^{-5}$ & $0.02054$ & $15.6$ \\
SXS:BBH:0169 & $2.0$ & $+0.00$ & $+0.00$ & $1.2\times 10^{-4}$ & $0.01799$ & $15.7$ \\
SXS:BBH:0170 & $1.0$ & $+0.44$ & $+0.44$ & $1.3\times 10^{-4}$ & $0.00842$ & $15.5$ \\
SXS:BBH:0172 & $1.0$ & $+0.98$ & $+0.98$ & $7.8\times 10^{-4}$ & $0.01540$ & $25.4$ \\
SXS:BBH:0174 & $3.0$ & $+0.50$ & $+0.00$ & $2.9\times 10^{-4}$ & $0.01337$ & $35.5$ \\
SXS:BBH:0180 & $1.0$ & $+0.00$ & $+0.00$ & $5.1\times 10^{-5}$ & $0.01227$ & $28.2$ \\
  \doubleline
%  \vspace{0.5cm}
  %% \caption{\numoriginal{} public SpEC waveforms used to calibrate the previous EOBNR model}
%\end{tabular}
\end{longtable}
\label{tab:original}
%% \caption{}
%\end{table}
\endgroup

\subsection{SXS waveforms from Ref.~\cite{Kumar:2015tha,Lovelace:2010ne,Scheel:2014ina}}
\label{sec:publicnew}
\begingroup
%\squeezetable
%\begin{table}[H]
%% \begin{tabular}{lllllll}
\begin{longtable}{ccccccc}
  \doubleline
ID & $q$ & $\chi_1$ & $\chi_2$ & $e$ & $M\omega_{22}$ & $N_\mathrm{orb}$\\
\hline
%% \endfirsthead
%% ID & $q$ & $\chi_1$ & $\chi_2$ & $e$ & $M\omega$ & $N_\mathrm{orb}$\\
%% \hline
%% \endhead
SXS:BBH:0177 & $1.0$ & $+0.99$ & $+0.99$ & $1.3\times 10^{-3}$ & $0.01543$ & $25.4$ \\
SXS:BBH:0178 & $1.0$ & $+0.99$ & $+0.99$ & $8.6\times 10^{-4}$ & $0.01570$ & $25.4$ \\
SXS:BBH:0202 & $7.0$ & $+0.60$ & $+0.00$ & $9.0\times 10^{-5}$ & $0.01324$ & $62.1$ \\
SXS:BBH:0203 & $7.0$ & $+0.40$ & $+0.00$ & $1.4\times 10^{-5}$ & $0.01322$ & $58.5$ \\
SXS:BBH:0204 & $7.0$ & $+0.40$ & $+0.00$ & $1.7\times 10^{-4}$ & $0.01044$ & $88.4$ \\
SXS:BBH:0205 & $7.0$ & $-0.40$ & $+0.00$ & $7.0\times 10^{-5}$ & $0.01325$ & $44.9$ \\
SXS:BBH:0206 & $7.0$ & $-0.40$ & $+0.00$ & $1.6\times 10^{-4}$ & $0.01037$ & $73.2$ \\
SXS:BBH:0207 & $7.0$ & $-0.60$ & $+0.00$ & $1.7\times 10^{-4}$ & $0.01423$ & $36.1$ \\
SXS:BBH:0306 & $1.3$ & $+0.96$ & $-0.90$ & $1.5\times 10^{-3}$ & $0.02098$ & $12.6$ \\
  \doubleline
  %% \caption{An additional \numpublicnew{} public SpEC waveforms with either high mass ratio or high spin}
%\end{tabular}
\end{longtable}
  %% \label{tab:publicnew}
  %% \caption{}
%\end{table}
\endgroup

\subsection{SXS waveforms from Ref.~\cite{Chu:2015kft}}
\label{sec:chu}
\begingroup
%\squeezetable
\begin{longtable}{ccccccc}
  \doubleline
ID & $q$ & $\chi_1$ & $\chi_2$ & $e$ & $M\omega_{22}$ & $N_\mathrm{orb}$\\
\hline
\endfirsthead
ID & $q$ & $\chi_1$ & $\chi_2$ & $e$ & $M\omega_{22}$ & $N_\mathrm{orb}$\\
\hline
\endhead
SXS:BBH:0290 & $3.0$ & $+0.60$ & $+0.40$ & $9.0\times 10^{-5}$ & $0.01758$ & $24.2$ \\
SXS:BBH:0291 & $3.0$ & $+0.60$ & $+0.60$ & $5.0\times 10^{-5}$ & $0.01764$ & $24.5$ \\
SXS:BBH:0289 & $3.0$ & $+0.60$ & $+0.00$ & $2.3\times 10^{-4}$ & $0.01711$ & $23.8$ \\
SXS:BBH:0285 & $3.0$ & $+0.40$ & $+0.60$ & $1.6\times 10^{-4}$ & $0.01732$ & $23.8$ \\
SXS:BBH:0261 & $3.0$ & $-0.73$ & $+0.85$ & $1.0\times 10^{-4}$ & $0.01490$ & $21.5$ \\
SXS:BBH:0293 & $3.0$ & $+0.85$ & $+0.85$ & $9.0\times 10^{-5}$ & $0.01813$ & $25.6$ \\
SXS:BBH:0280 & $3.0$ & $+0.27$ & $+0.85$ & $9.7\times 10^{-5}$ & $0.01707$ & $23.6$ \\
SXS:BBH:0257 & $2.0$ & $+0.85$ & $+0.85$ & $1.1\times 10^{-4}$ & $0.01633$ & $24.8$ \\
SXS:BBH:0279 & $3.0$ & $+0.23$ & $-0.85$ & $6.0\times 10^{-5}$ & $0.01629$ & $22.6$ \\
SXS:BBH:0274 & $3.0$ & $-0.23$ & $+0.85$ & $1.6\times 10^{-4}$ & $0.01603$ & $22.4$ \\
SXS:BBH:0258 & $2.0$ & $+0.87$ & $-0.85$ & $1.8\times 10^{-4}$ & $0.01612$ & $22.8$ \\
SXS:BBH:0248 & $2.0$ & $+0.13$ & $+0.85$ & $7.0\times 10^{-5}$ & $0.01552$ & $23.2$ \\
SXS:BBH:0232 & $1.0$ & $+0.90$ & $+0.50$ & $2.8\times 10^{-4}$ & $0.01558$ & $23.9$ \\
SXS:BBH:0229 & $1.0$ & $+0.65$ & $+0.25$ & $3.1\times 10^{-4}$ & $0.01488$ & $23.1$ \\
SXS:BBH:0231 & $1.0$ & $+0.90$ & $+0.00$ & $1.0\times 10^{-4}$ & $0.01487$ & $23.1$ \\
SXS:BBH:0239 & $2.0$ & $-0.37$ & $+0.85$ & $9.1\times 10^{-5}$ & $0.01478$ & $22.2$ \\
SXS:BBH:0252 & $2.0$ & $+0.37$ & $-0.85$ & $3.8\times 10^{-4}$ & $0.01488$ & $22.5$ \\
SXS:BBH:0219 & $1.0$ & $-0.50$ & $+0.90$ & $3.3\times 10^{-4}$ & $0.01484$ & $22.4$ \\
SXS:BBH:0211 & $1.0$ & $-0.90$ & $+0.90$ & $2.6\times 10^{-4}$ & $0.01411$ & $22.3$ \\
SXS:BBH:0233 & $2.0$ & $-0.87$ & $+0.85$ & $6.0\times 10^{-5}$ & $0.01423$ & $22.0$ \\
SXS:BBH:0243 & $2.0$ & $-0.13$ & $-0.85$ & $1.8\times 10^{-4}$ & $0.01378$ & $23.3$ \\
SXS:BBH:0214 & $1.0$ & $-0.62$ & $-0.25$ & $1.9\times 10^{-4}$ & $0.01264$ & $24.4$ \\
SXS:BBH:0209 & $1.0$ & $-0.90$ & $-0.50$ & $1.7\times 10^{-4}$ & $0.01137$ & $27.0$ \\
SXS:BBH:0226 & $1.0$ & $+0.50$ & $-0.90$ & $2.4\times 10^{-4}$ & $0.01340$ & $22.9$ \\
SXS:BBH:0286 & $3.0$ & $+0.50$ & $+0.50$ & $8.0\times 10^{-5}$ & $0.01693$ & $24.1$ \\
SXS:BBH:0253 & $2.0$ & $+0.50$ & $+0.50$ & $6.7\times 10^{-5}$ & $0.01397$ & $28.8$ \\
SXS:BBH:0267 & $3.0$ & $-0.50$ & $-0.50$ & $5.6\times 10^{-5}$ & $0.01410$ & $23.4$ \\
SXS:BBH:0218 & $1.0$ & $-0.50$ & $+0.50$ & $7.8\times 10^{-5}$ & $0.01217$ & $29.1$ \\
SXS:BBH:0238 & $2.0$ & $-0.50$ & $-0.50$ & $6.9\times 10^{-5}$ & $0.01126$ & $32.0$ \\
SXS:BBH:0288 & $3.0$ & $+0.60$ & $-0.40$ & $1.9\times 10^{-4}$ & $0.01729$ & $23.5$ \\
SXS:BBH:0287 & $3.0$ & $+0.60$ & $-0.60$ & $7.0\times 10^{-5}$ & $0.01684$ & $23.5$ \\
SXS:BBH:0283 & $3.0$ & $+0.30$ & $+0.30$ & $7.6\times 10^{-5}$ & $0.01646$ & $23.5$ \\
SXS:BBH:0282 & $3.0$ & $+0.30$ & $+0.00$ & $7.5\times 10^{-5}$ & $0.01629$ & $23.3$ \\
SXS:BBH:0281 & $3.0$ & $+0.30$ & $-0.30$ & $6.7\times 10^{-5}$ & $0.01618$ & $23.2$ \\
SXS:BBH:0277 & $3.0$ & $+0.00$ & $+0.30$ & $7.0\times 10^{-5}$ & $0.01595$ & $22.9$ \\
SXS:BBH:0284 & $3.0$ & $+0.40$ & $-0.60$ & $1.5\times 10^{-4}$ & $0.01656$ & $22.8$ \\
SXS:BBH:0278 & $3.0$ & $+0.00$ & $+0.60$ & $2.1\times 10^{-4}$ & $0.01623$ & $22.8$ \\
SXS:BBH:0256 & $2.0$ & $+0.60$ & $+0.60$ & $7.8\times 10^{-5}$ & $0.01598$ & $23.9$ \\
SXS:BBH:0230 & $1.0$ & $+0.80$ & $+0.80$ & $1.3\times 10^{-4}$ & $0.01542$ & $24.2$ \\
SXS:BBH:0255 & $2.0$ & $+0.60$ & $+0.00$ & $4.0\times 10^{-5}$ & $0.01580$ & $23.3$ \\
SXS:BBH:0276 & $3.0$ & $+0.00$ & $-0.30$ & $6.7\times 10^{-5}$ & $0.01559$ & $23.0$ \\
SXS:BBH:0251 & $2.0$ & $+0.30$ & $+0.30$ & $7.5\times 10^{-5}$ & $0.01514$ & $23.5$ \\
SXS:BBH:0250 & $2.0$ & $+0.30$ & $+0.00$ & $7.5\times 10^{-5}$ & $0.01503$ & $23.2$ \\
SXS:BBH:0271 & $3.0$ & $-0.30$ & $+0.00$ & $6.3\times 10^{-5}$ & $0.01508$ & $22.5$ \\
SXS:BBH:0249 & $2.0$ & $+0.30$ & $-0.30$ & $7.2\times 10^{-5}$ & $0.01478$ & $23.2$ \\
SXS:BBH:0275 & $3.0$ & $+0.00$ & $-0.60$ & $1.2\times 10^{-4}$ & $0.01569$ & $22.6$ \\
SXS:BBH:0254 & $2.0$ & $+0.60$ & $-0.60$ & $6.0\times 10^{-5}$ & $0.01541$ & $22.9$ \\
SXS:BBH:0269 & $3.0$ & $-0.40$ & $+0.60$ & $1.2\times 10^{-4}$ & $0.01563$ & $22.3$ \\
SXS:BBH:0225 & $1.0$ & $+0.40$ & $+0.80$ & $3.5\times 10^{-4}$ & $0.01536$ & $23.5$ \\
SXS:BBH:0270 & $3.0$ & $-0.30$ & $-0.30$ & $6.2\times 10^{-5}$ & $0.01482$ & $22.8$ \\
SXS:BBH:0245 & $2.0$ & $+0.00$ & $-0.30$ & $6.8\times 10^{-5}$ & $0.01441$ & $23.0$ \\
SXS:BBH:0242 & $2.0$ & $-0.30$ & $+0.30$ & $6.7\times 10^{-5}$ & $0.01417$ & $23.1$ \\
SXS:BBH:0223 & $1.0$ & $+0.30$ & $+0.00$ & $6.7\times 10^{-5}$ & $0.01402$ & $23.3$ \\
SXS:BBH:0241 & $2.0$ & $-0.30$ & $+0.00$ & $6.6\times 10^{-5}$ & $0.01394$ & $23.1$ \\
SXS:BBH:0240 & $2.0$ & $-0.30$ & $-0.30$ & $6.4\times 10^{-5}$ & $0.01359$ & $23.5$ \\
SXS:BBH:0222 & $1.0$ & $-0.30$ & $+0.00$ & $7.4\times 10^{-5}$ & $0.01324$ & $23.6$ \\
SXS:BBH:0228 & $1.0$ & $+0.60$ & $+0.60$ & $3.2\times 10^{-4}$ & $0.01543$ & $23.5$ \\
SXS:BBH:0247 & $2.0$ & $+0.00$ & $+0.60$ & $1.0\times 10^{-4}$ & $0.01530$ & $22.6$ \\
SXS:BBH:0263 & $3.0$ & $-0.60$ & $+0.60$ & $1.9\times 10^{-4}$ & $0.01526$ & $22.0$ \\
SXS:BBH:0266 & $3.0$ & $-0.60$ & $+0.40$ & $1.8\times 10^{-4}$ & $0.01488$ & $22.0$ \\
SXS:BBH:0227 & $1.0$ & $+0.60$ & $+0.00$ & $3.1\times 10^{-4}$ & $0.01452$ & $23.1$ \\
SXS:BBH:0221 & $1.0$ & $-0.40$ & $+0.80$ & $2.7\times 10^{-4}$ & $0.01440$ & $22.7$ \\
SXS:BBH:0237 & $2.0$ & $-0.60$ & $+0.60$ & $6.1\times 10^{-5}$ & $0.01433$ & $22.6$ \\
SXS:BBH:0244 & $2.0$ & $+0.00$ & $-0.60$ & $7.5\times 10^{-5}$ & $0.01422$ & $23.2$ \\
SXS:BBH:0217 & $1.0$ & $-0.60$ & $+0.60$ & $1.5\times 10^{-4}$ & $0.01421$ & $22.7$ \\
SXS:BBH:0215 & $1.0$ & $-0.60$ & $-0.60$ & $1.8\times 10^{-4}$ & $0.01189$ & $25.8$ \\
SXS:BBH:0262 & $3.0$ & $-0.60$ & $+0.00$ & $2.0\times 10^{-4}$ & $0.01473$ & $22.5$ \\
SXS:BBH:0213 & $1.0$ & $-0.80$ & $+0.80$ & $1.4\times 10^{-4}$ & $0.01435$ & $22.3$ \\
SXS:BBH:0265 & $3.0$ & $-0.60$ & $-0.40$ & $9.0\times 10^{-5}$ & $0.01422$ & $23.4$ \\
SXS:BBH:0264 & $3.0$ & $-0.60$ & $-0.60$ & $2.8\times 10^{-4}$ & $0.01410$ & $23.4$ \\
SXS:BBH:0224 & $1.0$ & $+0.40$ & $-0.80$ & $2.5\times 10^{-4}$ & $0.01361$ & $22.9$ \\
SXS:BBH:0236 & $2.0$ & $-0.60$ & $+0.00$ & $1.2\times 10^{-4}$ & $0.01361$ & $23.4$ \\
SXS:BBH:0216 & $1.0$ & $-0.60$ & $+0.00$ & $2.6\times 10^{-4}$ & $0.01300$ & $23.6$ \\
SXS:BBH:0235 & $2.0$ & $-0.60$ & $-0.60$ & $6.1\times 10^{-5}$ & $0.01274$ & $25.1$ \\
SXS:BBH:0220 & $1.0$ & $-0.40$ & $-0.80$ & $1.0\times 10^{-4}$ & $0.01195$ & $25.7$ \\
SXS:BBH:0212 & $1.0$ & $-0.80$ & $-0.80$ & $2.4\times 10^{-4}$ & $0.01087$ & $28.6$ \\
SXS:BBH:0303 & $10.0$ & $+0.00$ & $+0.00$ & $5.1\times 10^{-5}$ & $0.02395$ & $19.3$ \\
SXS:BBH:0300 & $8.5$ & $+0.00$ & $+0.00$ & $5.7\times 10^{-5}$ & $0.02311$ & $18.7$ \\
SXS:BBH:0299 & $7.5$ & $+0.00$ & $+0.00$ & $5.9\times 10^{-5}$ & $0.02152$ & $20.1$ \\
SXS:BBH:0298 & $7.0$ & $+0.00$ & $+0.00$ & $6.1\times 10^{-5}$ & $0.02130$ & $19.7$ \\
SXS:BBH:0297 & $6.5$ & $+0.00$ & $+0.00$ & $6.4\times 10^{-5}$ & $0.02082$ & $19.7$ \\
SXS:BBH:0296 & $5.5$ & $+0.00$ & $+0.00$ & $5.2\times 10^{-5}$ & $0.01668$ & $27.9$ \\
SXS:BBH:0295 & $4.5$ & $+0.00$ & $+0.00$ & $5.2\times 10^{-5}$ & $0.01577$ & $27.8$ \\
SXS:BBH:0259 & $2.5$ & $+0.00$ & $+0.00$ & $5.9\times 10^{-5}$ & $0.01346$ & $28.6$ \\
SXS:BBH:0292 & $3.0$ & $+0.73$ & $-0.85$ & $1.8\times 10^{-4}$ & $0.01749$ & $23.9$ \\
SXS:BBH:0268 & $3.0$ & $-0.40$ & $-0.60$ & $1.7\times 10^{-4}$ & $0.01473$ & $22.9$ \\
SXS:BBH:0234 & $2.0$ & $-0.85$ & $-0.85$ & $1.4\times 10^{-4}$ & $0.01147$ & $27.8$ \\
SXS:BBH:0273 & $3.0$ & $-0.27$ & $-0.85$ & $2.0\times 10^{-4}$ & $0.01487$ & $22.9$ \\
SXS:BBH:0210 & $1.0$ & $-0.90$ & $+0.00$ & $1.8\times 10^{-4}$ & $0.01248$ & $24.3$ \\
SXS:BBH:0260 & $3.0$ & $-0.85$ & $-0.85$ & $3.5\times 10^{-4}$ & $0.01285$ & $25.8$ \\
SXS:BBH:0302 & $9.5$ & $+0.00$ & $+0.00$ & $6.0\times 10^{-5}$ & $0.02366$ & $19.1$ \\
SXS:BBH:0301 & $9.0$ & $+0.00$ & $+0.00$ & $5.5\times 10^{-5}$ & $0.02338$ & $18.9$ \\
SXS:BBH:0272 & $3.0$ & $-0.30$ & $+0.30$ & $6.4\times 10^{-5}$ & $0.01521$ & $22.7$ \\
SXS:BBH:0246 & $2.0$ & $+0.00$ & $+0.30$ & $7.2\times 10^{-5}$ & $0.01514$ & $22.9$ \\
  \doubleline
  %% \caption{\numchu{} SpEC waveforms with moderate mass ratios and spins}
  %% \caption{}
\end{longtable}
\endgroup

\section{Comparing the nonspinning \texttt{SEOBNRv4HM} and \texttt{EOBNRv2HM} models}
\label{sec:EOBNRv2HM}

\begin{figure}[htb]
\centering
\includegraphics[width=0.5\textwidth]{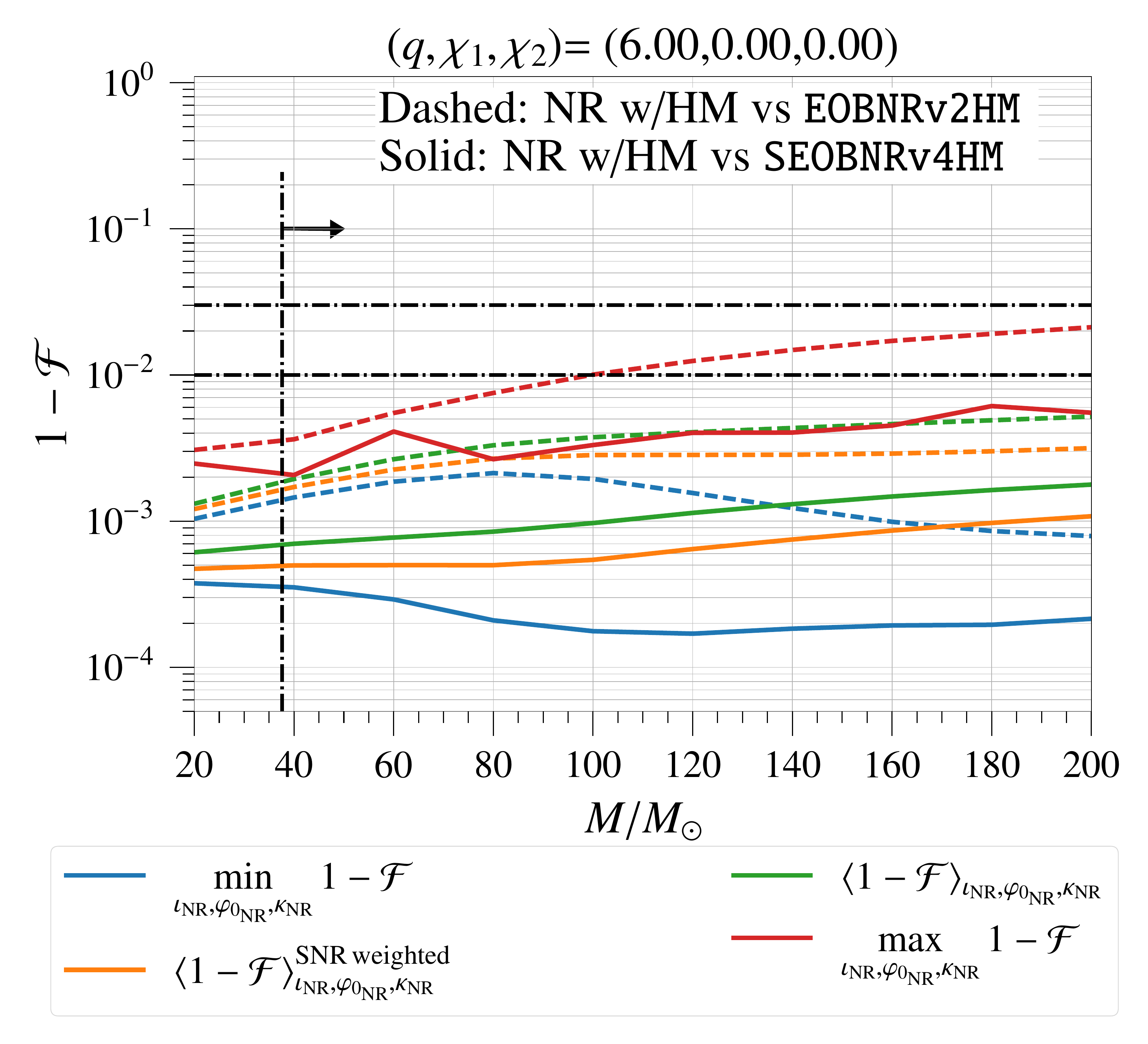}
\caption{Unfaithfulness $(1-\mathcal{F})$ in the mass range $20 M_\odot \leq M \leq 200 M_\odot$ for the configuration $(q = 6,\, \chi_1 = \chi_2 = 0)$. Dashed (plain) curves refer to results for \texttt{EOBNRv2HM} (\texttt{SEOBNRv4HM}). Plotted data as in Fig.~\ref{fig:unfaith_mass_q3chi085}.}
\label{fig:unfaith_mass_q6}
\end{figure}

Here we compare the nonspinning limit of \texttt{SEOBNRv4HM} to its predecessor,  
the \texttt{EOBNRv2HM} model developed in 2011~\cite{Pan:2011gk}, which is available 
in the LIGO Algorithm Library (LAL) and it has been used in Refs.~\cite{Capano:2013raa,Graff:2015bba,Harry:2016ijz}
to assess the importance of higher-order modes in Advanced LIGO searches and parameter estimation. The model \texttt{EOBNRv2HM} was also used 
to search for intermediate binary black holes~\cite{Aasi:2014iwa,Aasi:2014bqj,Aasi:2012rja,Virgo:2012aa}). The \texttt{EOBNRv2HM} model
includes the same higher-order modes as \texttt{SEOBNRv4HM}, that is 
$(2,2),(2,1),(3,3),(4,4),(5,5)$. Given that the
\texttt{EOBNRv2HM} model was calibrated against NR waveforms up to mass
ratio $q = 6$, we decide to compare first the two models for a configuration
with this mass ratio (\texttt{SXS:BBH:0166}). In
Fig.~\ref{fig:unfaith_mass_q6} we show the unfaithfulness results 
for maximum, minimum, average and SNR-weighted average with respect to the
angles $\iota_{\textrm{NR}},{\varphi_0}_{\textrm{NR}},\kappa_{\textrm{NR}}$ 
of the models against NR waveforms with the
modes $(\ell \leq 5,\, m\neq 0)$. The unfaithfulness is shown as a
function of total mass. The dashed (solid) lines
represent the results for \texttt{EOBNRv2HM} (\texttt{SEOBNRv4HM}).
The minimum of the unfaithfulness, reached for a face-on orientation,
is different for the two models and it is smaller for the
\texttt{SEOBNRv4HM} model. Since, for a face-on orientation, all the
higher-order modes included in the two models are exactly zero because of
the spherical harmonics, this difference is only due to a better
modeling of the dominant $(\ell, m) = (2,2)$ mode. This difference is
very small and both models yield a minimum of the unfaithfulness
much smaller than $1\%$ in the total mass range $20 M_\odot \leq M
\leq 200 M_\odot$. The most important quantity to compare is the
maximum of the unfaithfulness which is reached for an edge-on
orientation, where the higher-order modes are more relevant. Also in
this case the \texttt{SEOBNRv4HM} model has a lower unfaithfulness
against the NR waveform with respect to the \texttt{EOBNRv2HM}
model. In particular at a total mass of $M = 200 M_\odot$
\texttt{EOBNRv2HM} returns a maximum unfaithfulness $(1-\mathcal{F})
\sim 2\%$, while the \texttt{SEOBNRv4HM} model only $(1-\mathcal{F})
\sim 0.6\%$. This means that also the higher-order modes are better
modeled in \texttt{SEOBNRv4HM} with respect to \texttt{EOBNRv2HM}. 

We find that the model \texttt{SEOBNRv4HM} returns smaller values of the
unfaithfulness against the NR waveforms than the \texttt{EOBNRv2HM}
model for every nonspinning configuration in our NR catalog with $q \leq
6$. A comparison between the two models for mass ratio higher than $q
= 6$ is unfair because \texttt{EOBNRv2HM} is not calibrated in this
region. However it is worth mentioning that for the numerical
simulation with the largest mass ratio at our disposal $(q = 10)$ the
average unfaithfulness of \texttt{EOBNRv2HM} is larger than that of
\texttt{SEOBNRv4HM}, but still smaller than $1\%$ in the mass range
considered. For this configuration the value of the maximum of the
unfaithfulness is $(1-\mathcal{F}) \sim 3.5\%$ for \texttt{EOBNRv2HM}
at $M = 200 M_\odot$, while is $(1-\mathcal{F}) \sim 2\%$ for
\texttt{SEOBNRv4HM}.

\section{Comparing \texttt{SEOBNRv4HM} and numerical-relativity waveforms in time domain}
\label{sec:time_domain}

\begin{figure*}
  \centering
\includegraphics[scale=0.17]{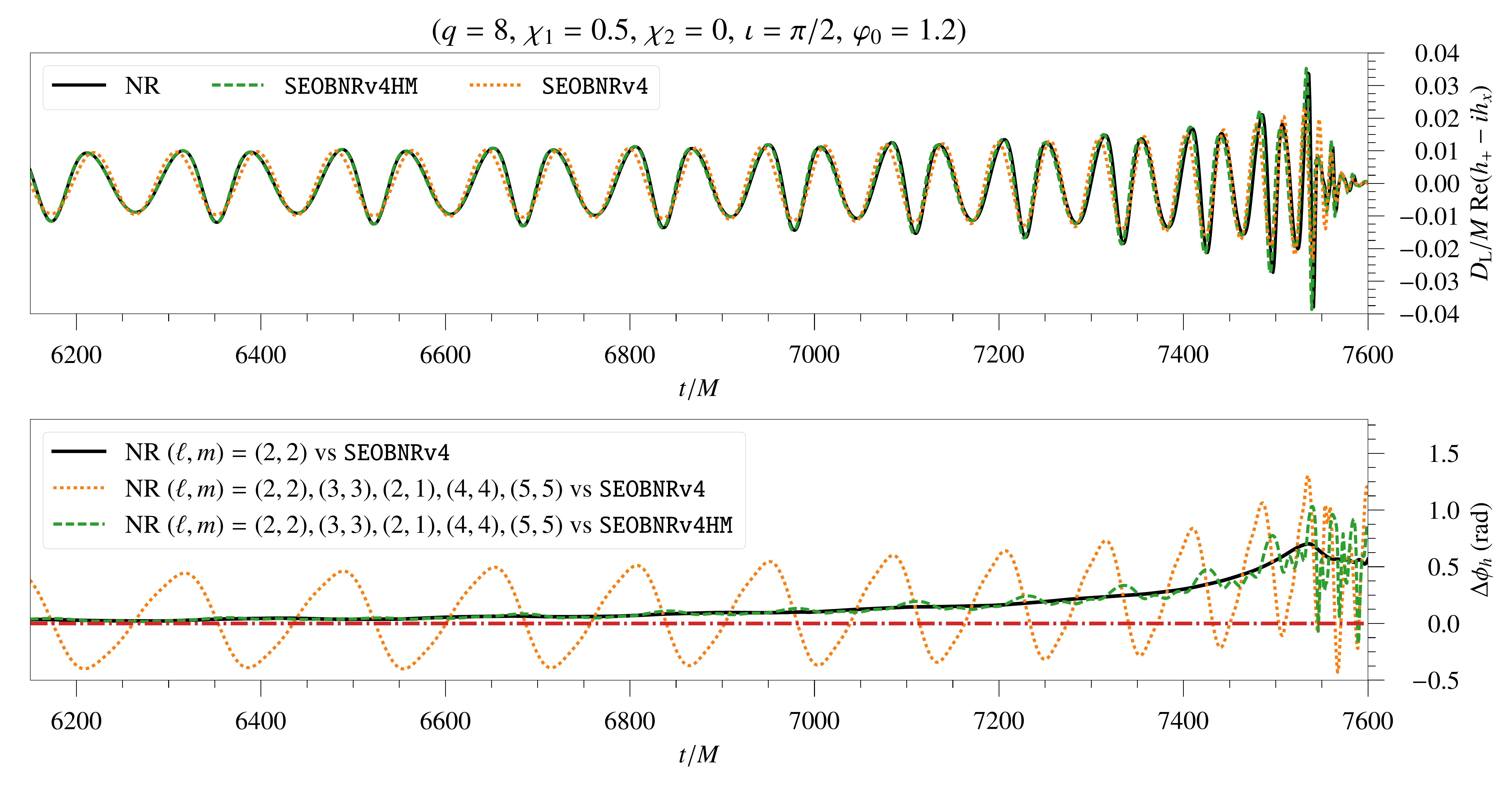}
\caption{Comparison between NR (solid black), \texttt{SEOBNRv4HM} (dashed green) and \texttt{SEOBNRv4} (dotted yellow) waveforms in an edge-on orientation $(\iota = \pi/2, \varphi_0 = 1.2)$ for the NR simulation \texttt{SXS:BBH:0065} $(q = 8,\, \chi_1 = 0.5,\, \chi_2 = 0)$. In the top panel is plotted the real part of the observer-frame's gravitational strain $h_+(\iota,\varphi_0;t) - i \ h_x(\iota,\varphi_0;t)$, while in the bottom panel the dephasing with the NR waveform $\Delta\phi_h$.The dotted-dashed red horizontal line in the bottom panel indicates zero dephasing with the NR waveform. Both \texttt{SEOBNRv4} and \texttt{SEOBNRv4HM} waveforms are phase aligned and time shifted at low frequency using as alignment window $t_{ini} = 1000 M$ and $t_{fin} = 3000 M$.}
\label{fig:q8v4vsHMwaveedgeon}
\end{figure*}

The improvement in waveform modeling obtained by including higher-order modes, 
can also be seen from a direct comparison of NR waveforms to \texttt{SEOBNRv4} and \texttt{SEOBNRv4HM} 
waveforms in time domain. We present this comparison in Fig.~\ref{fig:q8v4vsHMwaveedgeon} for the simulation
\texttt{SXS:BBH:0065}. We show the NR waveform with $(2,2),(2,1),(3,3),(4,4),(5,5)$ modes (solid black), the \texttt{SEOBNRv4HM} 
(dashed green) and \texttt{SEOBNRv4} (dotted yellow) waveforms in an edge-on 
orientation. The effect of neglecting higher-order modes results in an
oscillatory phase difference (dotted yellow curve of the bottom panel in Fig.~\ref{fig:q8v4vsHMwaveedgeon}) around the mean dephasing due to
the dominant $(2,2)$ mode (solid black curve of the same
panel). These oscillations in the dephasing are almost totally removed
up to merger when we include higher-order modes (dashed green of the
bottom panel in Fig.~\ref{fig:q8v4vsHMwaveedgeon}) where now the phase
difference with the NR waveform is dominated again by the discrepancy of the
$(2,2)$ mode. The residual oscillations of the dashed green curve
around the dephasing of the dominant $(2,2)$ mode is due
to the superposition of the different dephasing of the various
higher-order modes. The effect of the inclusion of higher-order modes
can be seen also in the amplitude of the waveform, in particular in
the last five cycle of the waveform there is an evident amplitude
difference between \texttt{SEOBNRv4} and NR waveforms, which is not present 
when the \texttt{SEOBNRv4HM} waveform is used.

\FloatBarrier

\bibliography{paper} 
\end{document}